\documentclass[11pt,a4paper,twoside,openright,titlepage,
]{report
}

\usepackage[signatures,norules]{frontespizio}
\usepackage[utf8]{inputenc}
\usepackage{amsmath}
\usepackage{amsfonts}
\usepackage{amssymb}
\usepackage[retainorgcmds]{IEEEtrantools}
\usepackage{verbatim}
\usepackage{hyperref}
\usepackage{feynmf}
\usepackage{makeidx} 
\usepackage{bbm}

\newcommand{\diff}[1]{\mathrm{d} #1}
\newcommand{\ddiff}[2]{\mathrm{d}^{#1} #2}
\newcommand{\bra}[1]{\langle #1 |}
\newcommand{\ket}[1]{| #1 \rangle}
\newcommand{\pardev}[2]{\frac{\partial #1}{\partial #2}}
\newcommand{\lrdev}[1]{\overset{\leftrightarrow}{\partial_{#1}}}
\DeclareMathAlphabet{\mathbbmsl}{U}{bbm}{m}{sl}
\newcommand{\transverse}[1]{\mathbf{#1}}

\DeclareMathOperator{\Res}{Res}
\DeclareMathOperator{\arcsinh}{arcsinh}
\renewcommand{\Im}[1]{\mathfrak{Im}(#1)}
\renewcommand{\Re}[1]{
\mathfrak{Re}(#1)
}

\title{Campi\hspace{15pt}gravitazionali\hspace{15pt}efficaci\\ in urti oltre la scala di Planck\\ \vspace{1cm} \textbf{Effective\hspace{10pt}gravitational\hspace{10pt}fields\\ in transplanckian scattering}}
\date{draft version 3.5---\today}

\begin{document}

\begin{titlepage}

\begin{tabular} {ll}

\begin{minipage}{.48\textwidth}
\vspace*{12pt}
\includegraphics[height=0.15\textheight]{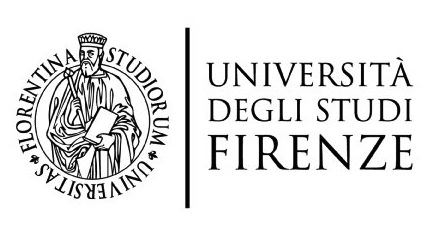}
\end{minipage}

\begin{minipage}{.42\textwidth}
\textnormal{\begin{flushleft}
 Scuola di Scienze Matematiche,
 Fisiche e Naturali\\ 
 \vspace*{12pt}
 Corso di Laurea 
 in
 Fisica
 \end{flushleft}}
\end{minipage}

\end{tabular}

\vspace*{3cm}

\begin{center}
{
\textbf{\huge Campi gravitazionali efficaci in urti oltre la scala di Planck}}

\vspace*{1cm}

{
\textbf{\huge Effective gravitational fields in transplanckian scattering}}
\end{center}

\vspace*{1cm}


\vspace*{1cm}

\begin{flushleft}
{
\textbf{Relatore:}\\
Dott. Dimitri Colferai\\
}

\vspace{1cm}



{
\textbf{Candidato:}\\
Luca Simone Giovanni Betti \\
}

\end{flushleft}

\vspace*{2cm}

\begin{center}
Anno Accademico 2013--2014
\end{center}
\end{titlepage}

\tableofcontents
\setlength{\baselineskip}{1.1\baselineskip}
\unitlength = 3pt 

\chapter*{Introduction}
\addcontentsline{toc}{chapter}{Introduction}


Gravitation and its related issues represent a major open problem in contemporary physics.
Gravity has been the first force to be successfully investigated  in  terms of modern science, right at its birth in 17\textsuperscript{th} century, thanks primarily  to Kepler's observations and Newton's theory of universal gravitation. As we know, that was an effective theory that was to be amended only in 1915 by Einstein's General Relativity (GR), which at practical level corrects Newton's law in extreme (by a human's point of view) situations. A remarkable point is that Einstein's theory is completely classic, as opposed to quantum, and also somewhat interprets gravitational effects in a radically different way than those due to other fundamental forces, giving them a geometric interpretation that other interactions lack. A century later, and three centuries after opening the road towards modern science, gravity is somehow 
 the least understood interaction, and the last one to remain to be successfully quantized.

		\subsection*{Quantum Gravity}
		
		At the fundamental level, the problem with gravity is the lack of a quantum theory that successfully describes it. 
Gravitation does not fit well in the Standard Model  (SM) of fundamental interactions  context.
The SM successfully describes three of the four fundamental interactions, that is electromagnetic, weak and strong forces, substantially with the introduction of exchanged particles that represent  interaction quanta: photons, $W^+$, $W^-$ and $Z^0$ boson vectors, and gluons. The so--called ``gravitons'' 
 that would quantize gravity seem to be elusive instead.
	
	Difficulties encountered in the path to a satisfactory Quantum Gravity (QG) theory can be ascribed to a number of intervening factors.
	
	One of the problems is the lack of experimental observation of quantum gravity effects that are expected at energies of Planck's mass scale, that is, around $10^{19}~GeV$. That is a fundamental obstacle in pursuing a valid theory with the galilean scientific method. Unfortunately, those energies not only are not accessible today even by the most advanced machines, but will presumably remain so  even in the far future, unless some unexpected very big leap in technology would  occur soon during the course of scientific progress.
	Thus, to conduct sensible investigations on gravity, scientists must find some other kind of clues. But, issues with gravitation theories are more fundamental than a lack of observations, as we explain in the following.
	A natural approach is trying to bend gravitation under the language of Quantum Field Theory (QFT). But, as we already noted, gravity has not been successfully described in terms of quantum fields yet. That actually represents the main problem: there are candidate theories for other interactions, which could be tested at higher energies. Gravity instead lacks even one of such candidates, that is, a theory which is consistent with GR and QFT.
		 However it can be argued that, analogously with other forces, gravity could be described in terms of field quanta, the already introduced particles labeled ``gravitons''.
		 
	That kind of speculations can pass from a philosophical status to scientific methodology when confronted with \emph{falsifiability}. That is a concept formalized by science philosopher K.~Popper \cite{pop} which can be briefly resumed here as follows: a scientific theory must be \emph{falsifiable}, that is, admit at least one experiment that could prove it wrong. We can add that a valid theory must yield predictions which are coherent with other theories which are considered valid themselves. In other words, even without the possibility of checking a theory's predictions with experimental observations, if we find that two theories lead to predictions which contradict each other, than we conclude that at least one of such two theories is wrong.
	In the case of gravity, Einstein's general relativity is obviously the theory of reference. Even if it is purely classical and thus lacks predictions on quantum effects, GR should represent the classical limit of a valid quantum theory of gravity. We can say we are just invoking nothing short of the correspondence principle introduced among quantum mechanics postulates by Niels Bohr \cite{sak}.
	
	The lack of experimental data has given space to a flourishing development of 
speculative theories. Different approaches have been pursued and are nowadays discussed; in these very years we have a rather complex situation given by a richness of alternative detailed theories. However we could very roughly identify two main branches among quantum gravity theory candidates: String Theory (ST) and Loop Quantum Gravity (LQG) \cite{ori}.
They both have appealing aspects and share a number of defenders and detractors, on the basis of different virtues and shortcomings whose exposition is beyond the scope of this work; we just resume some basilar aspects to give the reader and idea of their main differences.

\pagebreak

String theory has a perturbative approach and includes gravitons; it also aspires to be a ``theory of everything'', that is, a fundamental theory capable of supplying a holistic explanation of all kind of matter and interactions. LQG is strictly a theory of gravity, it has a non--perturbative approach and does not include gravitons. In fact, instead of quantizing  fields, it quantizes geometric space, supplying discrete operators for areas and volumes \cite{ori,rov}; curvature itself can be discretized in terms of Regge calculus \cite{reg,b-g}. That approach can appear natural considering the fact than in  GR space--time curvature is considered to assume the role of the gravity field; on the other hand in perturbative gravitation, such as in ST and in our work, gravitons and field quanta $h_{\mu\nu}$ in general are associated to an expansion of the metric tensor as $g_{\mu\nu}=\eta_{\mu\nu}+h_{\mu\nu}$, where $\eta_{\mu\nu}$ represent unperturbed flat space--time. Incidentally, GR's peculiar treatment of gravity as a geometric effect, not shared by any other force, suggests gravitation is a fundamentally different interaction from the other ones. That difference could be the basilar motivation of our inability to include it in the standard model.

Both theories introduce a minimal observable length, which in LQG is related to space quantization and in ST is the string length (see Section \ref{sectionacv}). That is related to the \emph{Planck's length}
\[
\lambda_P = \frac{\hbar}{m_P c}\approx 10^{-20}~fm
\]
which represents a classical minimal observable length. Simplifying, the motivation can be justified as follows. Trying to observe a smaller length would require a photon whose energy is greater than $m_P$, but that could cause a classical collapse in a black hole of the region intended to be observed. The consequent event horizon would casually disconnect the inner physics from the outer observer. That makes \emph{de facto} the Universe unobservable at lengths smaller than $\lambda_P$.

There are other technical problems to be faced when attempting to quantize gravity.

First of all, the coupling $G$ has dimensions $E^{-2}$ with respect to energy $E$, and that makes the theory not  renormalizable 
by power counting.

Then, there are issues concerning information
.
One of those issues concerns the information loss related to black holes. When a physical entity crosses an event horizon, it is no longer observable; however it is thought of as still belonging to the Universe, along with the informations it carries
. But, it is predicted \cite{haw} that black holes could evaporate and consequently destroy those informations.

\pagebreak

Lastly, general relativity as expressed by Einstein's field equations
\[
G_{\mu\nu} = 8\pi G~T_{\mu\nu}
\]
is highly non linear. That is a technical difficulty due to the fact that while stress--energy tensor $T_{\mu\nu}$ contains source terms, Einstein's tensor $G_{\mu\nu}$ contains the fields and also, in a non--separable way, energy terms which take the role of sources themselves. 
That is, it is not possible to  separately give energy and masses as initial data and then compute the fields as a solution; that can be done 
 linearizing the theory, which has the physical meaning of considering weak field perturbations $h_{\mu\nu}$ with respect to a flat Minkowskian space--time $\eta_{\mu\nu}$. 
Such a non--linearity has a rather simple physical interpretation: gravity couples masses and every kind of energy, including itself. In other words, gravitons transport, and also generate and are subject to gravitational interaction. It can be argued, that is not a too demanding issue because also gluons, the strong interaction quanta described by Quantum Chromo--Dynamics (QCD) are ``colored'', that is, charged with respect to the interaction they mediate; however in the case of gravitation the bigger issue is that \emph{every} kind of matter and radiation has a non vanishing mass or energy, and thus is charged with respect to the interaction.

	\section*{Outline of this work}
	
	For the reasons explained above, several studies are oriented to the analysis of gravitational processes in which quantum effects are influential,  but can be treated in semiclassical terms. Our aim is to introduce one of those approaches and derive a result to humbly contribute to those studies.
	In this work we  analyze a semiclassical model for high energy gravitational interactions and compute the expectation value of an effective gravitational field using the model's prescriptions.

We recall that a semiclassical model describes 
a system in terms of quantum mechanics (QM), in the approximation of ``high'' quantum numbers, \emph{i.e.} small wave length, which can let some of the classical concepts, like that of trajectory, to be recovered.

The model itself is an $S$--matrix description due to Amati, Ciafaloni and Veneziano (ACV in the following) \cite{acv87,acv}  for the gravitational interaction. In standard quantum scattering theory, the $S$--matrix expresses the relation between initial and final states, in a manner we will recall and detail later.

By the expression ``high energy'' throughout  this work we mean energies $E=\sqrt{s}$, measured in the  center--of--mass of the elastic scattering of two particles,  which are of order, or higher, than Planck's mass\footnote{When otherwise not stated, we use the convention $\hbar=c=G=1$. Sometimes, like in this case, we will write the constants explicitly, if useful.} \[m_P = \sqrt{\hbar c /G}\approx 10^{19}~GeV\lesssim \sqrt{s}\qquad.\]
For the most part, in this work we will actually consider transplanckian energies which, strictly speaking, are much higher than $m_P$: $E \gg m_P$.


We also would like to recall that the term ``effective'' refers to a theory or model that describes some phenomena in a physical system under some implicit or explicit limits of validity, without necessarily introducing a detailed underlying model of the effects that produce the phenomena's dynamics.
While standard field theory can be considered the effective limit of some underlying and yet to be discovered theory of everything, in the case of gravity an effective description is particularly necessary,  given the intrinsic quantization issues.

	As stated above, in our derivation we choose as a starting point the semiclassical approach to gravity due to ACV. Their S-matrix description for gravitational interactions between particles is in turn based on string theory. 
The model can however be derived from other starting points. For example,  Lipatov \cite{lip} derived the reduced action (\ref{acvaction}) independently from ST and Kirschner and Szymanowsky \cite{k-s} more recently were able to reproduce the resulting effective description starting from Einstein's equations in high--energies regime
.

ACV studied in terms of the $S$--matrix the possibility of a gravitation collapse from high-energy scattering
.
That is a key issue, given that gravitational collapse represents another fundamental incongruence between quantum mechanics and general relativity, which predicts a classical curvature singularity. On the QM front, the collapse seems to be related to a lack in unitarity of the $S$--matrix \cite{fal}.

Constructing an $S$--matrix for a particular interaction requires to know the effect of that interactions 
on plane wave states. That is because those states form a basis for the space of ingoing and outgoing states, which we would like to link to the $S$-matrix. 
A simple case is a derivation due to 't~Hooft \cite{tho}, which describes the  scattering of a plane wave and a classical gravitational shock wave. The shock wave's description is due to Aichelburg and Sexl \cite{a-s} (AS thereafter), which derived it as the metric produced by a source moving at the speed of light.
On the other hand, the $S$--matrix for the scattering of two quantum particles in transplanckian collisions has been derived by ACV. 

We sketch here the logical path we follow in this work. 

In Chapter \ref{chapterclassical}, we derive AS' result for the metric of a single source moving at speed $c$. That metric has the peculiar form of a plane shock wave moving with the particle which generates it. The functional profile of the metric w.r.t.\ transverse directions can be recovered in a one of the quantum fields in ACV's approach. We illustrate the behavior of geodesics crossing the shock, which are subject to a shift in space and time.

We also resume two other results. The first one approximates the metric of the scattering of two null particles moving almost at the speed of light. The second one concerns the metric of two colliding infinite planes of matter.

In Chapter \ref{chaptersemiclassical}, we describe 't Hooft's result for the collision of a plane wave on an AS--like wave--front and  generalize it to wave packets.
We then resume ACV's approach, particularly for eikonal graviton exchanges, and introduce an analytical  result relative to the axisymmetric scattering of a particle and a ring of matter. We then introduce a critical impact parameter below which quantum effects are suspected, and that can play a role in gravitational collapse.

In Chapter \ref{chapterresult}, we show how to compute gravitational scattering Feynman diagrams in general, then we detail the computation of the eikonal exchange of zero and one graviton.
Then, we present the main result we achieved. We calculate the expectation value of the gravitational field related to an infinite exchange of gravitons with finite total transferred momentum. We find a result that we interpret in terms of perturbation theory as the resummation of eikonal diagrams with $n$ gravitons exchange. Remarkably, our result reproduce the AS classical shifts.
Then, we also suggest  the possible effects of considering some approximations' relaxation in our computation.

In Chapter \ref{chapterconclusions}, we discuss our result and remark many important aspects of the studies we analyzed. At the end, we suggest a few possible additional investigations. 



	\section*{Experimental horizons}
	
	We would like to cite some experimental researches in gravitation, including possible developments.
	As already stated, phenomenological interest for high--energy scattering and gravitational quantum effects, including a quantum version of gravitational collapse, seems confined to the far future. That is because the order of magnitude of the Plank scale, at which those effects are expected to take place, is too high for the state--of--the--art technology.
	However, extra--dimensional theories can admit 
	some quantum effects in the few--$TeV$ range 
\cite{dim}. Were they correctly predictive, the experimental interest for some measurable QG effect could be nearer and would rise with the energy accessible by last--generation machines.

In any case, gravitation experiments at lower scales can still  supply useful clues. For example, predicted graviton properties are related to the general behavior of gravity: the $r^{-2}$ dependence on distance in Newton's law is related to mediators with predicted zero mass; the attractive behavior is related to even--spin exchanged particles, and the prediction of a spin $2$ for gravitons. The two states of polarizations of gravitational waves predicted by GR can be investigated by observation. Experiments like VIRGO \cite{vyr}, to cite just one, are developed to reveal and study those waves.

Despite being difficult to reveal, given its low magnitude with respect to other forces, gravitation  cannot be turned off in any physical system, because  every physical entity, having non vanishing mass or energy, generates and is subject to it. Problem is, gravity is difficult to reveal given its low intensity with respect to other forces. But, if very carefully investigated, its effects could in principle manifest in the wider number of physical research contexts.

Recent laboratory experiments like MAGIA 
and an observation of gravitational interaction between neutrons, performed by Abele and Leeb \cite{abe}, explore the possibility to study the behavior of Newton's law beyond the scales  until now observed and could  shed some light on the subject in future. The former measured a value for $G$ with a precision of $0,015\%$ analyzing atom fountains with spectroscopic methods. Next generation radiogravimeters could reach the relative uncertainty of 10 parts per million \cite{tin}. Their future accuracy range is of potential interest, because it could lead to a study of gravitation at smaller ranges than any studied before. If any discrepancies  from $r^{-2}$ dependence would be found, it would open the road to possible massive gravitons and a rethinking of gravitational collapse.

On the other hand, at very large scales, that is cosmological scales, there is the issue with dark matter. That issue concerns  gravitational anomalies in the observed Universe that could be explained by some form of unobserved mass or energy. Conjectures have been proposed about a possible repulsive behavior of gravity at very large distances. Among them,  Hajdukovic \cite{haj} suggested the possibility of a repulsive behavior between matter and anti--matter, that could manifest at cosmological scales. 
While the conjecture seems to be proved wrong by LHC's ALPHA experiment in year 2013 \cite{alp}, that is just another example of where and how gravity can be 
 investigated
 .


Moreover, not all physics must necessarily be reproduced in laboratory; astrophysics is by nature observational and its phenomena, which encompass the greater range of scales of all physics, are dominated by gravity. We already cited VIRGO; Bambi and others \cite{bam} suggested star collapse as a possible observation method of quantum gravity effects, alternative to laboratory high--energy scattering.
As stated, GR predicts unstable states and singular states like black holes, which clash with quantum notions.  However it is conjectured the possibility that quantum effects could remove the classical singularity (studying the behavior of gravity at small distances  could be suggestive also in that sense). Bambi studied that conjecture in a model of star collapse. Such a model could possibly yield the development of an observational method to research signatures of quantum effects in collapsing stars. 

As we see there is much research that can be done at the experimental and observational level even below planckian energies.

\chapter{Classical relativity results}\label{chapterclassical}

In this chapter we will introduce the classical metric generated by a single particle moving at the speed of light, following a derivation 
 by Aichelburg and Sexl.

We will see that such a particle produces a gravitational shock wave that moves along with it, showing a logarithmic profile in transverse direction.
That result will be the base for a semiclassical treatment of a quantum scattering theory for gravity that will be detailed in Chapter \ref{satqg}, and the logarithmic form itself will appear in gravitational quantum fields $a$, $\bar a$ that will be contextually introduced.\footnote{Gravitons, the hypothetical gravitational fields quanta, will be interpreted as related to quantum fields $\phi$, introduced later in the same context of $a$ and $\bar a$.} More generally, gravitational fields generated by sources moving at the speed of light also have the notable property of being closely related to gravitational waves \cite{a-s}.

We pursue the derivation of AS' result on the following logical path:
\begin{enumerate}
\item consider a particle $P$ at rest with respect to a reference system and introduce its metric, that is the Schwarzschild solution. Coordinates are related to measures performed by an observer $O$ at rest in an RS fixed with $P$;
\item consider the motion of the $P$ particle with a RS boost along $z$. The relative coordinates transformation yields a new metric solution. New coordinates are related to measures performed by an observer $O'$ which measures a $\beta$ speed along $z$ for $P$;
\item show a technical problem related to the metric in the limit $\beta \to 1$, and address it with a procedure that let it be performed, yielding the AS metric.
\end{enumerate}

We  then describe other related classical results for gravitational collisions between different objects, namely an approximate metric derived by D'Eath and Payne \cite{dep} for the collision of two ultra\-relativistic black holes and a metric derived by Dray and 't Hooft \cite{d-t} for the collision of two planes of matter.

\section{Metric of one particle source moving at the speed of light}

Aichelburg and Sexl \cite{a-s} determined the metric due to a gravitational field generated by a massless particle moving at speed $c$ along $z$ direction in a cartesian reference system   
as
\begin{equation}
\diff{s}^2= \diff{ t}^2-\diff{ x}^2-\diff{ y}^2-\diff{ z}^2+4\varepsilon
~\delta ( t -  z)\ln ({x}^2 + {y}^2)^{1/2}
(\diff{ t} - \diff{ z})^2
\label{asmetric}
\end{equation}
with $\varepsilon=2GE$, $E$ being the particle's energy in the reference frame.

\begin{figure}
\begin{center}
	\begin{picture}(59,69)(18,-35)
\put(20,0){\line(1,0){15}}\put(45,0){\vector(1,0){20}}\put(65,2){$z$}
\put(20,0){\vector(0,1){10}}\put(20,0){\vector(1,1){5}}\put(20,12){$x$}\put(25,7){$y$}
\put(35,12.5){\line(0,-1){50}}\put(35,12.5){\line(3,4){20}}
\put(55,39){\line(0,-1){50}}\put(35,-37.5){\line(3,4){20}}
\put(45,0){\circle*{1}}
\put(50,-25){\vector(1,0){5}}\put(52,-23){$c$}
\put(40,10){\vector(1,-2){4}}\put(45,20){\vector(0,-1){15}}\put(50,20){\vector(-1,-4){4}}
\put(40,-10){\vector(1,2){4}}\put(50,10){\vector(-1,-2){4}}
\put(40,-20){\vector(1,4){4}}\put(45,-20){\vector(0,1){15}}\put(50,-10){\vector(-1,2){4}}
	\end{picture}
		\begin{picture}(59,69)(2,0)
\put(0,0){\vector(1,1){60}}\put(0,62){$x^-$}
\put(60,0){\vector(-1,1){60}}\put(60,62){$x^+$}
	\put(15,37.5){\vector(2,-1){30}}\put(48,20){$\transverse{x}$}
\put(30,30){\circle*{1}}
\put(-4,7){\line(1,1){55}}\put(65,55){\line(-2,1){14}}
\put(10,0){\line(-2,1){14}}\put(10,0){\line(1,1){55}}
	\end{picture}
	
\end{center}
\caption{A depiction of Aichelburg--Sexl metric in 3--dimensional and light--cone coordinates. The plane containing the source is the region of non--Minkowskian metric that extends to infinity;  square borders and a few field lines are shown for ease of visualization.\label{asplane}}
\end{figure}
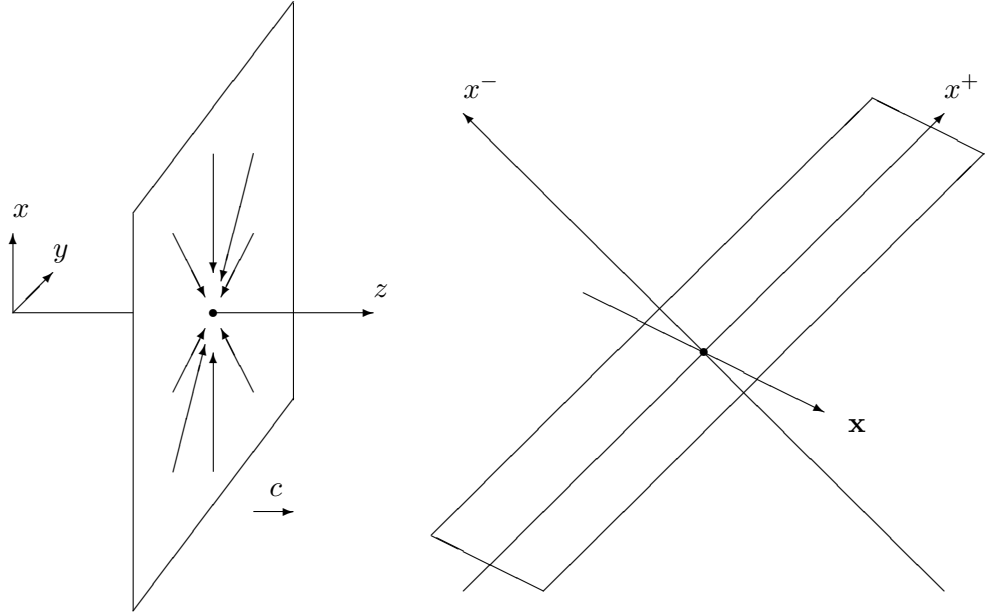

As can be seen, that metric can be visualized as a plane--shaped attractive field that moves along the $z$ direction at speed $c$ (or, equivalently, congruent with the hyper--surface at rest in ${x}^-=t -  z=0$) within a flat Minkowski space--time---see Fig.~\ref{asplane}.

AS derive their result in two ways. With the first one they find a linearized solution for a particle with rest mass $m$ moving uniformly with speed $\beta=v$ and then take the limit with $v\to 1$; that solution however diverges in the hyper--surface $t-z=0$, where the linearized theory cannot be applied due to the $\delta$--like behavior of gravitational potentials and curvature tensors. Then with the second method  they find an exact solution valid in $t=z$, too, and note that elsewhere that solution completely agrees with the linearized one. 
However, AS treatment actually has a limited validity in space--time; we will show both linearized and exact solutions, extending their validity using a derivation due to Colferai  \cite{col} which also stresses the physical meaning of the formal mathematical relation (\ref{asrelation}) used by AS in their derivation.

			\subsubsection{Linearized solution}
			
			By the term \emph{linearized solution} we actually mean a solution of a set of  linearized equations. Generally, linearizing  is a method to treat non--linear equations in an approximate way which makes them linear in a mathematical sense and then easier to solve. Being it related to some kind of underlying approximation, a linearized solution usually is naturally subject to a limited validity, directly deriving from that approximation's conditions.
As we recalled in the Introduction, Einstein's field equations ($c=G=1$)
\begin{equation}\label{einsteinsequations}
G_{\mu\nu}=8\pi ~T_{\mu\nu}
\end{equation}
 are highly non--linear because of the fact that the Einstein's tensor $G_{\mu\nu}$ contains the Riemann tensor $R_{\mu\nu\rho\sigma}$ which expresses curvature and the metric $g_{\mu\nu}$, \emph{i.e.} the fields, and also, in a non--separable way, energy terms.
Those terms represents auto--interaction energies that persist even with a vanishing energy--momentum tensor $T_{\mu\nu}$. In other words, in general it is not possible to supply source terms to the equations and compute the metric fields as its solutions.
 The linearizing condition consists in considering the physical situation of gravitational fields which are weak enough to let us take, in first approximation, just the linear part of the equations, considering the rest as higher--order perturbations to be ignored at that level.

In the case of a source particle moving at speed $v$ along $z$ coordinate, expressed by a source term
\footnote{We abuse the 4--vector notation to avoid confusions between $x^{\lambda}$ and the $x$ coordinate.}
\begin{equation}\label{assource}
T^{\mu\nu}(x^{\lambda})=m(1-v^2)^{-1/2}\delta(z-vt)\delta(x)\delta(y)s^{\mu}s^{\nu}
\end{equation}
and $s^{\mu}=\delta_0^{\mu}+v~\delta_1^{\mu}$, 
a linearized solution
\[
\psi^{\mu\nu} = \left[ \left\lbrace \left(z-vt\right)^2+\left(1-v^2\right)\left(x^2+y^2\right)\right\rbrace \left(1-v^2\right)\right]^{-1/2} m~s^{\mu}~s^{\nu}
\label{linear}
\]
is found. Here\[
\psi^{\mu\nu} =h^{\mu\nu}-\frac{1}{2}\eta^{\mu\nu}h_{\phantom{\lambda}\lambda}^{\lambda}
\]
given the standard approximation $g_{\mu\nu}=\eta_{\mu\nu}+2h_{\mu\nu}$, where $\eta_{\mu\nu}$ is the flat Minkowski metric.

When $v=0$, Eq.~\ref{linear} yields Schwarzschild solution
\[
\psi^{\mu\nu}=\frac{2m}{r}	\delta_0^i\delta_0^k
\]
with $r^2=x^2+y^2+z^2$.

When $v\to 1$ the energy would diverge because of the finite mass $m$. The solution is taking for the mass a form
\[
m = 2\varepsilon \left( 1-v^2\right)^{+1/2}
\]
in the limit $v\to 1$, keeping $\varepsilon$ constant. However,
\begin{equation}\label{limv1}
\lim_{v\to 1} \left\lbrace \left(z-vt\right)^2+\left(1-v^2\right)\left(x^2+y^2\right)\right\rbrace^{-1/2}
\end{equation}
is not a tempered distribution, and the limit in Eq.~\ref{limv1} is not defined for all coordinate values
.

Taking directly $v=1$ in $T^{\mu\nu}
$ would not lead to the expected correct solution, that is, the one which will be found later with exact treatment. Then an ansatz involving a $\delta(z-t)$ split of the solution is made as follows:
\begin{equation}\label{assplit}
\psi^{\mu\nu}(x) = p~\delta(z-t)~G_2(x,y)~\bar{s}^{\mu}~\bar{s}^{\nu}+\psi_H^{\mu\nu}
\end{equation}
with
\begin{itemize}
\item $\bar{s}^{\mu} = s^{\mu}(v=1)$,
\item $\psi_H^{\mu\nu}$ homogeneous solution of $\Box \psi^{\mu\nu}=8\pi T^{\mu\nu}$ and
\item $G_2(x,y)$ Green function of 2--d Poisson equation
\[
\left( \pardev{{}^2}{x^2}+\pardev{{}^2}{y^2} \right) G_2(x,y) = 2\pi~\delta(x)\delta(y)\qquad.
\]
\end{itemize}

Aichelburg and Sexl describe a prescription for a meaningful limiting procedure for (\ref{limv1}) that leads to (\ref{assplit}). The result is all $x^{\mu\nu}$ being proportional to $\delta(z-t)$. That is equivalent to a Minkowski metric  and $R_{\mu\nu\rho\sigma}=0$  for $z\neq t$, while in $z=t$ linearized theory cannot be applied. 

The result of the linearized theory is then just the Minkowski part of AS' metric (\ref{asmetric}), and the fact that in $z=t$ metric diverges, with no more details. 
We then focus on the exact solution.

			\subsubsection{Exact solution}



We start with the Schwarzschild metric, which is the exact exterior solution for a mass $m$ at rest,  that in isotropic coordinates is given by \cite{wal,wei}
\[
\diff{s}^2 = \frac{(1-A)^2}{(1+A)^2}\diff{t}^2 - (1+A)^4(\diff{x}^2+\diff{y}^2+\diff{z}^2)
\]
with $A\equiv m/(2r)$ and $r^2=x^2+y^2+z^2$.

Then we make a Lorentz transformation
\[
\left\lbrace\begin{array}{rcl}
\bar t 	&=&	(1-v^2)^{-1/2}(t+vz),\quad \\
\bar z 	&=&	(1-v^2)^{-1/2}(z+vt),\quad \\
\bar x &=& x\\
\bar y &=& y
\end{array}\right.
\]
and  expand the metric in powers of $A$ as
\[
\left( \frac{1-A}{1+A} \right)^2 - (1+A)^4 \simeq -4A +2A^2-16A^4+\ldots
\]
so that		
		Schwarzschild metric expressed in isotropic coordinates and boosted by $\beta = v$ along $z$, at first order in $A$, is
\[
\diff{s}^2\simeq(1-4A)~\diff{s_0}^2-8A\left[\diff{x}^2+\diff{y}^2+\gamma^2(\diff{z}-\beta\diff{t})^2\right]+o(A)
\]
with $\diff{s_0}^2 \equiv \diff{t}^2 - \diff{\vec x}^2$
.

Missing terms are suppressed by a $\gamma^{-2}$ factor, while
\[
\lim_{\gamma\to +\infty} A\gamma^2 = \frac{\varepsilon}{|t-z|}
\]
 is a distribution undefined at $t=z$, where it cannot be written even in terms of generalized functions.
 That is the problem to be addressed in the following.

AS formally carry out limit $v\to 1$ for $\bar{z}=\bar{t}$ using the following relation:
\begin{equation}
\lim_{\beta\to 1} \{[(\bar z-\beta \bar t)^2+(1-\beta^2)\rho^2]^{-\frac{1}{2}}-[(\bar z-\beta \bar t)^2+(1-\beta^2)]^{-\frac{1}{2}}\}=-2\delta(\bar z-\beta \bar t)\ln\rho
\label{asrelation}
\end{equation}
with $\rho^2=\bar x^2+\bar y^2$,
then
 generate a term equal to l.h.s.\ of (\ref{asrelation}) with a transformation $T(v)$ defined as follows:
	\[
	T(v):
	\left\lbrace
	\begin{array}{rcl}
	z'-vt' &=& \bar z - v \bar t\\
	z'+vt' &=& \bar z + v \bar t - 8\varepsilon \ln \left[ \sqrt{(\bar z-v\bar t)^2+(1-v^2)}-(\bar z - \bar t)\right]
	\end{array}
	\right.\label{astransformation}
	\]
and carry  out resulting metric's limit for $v\to 1$ using relation \ref{asrelation}, obtaining
\[
\diff{s}^2=\diff{t'}^2-\diff{x'}^2-\diff{y'}^2-\diff{z'}^2+16\varepsilon~\delta (t'-z') \ln (x'{}^2+y'{}^2)^{1/2}(\diff{t'}-\diff{z'})^2\qquad.
\]
Finally, they consider inverse transformation $\left[ T(v=1) \right]^{-1}$ leading to the same result obtained from linearized solution for $\bar{z}\neq \bar{t}$, and a $\delta$ for $\bar{z}=\bar{t}$, expressed by Eq.~\ref{asmetric} (omitting the barred notation).

That leads however to a solution not valid in the whole space--time: in AS' description, geodesics are linear except on the wave front, so the whole deflection does occur on the wave front itself. In other words, the deflection occurs 
in proper time intervals of order $\Delta\tau \sim b/\beta\gamma$. It is not restrictive to take a vanishing proper time $\tau =0$ in $t=z=0$; considering a proper time value $\bar \tau \gg b/\beta\gamma$, 
we note that for each finite $\gamma$ there is a critical value for $b$ beyond which AS' description is incomplete. Also we note that relation \ref{asrelation} is formally valid\footnote{A formal proof of relation \ref{asrelation} can be found in \cite{a-s}.} but its physical interpretation still needs to be clarified. Thus, as anticipated, we now turn on a derivation due to Colferai \cite{col} which addresses both matters, obtaining the result on a more physical point of view and extending it to the whole space--time.


Defining
\[
u\equiv \beta t-z,\qquad v\equiv \beta t+z
\]
we consider a coordinate transformation $\mathcal{T}_b(\beta)$ (to be justified in Section \ref{asfiniteb})
\[
\mathcal{T}_b(\beta):
\left\lbrace
\begin{array}{rcl}
u'= \beta t'-z' &=& u\\
\\
v'= \beta t'+z' &=& v-8\varepsilon~\arcsinh \left(\frac{\gamma}{b}u\right)
\end{array}
\right.\label{colferaitransformation}
\]
that depends on a fixed $b$ parameter but involves only longitudinal components $(t,z)$,
and
is substantially analogous to AS' $T(v)$ transformation. W.r.t.\ $T(v)$,  a logarithm has been replaced by a hyperbolic arcsine because 
\[
\arcsinh \left(\frac{\gamma}{b}u\right) = \ln \left[ \frac{\gamma}{b}u + \sqrt{\left(\frac{\gamma}{b}u\right)^2+1} \right] \qquad.
\]
When $b\ll \beta \gamma \bar \tau$,  $T(v)$ reaches a ``logarithmic regime'' where $T(v) \sim \mathcal{T}_b(\beta)$ and\footnote{See also AS geodesics in section \ref{asgeosec}.
}
\[
\arcsinh \left(\frac{\gamma}{b}u\right) \simeq \ln \left( 2\frac{\gamma}{b}u  \right) \qquad.
\]
The metric transforms as
\[
g' = B^T g B
\]
with
\[
B^{\mu}_{\phantom{\mu}\alpha} = \pardev{x^{\mu}}{x'^{\alpha}}\qquad.
\]
The transformation can be expressed as
\[
g=\mathcal{R}\left(\begin{array}{cc}1&0\\ 0&-1\end{array}\right)
+
\mathcal{P}\left(\begin{array}{cc}\beta^2&-\beta\\ -\beta&1\end{array}\right)
\]
with
\begin{eqnarray*}
\mathcal{R}&=&\frac{(1-A)^2}{(1+A)^2}=1+\mathcal{O}(\gamma^{-2})\\
\mathcal{P}&=&\left[(1+A)^4-\frac{(1-A)^2}{(1+A)^2}\right]\gamma^2=8A\gamma^2+\mathcal{O}(\gamma^{-2})\qquad.
\end{eqnarray*}
Neglecting $\mathcal{O}(\gamma^{-2})$ terms, 
\[
8A\gamma^2=\frac{8\varepsilon}{\sqrt{u^2+\left(\frac{b}{\gamma}\right)^2}}\equiv \mathcal{P}_0(b)\qquad.
\]
The last step 
to obtain AS' metric is noting that
$\mathcal{P}_0(b)-\mathcal{P}_0(b_0)$ defines a distribution:
\[
\mathcal{P}_0(b)-\mathcal{P}_0(b_0) = 8\varepsilon~\left[ \frac{1}{\sqrt{u^2+\left(\frac{b}{\gamma}\right)^2}}- \frac{1}{\sqrt{u^2+\left(\frac{b_0}{\gamma}\right)^2}}\right]\qquad.
\]
In fact,
\begin{itemize}
\item with $u\neq 0$,\[
\lim_{\gamma\to \infty} 
\left[ \frac{1}{\sqrt{u^2+\left(\frac{b}{\gamma}\right)^2}}- \frac{1}{\sqrt{u^2+\left(\frac{b_0}{\gamma}\right)^2}}\right]
= 0\qquad,
\]
\item with $u=0$,  
\[
\lim_{\gamma\to \infty} \gamma \left(\frac{1}{b}-\frac{1}{b_0}\right) = \infty\qquad,
\]
\item and it measures
\begin{eqnarray*}
\int_{-\infty}^{+\infty} \diff{u}~\left[ \frac{1}{\sqrt{u^2+\left(\frac{b}{\gamma}\right)^2}}- \frac{1}{\sqrt{u^2+\left(\frac{b_0}{\gamma}\right)^2}}\right]
&=&\\
2~\lim_{\xi\to \infty} \int_0^{\xi}  \diff{u}~\left[ \frac{1}{\sqrt{u^2+\left(\frac{b}{\gamma}\right)^2}}- \frac{1}{\sqrt{u^2+\left(\frac{b_0}{\gamma}\right)^2}}\right]
&=&
2\ln\frac{b_0}{b}
\end{eqnarray*}
\end{itemize}
so that
\[
\lim_{\gamma\to +\infty} \left[ \mathcal{P}_0(b)-\mathcal{P}_0(b_0) \right] = -16\varepsilon~\delta(u)~\ln\frac{b}{b_0}
\]
and
\[
\diff{s}^2_L \simeq \diff{u'}\diff{v'} + 8\varepsilon~\delta(u')\ln\left(\frac{b_0}{b}\right)^2(\diff{u'})^2\qquad.
\]
In $(t',x')$ coordinates:
\[
\diff{s}^2 = \diff{s}^2_0 + 8\varepsilon~\delta(t'-x')\ln\left(\frac{b}{b_0}\right)^2\left[\diff{(t'-z')}\right]^2\qquad.
\]


		\subsection{Aichelburg and Sexl's geodesics}\label{asgeosec}
		
		Aichelburg--Sexl metric has a delta--like behavior that could in principle be thought as a limit of the gravitational field of a particle with increasing speed, that is for $\gamma\to\infty$.

		Starting as usual from a Schwarzschild metric in standard form with $c=1$, introducing capital letter space--time coordinates $(T,X,Y,Z)$ related to polar coordinates $(T,r,\theta,\phi)$ in the usual way, \cite{str}
		
		\[
		\diff{s}^2 = B(r)~\diff{T}^2 - B^{-1}(r)~\diff{r}^2-r^2~\left(\diff{\theta}^2+\sin^2\theta~\diff{\phi}^2\right)
		\]
with
\[
B\equiv 1-\frac{2Gm}{r}
\]
	we consider	a geodesic, which is the natural motion trajectory, \emph{i.e.} a solution for the equations of motion (EOM) of the considered physical system. In that case a geodesic has the following properties: \cite{wei}
		\begin{eqnarray*}
		 \frac{1}{B(r)} &=& \frac{\diff{T}}{\diff{\eta}} \\
		 C_1 &=& r^2\frac{\diff{\phi}}{\diff{\eta}}\\
		 C_2 &=& 
		\frac{1}{B(r)}\left[\left(\frac{\diff{r}}{\diff{\eta}}\right)^2-1\right]+\frac{C_1^2}{r^2} 
		\end{eqnarray*}
		with $\eta$ affine parameter and $C_i$ constants of motion.
		For trajectories coming and going to infinity, $B\to 1$, $\diff{\eta}\simeq \diff{T}$ and
		\begin{eqnarray*}
		-(1-\beta)^2 \simeq \left(\frac{\diff{r}}{\diff{\eta}}\right)^2 - 1 &\to& C_2\\
		\pm | \vec r \times \vec v | \simeq r^2 \frac{\diff{\varphi}}{\diff{\eta}} &\to& C_1\qquad.
		\end{eqnarray*}
		
		We now consider linearized equations to find geodesics at first order in $Gm \propto A$.
		Proper time $\tau$ results
		\[
		\diff{\tau} \equiv \diff{s} =\sqrt{1-\beta^2}~\diff{\eta}\qquad.
		\]
		Consider now a test particle, without generality loss, coming from $Z>0$ region with $X=0$ 
		and 
		with impact parameter $b=Y_0$. Its geodesic would be
		\[
		\left\lbrace
		\begin{array}{rcccl}
		T &=& \gamma(\tau-\tau_0) &+& mG\frac{2}{\beta}\arcsinh \left(\frac{\beta\gamma\tau}{b}\right)\\
		\\
		Z &=& -\beta\gamma\tau &-& mG\frac{1}{\beta\gamma}\arcsinh \left(\frac{\beta\gamma\tau}{b}\right)\\
		\\
		Y  &=& Y_0 &-& mG \left[a\beta\gamma\tau-\frac{1+\beta^2}{\beta^2Y_0}\sqrt{Y_0^2+(\beta\gamma\tau)^2}\right]
		\end{array}
		\right.
		\]
		where we assumed for simplicity that $\tau(Z=0)=0$ and $a$ is a real parameter appearing from the homogenous differential equation ($h$ standing for \emph{homogenous})
		\[
		\dot{Y}_1^h Z_0 = \dot{Z}_0 Y_1^h \Rightarrow Y_1^{h} = a Z_0
		\]
		 and thus will be fixed by physical boundary conditions in the  following; its role is to parametrize the initial direction in the $(y,z)$ plane.
		
					\subsubsection{Geodesics in the reference system of an observer at infinity}
		
		To study that situation from the point of view of a test particle which sees the field--generating particle approaching at speed $\beta=1$, we now boost the system with the particular Lorentz transformation which poses the test particle  at rest when $\tau\to\infty$
		. That is a $\beta$--speed boost, and initially we take $\beta < c$.
		The resulting geodesics are, arbitrarily imposing $t(z=0)=0$,
		\[
		\left\lbrace
		\begin{array}{rcl}
		t(z)	&=&	\frac{b}{\beta \gamma}\sinh \left[\frac{z}{2\epsilon \left(3-\frac{1}{\beta^2}\right)}\right]+
					\frac{1+\beta^2}{\beta \left(3-\frac{1}{\beta^2}\right)}z\\
					\\
		y(z)	&=&	Y_0 + \frac{p}{\gamma}\frac{1+\beta^2}{\beta^2}\frac{Y_0}{|Y_0|}
					\left\lbrace \sinh \left[\frac{z}{2\epsilon \left(3-\frac{1}{\beta^2}\right)}\right] - \chi \cosh \left[\frac{z}{2\epsilon \left(3-\frac{1}{\beta^2}\right)}\right] \right\rbrace
		\end{array}
		\right.
		\]
		where parameter $a$ assumes value $\chi \frac{1+\beta^2}{\beta^2}Y_0$, and the factor $\chi=\pm 1$ distinguishes trajectories that tend to be parallel to the $Z$ axis in their future or in their past, that is with $\tau \to \pm \infty$
.
		Without loss of generality we can focus on $\chi=+1$ geodesics; at large $z$ we find $\diff{z}/\diff{t}\to 0$ as expected.
		
		Even if 
an observer at infinity would see geodesics with finite $b$ and increasing $\gamma$ increasingly more dragged by the field--generating particle, the proper time $\Delta\tau$ between starting point $(-\bar \tau,0)$ and crossing point $(t_c,z_c)$ is finite and equal to $\bar \tau$, because of an opportunely increasing times dilation which perfectly balance the space--time shift.

\begin{figure}
\begin{center}
	\begin{picture}(120,69)(0,0)
		\put(5,25){\vector(1,0){100}}\put(105,20){$z$}
		\put(20,0){\vector(0,1){70}}\put(15,70){$t$}
		\put(5,10){\line(1,1){60}}\put(0,5){$t=z$}
		\put(20,10){\circle*{1}}\put(22,8){$-\bar \tau$}
		\thicklines
		\qbezier(20,10)(19,6)(19,2)\qbezier(20,10)(21,20)(25,25)\qbezier(25,25)(29,30)(35,40)\qbezier(35,40)(43,50)(45,70)\put(40,70){$\gamma_1$}
		\qbezier(20,10)(18,6)(18,2)\qbezier(20,10)(23,20)(28,25)\qbezier(28,25)(38,35)(55,60)\qbezier(55,60)(57,63)(58,70)\put(52,70){$\gamma_2$}
		\qbezier(20,10)(18,7)(17,2)\qbezier(20,10)(22,15)(31,25)\qbezier(31,25)(44,37)(70,70)\put(70,65){$\gamma_3 > \gamma_2 > 	\gamma_1$}
		\multiput(55,25)(0,2){18}{\line(0,1){1}}\put(55,20){$z_c$}
		\multiput(20,60)(2,0){18}{\line(1,0){1}}\put(15,60){$t_c$}
	\end{picture}
\end{center}
\caption{Geodesics in weak Schwarzschild metric boosted with increasing $\gamma$ values, crossing field--generating particle trajectory at increasing $(z,t)$ values.\label{bartgeo}}
\end{figure}
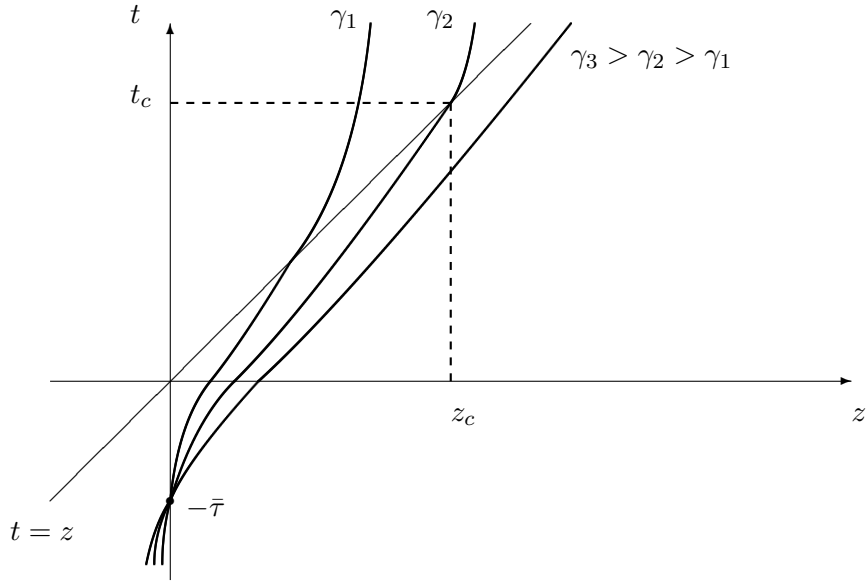

		We also see a peculiar behavior of geodesics for $\gamma \to \infty$.
		In light--cone coordinates, the geodesic of a test particle hit by a shock wave moving with finite $\gamma$ 
		in this reference frame can be roughly described as an almost straight line that, upon approaching the field--generating particle, curves as if the test particle would get 
		dragged along with the source particle for a certain amount of time, continuing to get nearer to the source's trajectory and eventually crossing it; subsequently it is released, substantially moving free again in a straight line.
		Increasing $\gamma$, the test particle is  dragged for increasingly longer amounts of time and space, and the shock--wave crossing point $z_c=\beta t_c$ is found at increasingly high $t_c$ and $z_c$. Moreover, $z_c, t_c \to +\infty$ with $\gamma \to \infty$.
		Also, the limit is non--uniform but
		in confined regions of the $(t,z)$ plane  where the test particle trajectory coincides with that of the field--generating particle.
		That is more easily visualized by 
				considering various geodesics passing by $(-\bar \tau,0)$ 
				 with different values of $b$ %
				(see Fig.~\ref{bartgeo}).

The peculiar behavior we just described is nothing short that the inevitable impossibility to define an observer at infinity for $\beta \to 1$. In fact, gravity has an infinite range and when approaching infinity its effects does not vanish, but cumulates leading to a logarithmic divergence when reaching coordinates' infinity. In order words, the integral of the potential form $1/r$ is a $\ln r$ which diverges in the limit $r\to \infty$. For that reason, we need to introduce a transformation leading to a reference frame that actually succeeds in putting the particle at rest at infinity. That is achieved by considering the RS of an observer at finite impact parameter $b$, as explained in next Section.

			\subsubsection{
			Geodesics in reference system of an observer with a finite $b$}\label{asfiniteb}

\begin{figure}
\begin{center}
	\begin{picture}(120,69)(-5,0)
	\thinlines
		\put(30,0){\vector(1,1){60}}\put(90,55){$x^+$}		\put(50,0){\vector(-1,1){50}}\put(5,50){$x^-$}
	\thicklines
		\put(60,10){\line(0,1){20}}	\put(55,25){\line(1,1){5}}	\put(55,25){\line(0,1){30}}\put(55,60){$b_1>$}
		\put(65,5){\line(0,1){50}}\put(65,60){$b_0>$}

		\put(70,0){\line(0,1){40}}	\put(75,45){\line(-1,-1){5}}	\put(75,45){\line(0,1){10}}\put(75,60){$b_2$}
	\end{picture}
\end{center}
\caption{Geodesics (shown in thick lines) in Aichelburg--Sexl metric with different impact parameters $b_i$, where $b_0$ is the observer's. Each geodesic section shown measures the same time--length.}
\end{figure}
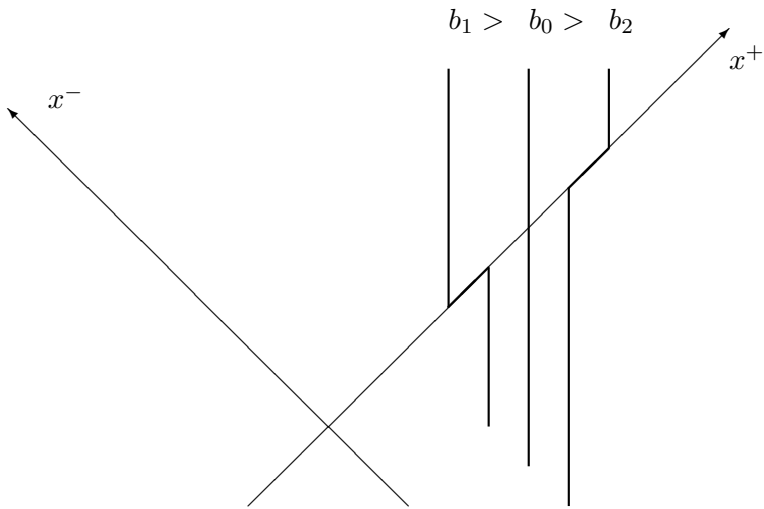

Now we would like to describe the situation of the passage of a particle of energy $E=\varepsilon/2G$ from the point of view of an observer with finite impact parameter $b=b_0$ and initially at rest.

For this purpose we make a coordinate transformation to a reference frame where geodesics with $|Y_0|=b_0$ have a vanishing speed, and the new time variable coincides with the proper time of such observer.

It also would be useful to require that at $\tau=0$, $\beta t'=z'$. Then, using $\beta\tau = \beta t -z$,
\[
\left\lbrace
\begin{array}{rcl}
t' &=& t - \frac{4\varepsilon}{\beta}\arcsinh \left( \frac{\gamma}{b_0}(\beta t-z)\right)\\
\\
z' &=& z-4\varepsilon\arcsinh \left( \frac{\gamma}{b_0}(\beta t-z)\right)\\
\end{array}
\right.
\]
is the sought for coordinate transformation, with the additional relation $(x',y')=(x,y)$. That is the justification of the form of the transformations $\mathcal{T}_b(\beta)$ and AS' $T(v)$ introduced above
.

The combination $\beta t'-z'=\beta t-z$ is conserved as required, while
\[
\beta t'+z' = \beta t + z -8\varepsilon~\arcsin \left( \frac{\gamma}{b}\left( \beta t-z\right)\right)\qquad.
\]
A geodesic trajectory with $b\neq b_0$ is now
\[
\left\lbrace
\begin{array}{rcl}
t' &=& \tau+\frac{z_0}{\beta}+\frac{4\varepsilon}{\beta}\left[\arcsinh\left(\frac{\beta\gamma\tau}{b}\right)-\arcsinh\left(\frac{\beta\gamma\tau}{b_0}\right)\right]\\
\\
z' &=& z_0+4\varepsilon\left[\arcsinh\left(\frac{\beta\gamma\tau}{b}\right)-\arcsinh\left(\frac{\beta\gamma\tau}{b_0}\right)\right]
\end{array}
\right.
\]
that, for $\gamma\to\infty$ at fixed $\tau\neq 0$, goes to
\[
\left\lbrace
\begin{array}{rcl}
t' &\to& \tau+\frac{z_0}{\beta}-\frac{4\varepsilon}{\beta}\frac{\tau}{|\tau|}\ln\frac{b}{b_0}\\
\\
z' &\to& z_0-4\varepsilon~\frac{\tau}{|\tau|}\ln\frac{b}{b_0}
\end{array}
\right.
\]
coherently with AS shifts
\begin{equation}\label{asshifts}
\left\lbrace
\begin{array}{rcl}
\Delta t' = t'(\tau>0) - t'(\tau<0) &=& -\frac{8\varepsilon}{\beta}\ln\frac{b}{b_0}\\
\\
\Delta z' = z'(\tau>0) - z'(\tau<0)&=& -8\varepsilon~\ln\frac{b}{b_0}\qquad.
\end{array}
\right.
\end{equation}


Measured in test particle's proper time, crossing time is zero. In that RS a test particle would appear to be instantaneously shifted and deflected upon traversing the plane--shaped delta--like gravitational field.
The whole deflection happens during the crossing, while elsewhere the trajectory is linear.

That also has an effect on synchronization of geodesics. Consider an observer, heading towards the shock wave with impact parameter $b_0$, along with other observers moving along geodesics at different impact parameters $b_1$, $b_2$, with $b_1>b_0>b_2$. Let us call an observer with impact parameter $b_i$ ``observer $i$''. Suppose observers' clocks are all synchronized prior to the crossing. Consider now a time surface $t=constant$ that intersects the three geodesics inside the future light cone; this means that, on the surface, all the three observers have already emerged on the other side of the shock. On that surface the clocks of the three observers are no more synchronized, being early or late with respect to each other depending on the relation between their impact parameters. Recalling the relativistic twin paradox, we can say in an intuitive way that observer $0$ would ``see'' observer $1$ emerging older and $2$ younger than they were before the crossing
.

The distance between two geodesics is also constant for $|\tau |>\bar \tau$, then definitely (at large $\tau$), $\Delta(x^{\pm})=x^\pm_2-x^\pm_1$ is constant for any given geodesics pair
.

It is important to note a physical limit of validity of this model, expressed by a length unit
\[
\zeta = mG\gamma
\]
relative to linearized model validity. For impact parameters $b<b_{min}\sim\zeta$, test particle would gain a transverse speed $|\diff{y}/\diff{t}|$ which is no longer negligible, as it would  be necessary for consistency with the linearized approximation. That is interesting because small impact parameters pose questions also about the quantum regime---see Section \ref{impactparameter}.
That also seems an incompleteness of the AS approach, because for each fixed finite $\gamma$, it fails to describe the occurring deflection beyond a minimum value of $b$, $b_m(\gamma)$ \cite{col}.


\subsubsection{Geodesics determination with Fermat's principle}\label{asgeo} 

Here we would like to find Aichelburg--Sexl geodesics with a heuristic approach, considering a finite--thickness ``slab'' region of AS metric immersed in a Minkowski space, a geodesic passing through the slab, and its limit when then slab's thickness vanishes.

To derive the finite--thickness geodesics we could \emph{e.g.} develop an affine connections calculation, 
 or use a form of Fermat principle 
as follows: take a geodesic passing by fixed points $A$ and $B$, standing outside the slab on either side, and variable incident and emerging points on the slab's surfaces (call them $1$ and $2$). Then take the limit $h\to 0$, with $2h$ expressing slab's thickness along $u$ direction, of the interval
\[
s \equiv s_{BA} = s_{B2}+s_{21}+s_{1A}\qquad.
\]

We opt for the latter method; we have
\[
\left\lbrace
\begin{array}{rcl}
s_{1A}^2 &=& (u_1-u_A)(v_1-v_A)-(\vec r_1 - \vec r_A)^2 \\
\\
s_{21}^2 &=& (u_2-u_1)\left(v_2-v_1+\alpha(\vec r_M)\right)-\vec r^2+O(h^2)\\
\\
s_{B2}^2 &=& (u_B-u_2)(v_B-v_2)-(\vec r_B - \vec r_2)^2
\end{array}
\right.
\]
where $\alpha (\vec r) \equiv 4GE\ln(r^2/L^2)$, $2~\vec r_M \equiv \vec r_1 +\vec r_2$ and  $O(h^2)$ are corrections of relative order $h$ proportional to $\nabla{\alpha}(\vec r_M)/\alpha(\vec r_M)$.

Light--like geodesics along $u$ will still stay on a light cone after traversing the shock, and that implies a small deviation along transverse directions and an even smaller one along $v$.

From $\nabla_{(v,r)}s_{ij}=0$ (all points are fixed along $u$) we get
\begin{equation}\label{geosystem}
\left\lbrace
\begin{array}{rcl}
(u_1-u_A) ~s_{21}&=& 2~ h~s_{1A}\\
\\
(u_B-u_2) ~s_{21}&=& 2~ h~s_{B2}\\
\\
\vec r_{1A}~s_{21} &=& s_{1A}~\left( (\vec r_{2}-\vec r_{1})+\vec\nabla_{r_1} \alpha\right) \\
\\
\vec r_{B2}~s_{21} &=& s_{B2}~\left( (\vec r_{2}-\vec r_{1})-\vec\nabla_{r_2} \alpha\right) \qquad.
\end{array}
\right.
\end{equation}

\begin{figure}
\begin{center}
	\begin{picture}(120,69)(-25,-35)
		\put(-15,0){\vector(1,0){100}}\put(85,-3){$v$}
		\put(0,-30){\vector(0,1){60}}\put(-3,30){$u$}
				\multiput(-5,-10)(0,20){2}{\line(1,0){80}}\put(-15,10){$+h$}\put(-15,-10){$-h$}
				\put(60,20){\circle*{1}}\put(60,15){$B$}
				\put(30,10){\circle*{1}}\put(25,5){$2$}
				\put(40,-10){\circle*{1}}\put(40,-15){$1$}
				\put(10,-20){\circle*{1}}\put(10,-25){$A$}
				\qbezier(10,-20)(25,-15)(40,-10)
				\qbezier(30,10)(35,0)(40,-10)
				\qbezier(30,10)(45,15)(60,20)
				\put(10,20){flat metric}
								\put(50,5){AS metric}				\put(50,-5){AS metric}				\put(50,-20){flat metric}
	\end{picture}
\end{center}
\caption{Aichelburg--Sexl ``slab'' of finite thickness $2h$.}
\end{figure}
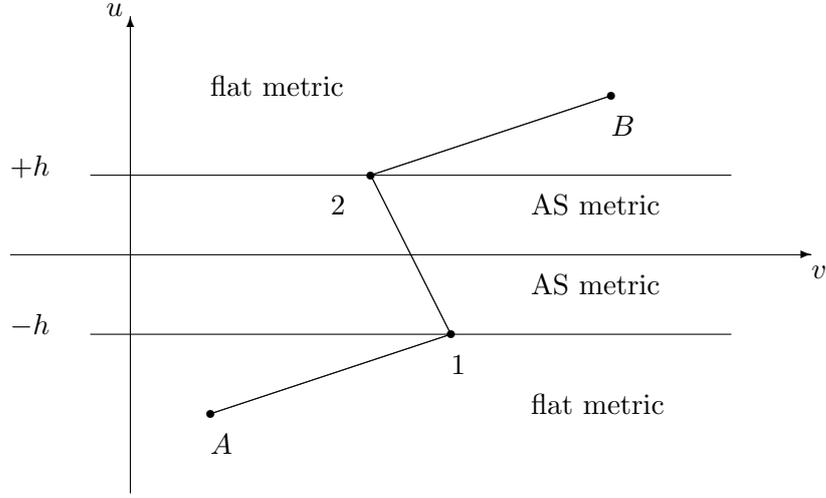

Massive test particles notoriously have non--vanishing finite $\diff{s}_{1A}$ and $\diff{s}_{B2}$ in the limit $h \to 0$; then $s_{21}=\mathcal{O}(h)$ and $|r_2-r_1|=\mathcal{O}(h)$. Consequently from
\[
s_{21}^2 = 2h\left[ v_2-v_1+\alpha(r)\right]-(r_2-r_1)^2 = \mathcal{O}(h^2)
\]
derives
\[
v_2-v_1+\alpha(\vec r_M) = \mathcal{O}(h)
\]
that is
\[
\lim_{h\to 0}~( v_2-v_1) = -\alpha(r_M)\qquad.
\]
Equations \ref{geosystem} then yields  geodesics that can be parametrized as
\[
\left(
\begin{array}{c}
u\\ v\\ \vec r
\end{array}
\right)
 =
\left(
\begin{array}{c}
u\\
\\
v_0+\dot{v}_0~u +\Theta (u) \left[ \left(\frac{1}{4}\left(\nabla \alpha\right)^2-\dot{\vec{r}}_0 \cdot\vec\nabla \alpha\right)u-\alpha (\vec{r}_0) \right]\\
\\
 \vec{r}_0 +\dot{\vec{r}}_0 ~u +\Theta (u) \left[ \left(-\frac{1}{2}\dot{\vec{r}}_0\cdot\vec\nabla \alpha\right) u\right] \\
 \end {array}
 \right)
\]
where quantities with a $0$ subscript represent geodesic's parameters prior to the crossing.

\begin{figure}
\begin{center}
	\begin{picture}(120,69)(-25,-35)
		\put(-15,0){\vector(1,0){100}}\put(85,-3){$v$}
		\put(0,-30){\vector(0,1){60}}\put(-3,30){$u$}
				\put(60,20){\circle*{1}}\put(60,15){$B$}
				\put(35,0){\circle*{1}}\put(35,-5){$2$}
				\put(25,0){\circle*{1}}\put(25,-5){$1$}
				\put(10,-20){\circle*{1}}\put(10,-25){$A$}
				\thicklines
				\qbezier(10,-20)(17.5,-10)(25,0)
				\qbezier(25,0)(28,0)(35,0)
				\qbezier(35,0)(48.5,10)(60,20)
				\put(10,20){flat metric}
								\put(50,-20){flat metric}
	\end{picture}
\end{center}
\caption{Aichelburg--Sexl geodesics obtained with $h\to 0$.}
\end{figure}

Let us now consider the incident and emerging angles  along transverse direction $x$:
\[
\theta = \arctan \frac{\Delta x}{\Delta z}\qquad.
\]
For light--like geodesics, it holds
\[
\tan \frac{\theta}{2} = \frac{\Delta x}{\Delta u}\qquad.
\]
 We note the attractiveness of the field, expressed by the difference between the tangents of emerging (denoted by a prime) and incident angles: 
\[
\tan \frac{\theta'}{2}-\tan \frac{\theta}{2}=\frac{x_B-x_2}{u_B-u_2}-\frac{x_1-x_A}{u_1-u_A}=-\frac{1}{2}\partial_r \alpha <0\qquad.
\]

In the special case of light--like geodesics, we are faced with a technical difficulty, being the vanishing of $\diff{s}^2$ in denominators; however carrying out the massless limit these quantities cancel out and the previous result remains valid. Alternatively,  the  spatial shift can be assumed and  the  procedure repeated extremizing just (proper) time coordinate, as with standard Fermat principle method.

An alternative treatment with affine connections $\Gamma_{\rho}^{\mu\nu}$, that will be non--vanishing only in the slab, would instead result in the appearance of derivatives of delta distribution, that should be handled with care. A sensible approach would be treating these derivatives in the sense of distributions and applying them to test functions.

	\section{Metric of 
	colliding ultrarelativistic black holes}
	\label{death}

	Because of our intention to study semiclassical gravitational scattering, that is, the transplanckian collision of two particles, it is sensible to take in account a result by 
	D'Eath and Payne (DP thereafter) \cite{dep}, which solve up to the third order in $\gamma^{-2}$ the metric ``after'' the collision of two (almost) light--like particles, ingoing with a transverse AS--like shock wave each. Being field--generating particles, we can call them ``black holes''.
	
\begin{figure}
\begin{center}
	\begin{picture}(120,69)(-30,-35)
		\thicklines
		\put(0,-35){\line(1,1){20}}		\put(35,-35){\line(1,1){20}}
		\thinlines					\put(55,-15){\vector(1,1){10}}\put(67,-3){$x^+$}
	\put(40,-35){\vector(-1,1){50}}\put(-10,17){$x^-$}		\put(75,-35){\line(-1,1){20}}\multiput(55,-15)(-4,4){3}{
	\qbezier(0,0)(-1,1)(-2,2)
	}
		\put(12,-15){\vector(1,0){55}}\put(70,-15){$\transverse{x}$}
		\put(55,-15){\vector(0,1){10}}\put(55,-2){$t$}
		\put(55,-15){\vector(4,-1){10}}\put(66,-21){$z$}
						\qbezier(37.5,11)(20,15)(0,30)\put(5,30){caustic}
\thicklines			\qbezier(20,-15)(22,-5)(1,20)		\qbezier(55,-15)(33,-5)(3,20)
				\put(5,-35){shock wave 2}			\put(45,-35){shock wave 1}
\thicklines
				\multiput(25,-15)(5,0){6}{\line(-1,-1){14}}
										\put(20,-15){\qbezier(0,0)(8.75,13)(17.5,26)}
				\put(25,-15){\line(-1,1){5}}		\put(20,-10){\qbezier(0,0)(7,11)(14,22)}
				\put(30,-15){\line(-1,1){14}}	\put(16,-1){\qbezier(0,0)(4.5,8)(9,16)}
				\put(35,-15){\line(-1,1){28}}	\put(7,13){\qbezier(0,0)(2,4.5)(4,9)}
				\put(40,-15){\line(-1,1){28}}	\put(12,13){\qbezier(0,0)(-0.5,4.5)(-1,9)}
				\put(45,-15){\line(-1,1){14}}	\put(31,-1){\qbezier(0,0)(-3,8)(-6,16)}
				\put(50,-15){\line(-1,1){5}}		\put(45,-10){\qbezier(0,0)(-5.5,11)(-11,22)}
										\put(55,-15){\qbezier(0,0)(-8.75,13)(-17.5,26)}
	\end{picture}
\end{center}
\caption{Shock waves collision. Shock wave $2$ hits shock wave $1$ and gets deformed in the future light--cone in logarithmic profile shock waves (shown in thick lines). Shock wave 2 generators (also shown in thick lines) converge in a caustic (topmost curved line).
\label{caustic}}
\end{figure}
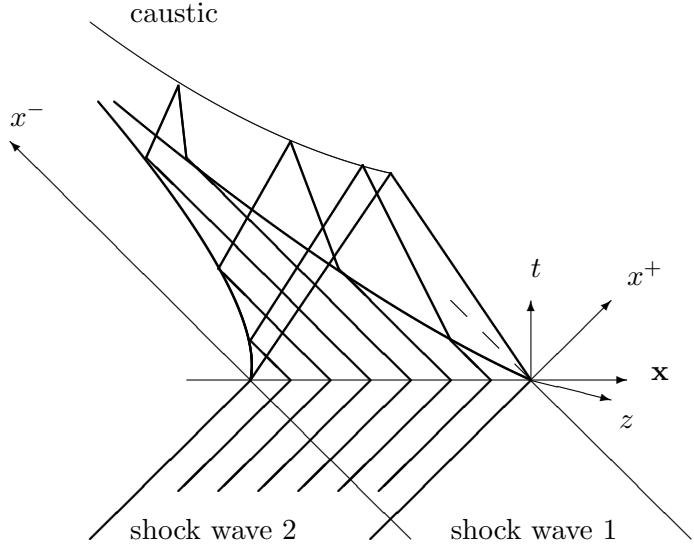

	Consider two such particles traveling in opposite directions on the $z$--axis, and colliding at $z=0$. The shock waves they generate are parallel before the collision and propagate freely. Let us denote the energy of each wave in the center--of--mass frame with $\mu$. A coordinate system is chosen where trajectory of shock 1 before the collision has coordinate $\hat u =0$, while that of shock 2 has $-\hat v =0$. Each shock wave suffers  an instantaneous AS--like shift and bending due to the collision with the other one. The situation can be visualized focusing on the fate of world lines of single time--space points on one of the two shock waves (its \emph{generators}) upon colliding with the other shock's plane wave--front (see Fig.~\ref{caustic}).

	In order to study that situation quantitatively, consider
	a $\beta$ boost along $z$ direction, with $(1-\beta)\ll 1$.
Let us denote the particles' energies in that boosted frame by
$\nu=\mu e^{\alpha}$, $\lambda=\mu e^{-\alpha}$, with
\[
e^{\alpha} = \frac{(1+\beta)^{1/2}}{(1-\beta)^{1/2}}
\]
so that $\lambda / \nu \ll 1$.
That allows us to study the evolution of a weak 
shock  wave in a region behind another, strong, shock wave, basing our considerations on AS result and its finite limit explained earlier. That is, we consider an expansion in terms of 
\[
\frac{\lambda}{\nu} = e^{-2\alpha} =  \frac{(1-\beta)}{(1+\beta)} =  \frac{1-\beta^{1/2}}{(1+\beta)^{1/2}}
\simeq \frac{1}{(2\gamma)^2}\qquad.
\]

\begin{figure}[tbp]
\begin{center}
	\begin{picture}(60,69)(0,0)
\put(5,5){\line(1,1){30}}
\put(35,35){\vector(-1,1){30}}
\put(35,5){\line(-1,1){15}}
\put(20,50){\vector(1,1){15}}
\put(20,5){$\mathcal{P}$}
\put(10,25){$\mathcal{L}$}
\put(24,62){$\mathcal{F}_L$}
\thicklines		\put(15,5){\line(0,1){10}}\put(15,15){\line(1,1){15}}
			\put(13,-23){\qbezier(17,53)(17.5,58)(18,62)}
\put(31,39){\line(-1,1){14}}\qbezier(17,53)(17.5,58)(18,63)
	\end{picture}
	\put(-15,34){$\longrightarrow$}
			\begin{picture}(60,69)(0,0)
\put(55,5){\line(-1,1){30}}\qbezier(25,35)(5,55)(2,64)\put(2.75,62){\vector(-1,3){1}}
\put(-5,5){\line(1,1){30}}\qbezier(25,35)(45,55)(48,64)\put(47.5,62){\vector(1,3){1}}
\put(25,20){$\hat{\mathcal{P}}$}
\put(0,25){$\hat{\mathcal{L}}$}\put(50,25){$\hat{\mathcal{R}}$}
\put(28,47){$
\hat{\mathcal{F}}_L$ }
\thicklines		\put(20,20){\line(0,1){10}}\qbezier(20,30)(20.5,36)(21,39)\qbezier(21,39)(21.5,44)(22,49)
	\end{picture}
\end{center}
\caption{Geodesics in D'Eath and Payne coordinates; one shown in thick line, before and after coordinate transformation. 
The rightmost region, $\hat{\mathcal{R}}$, is causally disconnected from $\hat{\mathcal{L}}$, while $\mathcal{P}$ and $\hat{\mathcal{P}}$ are congruent. \label{depgeo}}
\end{figure}
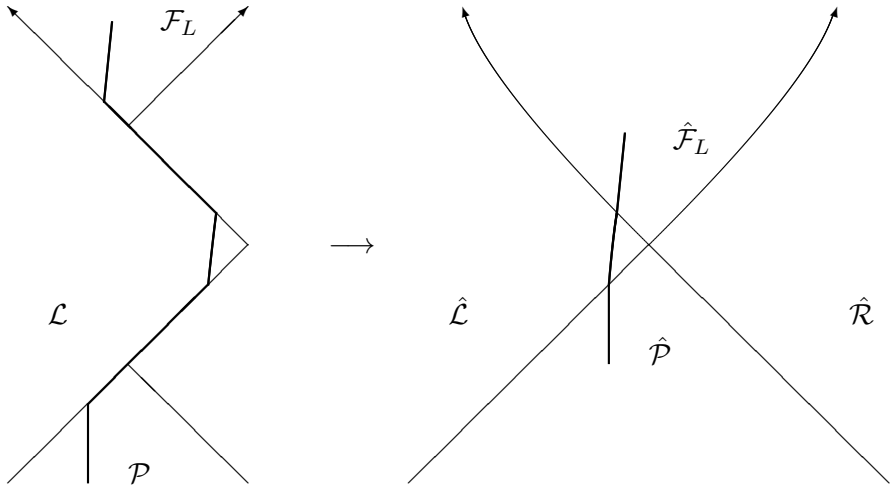

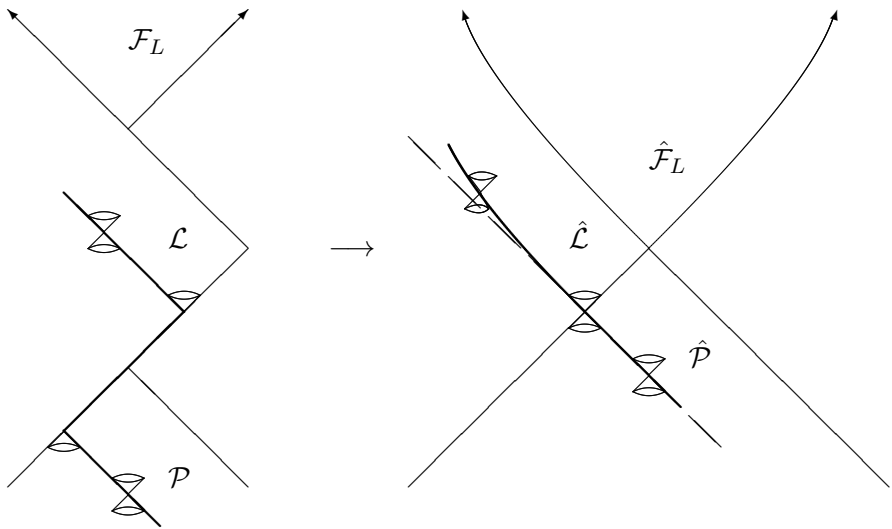
\begin{figure}[tbp]
\begin{center}
	\begin{picture}(60,69)(0,0)
\put(5,5){\line(1,1){30}}
\put(35,35){\vector(-1,1){30}}
\put(35,5){\line(-1,1){15}}
\put(20,50){\vector(1,1){15}}
\put(25,5){$\mathcal{P}$}
\put(25,35){$\mathcal{L}$}
\put(20,60){$\mathcal{F}_L$}
	 \newsavebox{\pastcone}
 	\savebox{\pastcone}(0,0)[tl]{\qbezier(0,0)(1,1)(2,2)\qbezier(2,2)(3,1)(4,0)\qbezier(0,0)(2,1)(4,0)\qbezier(0,0)(2,-1)(4,0)}
		 \newsavebox{\futurecone}
 	\savebox{\futurecone}(0,0)[bl]{\qbezier(0,0)(1,-1)(2,-2)\qbezier(2,-2)(3,-1)(4,0)\qbezier(0,0)(2,1)(4,0)\qbezier(0,0)(2,-1)(4,0)}
 	\put(15,37){\usebox{\pastcone}}\put(15,37){\usebox{\futurecone}}
 			\put(18,4){\usebox{\pastcone}}\put(18,4){\usebox{\futurecone}}
			\put(10,12){\usebox{\pastcone}}\put(25,27){\usebox{\futurecone}}
	\thicklines		
			\put(12,12){\line(1,-1){12}}\put(12,12){\line(1,1){15}}\put(27,27){\line(-1,1){15}}
	\end{picture}
	\put(-15,34){$\longrightarrow$}
			\begin{picture}(60,69)(0,0)
\put(55,5){\line(-1,1){30}}\qbezier(25,35)(5,55)(2,64)\put(2.75,62){\vector(-1,3){1}}
\put(-5,5){\line(1,1){30}}\qbezier(25,35)(45,55)(48,64)\put(47.5,62){\vector(1,3){1}}
\put(30,20){$\hat{\mathcal{P}}$}
\put(15,35){$\hat{\mathcal{L}}$}
\put(25,45){
$\hat{\mathcal{F}}_L$ }
	\multiput(34,10)(-5,5){8}{\line(-1,1){4}}
 	\savebox{\pastcone}(0,0)[tl]{\qbezier(0,0)(1,1)(2,2)\qbezier(2,2)(3,1)(4,0)\qbezier(0,0)(2,1)(4,0)\qbezier(0,0)(2,-1)(4,0)}
		\savebox{\futurecone}(0,0)[bl]{\qbezier(0,0)(1,-1)(2,-2)\qbezier(2,-2)(3,-1)(4,0)\qbezier(0,0)(2,1)(4,0)\qbezier(0,0)(2,-1)(4,0)}
		\newsavebox{\pastconedp}
		\savebox{\pastconedp}(0,0)[tl]{\qbezier(0,0)(1,1)(2,2)\qbezier(2,2)(2.5,1)(3,0)\qbezier(0,0)(2,1)(3,0)\qbezier(0,0)(2,-1)(3,0)}
		\newsavebox{\futureconedp}\savebox{\futureconedp}(0,0)[bl]{\qbezier(0.5,0)(1.25,-1)(2,-2)\qbezier(2,-2)(3,-1)(4,0)\qbezier(0.5,0)(2,1)(4,0)\qbezier(4,0)(2,-1)(0.5,0)}
	\put(2,42){\usebox{\pastconedp}}\put(2,42){\usebox{\futureconedp}}
 			\put(23,19){\usebox{\pastcone}}\put(23,19){\usebox{\futurecone}}
			\put(15,27){\usebox{\pastcone}}\put(15,27){\usebox{\futurecone}}	
 \thicklines 			\put(17,27){\line(1,-1){12}}\qbezier(17,27)(4,40)(0,48)
	\end{picture}
\end{center}
\caption{Light cones in D'Eath and Payne coordinates. The effect of the transformation on a light--like geodesic is shown. In $\hat{\mathcal{L}}$ region, the more the distance between the light--cone and $(u,v)=(0,0)$, the higher its deformation. However, the description is valid in limited regions of space--time
. A dashed straight line is shown to make the deformation clearer. \label{deplc}}
\end{figure}

Within these notations, the metric is region $\hat{\mathcal{P}}$ and in a limited region inside $\hat{\mathcal{L}}$ and $\hat{\mathcal{R}}$ 
(see Fig.~\ref{depgeo})
is
\begin{IEEEeqnarray*}{rcl}
\diff{s}^2 &~=~& 
\diff{\hat u'}\diff{\hat v'} 
+ \left[ 1 + 4\nu\hat u' \theta(\hat u')\hat \rho '^{-2}\right]^2\diff{\hat \rho'}^2 \\
& &+\left[-8\lambda\hat v' \theta(-\hat v')\hat \rho'{}^{-2}+16\lambda^2 \hat \nu'{}^2  \theta(-\hat v')\hat \rho'{}^{-2}\right]\diff{\hat \rho'}^2 \\
& &+\hat \rho'{}^{2}\left[1-4\nu\hat u' \theta (\hat u') \rho'{}^{-2}\right]^2 \diff{\phi}^2\\
&&+ \hat \rho'{}^{2}\left[ 8\lambda \hat v '  \theta(-\hat v')\hat \rho'{}^{-2}+16\lambda^2 \hat \nu'{}^2  \theta(-\hat v')\hat \rho'{}^{-2}\right] \diff{\phi}^2\qquad.\IEEEyesnumber
\end{IEEEeqnarray*}
That metric has the advantage of making AS geodesics continuous at zeroth and first--derivative orders, while still accounting for a Minkowski metric, at the price of being discontinuous and curvilinear  in flat regions  because of the presence of various Heaviside's $\theta$ functions and the particular dependence of metric elements on various coordinates. In particular, we note that it is not valid in regions where
\[
\hat{u} \geq \frac{\rho'^2}{4 \nu}
\]
because the $\phi$ angle would not be determined, in a similar way to what occurs on the  pole of polar coordinates. Also the metric is valid in regions which are larger the larger is $|\transverse{x}|$, \emph{i.e.} the higher the transverse distance from the collision.
	D'Eath and Payne's transformation should be clearer looking at its effect on space--time as depicted in Figures \ref{depgeo} and \ref{deplc}.
	
	D'Eath and Payne  give an expansion in terms of $\lambda / \nu$ of the so--called \emph{news function} $c_0(\tau,\theta)$. That function is a real one, expressing, in this system, gravitational radiation (rather than the metric itself) \cite{bvm,dep}.

A valid definition of the news function in this context is\footnote{Strictly speaking, that is not valid in this context but yields  correct result here for Minkowskian coordinates \cite{dep}.}
\[
c_0(\tau,\theta) = -\frac{1}{2} \lim_{r\to\infty}\left[ \frac{1}{r~\sin^2(\theta)}\pardev{g_{\phi\phi}}{\tau} \right]
\]
where $\theta$ is the polar angle w.r.t.\ the $z$ axis.

The first--order news function after the shock is
\begin{equation}\label{dpresult}
c_0^{(1)} (\tau,\theta ) = \frac{\lambda}{\nu} \sec^4 \left(\frac{\theta}{2}\right) H_0(T)
\end{equation}
with
\[
T=\frac{\tau}{\nu}\sec^2\frac{\theta}{2}-8\ln \left( 2\tan \frac{\theta}{2\nu} \right)
\]
and\[
H_0(T)= \frac{4}{\pi}\int_D \frac{\diff{s}}{s^2}\left[ 2\left( \frac{T+8\ln s}{s} \right)^2 -1 \right]
\left[ 1-\left( \frac{T+8\ln s}{s} \right)^2 \right]^{-\frac{1}{2}}\qquad.
\]

That form is valid only for values of the $\theta$ angle not too close to $\pi$, and more generally all the expansion is valid at least within a region delimited by a caustic, which is the locus of interception point of geodesics pairs after the shock (see Fig.~\ref{caustic}). 
The result should give a good description of the parts of the space--time near the forward and backward directions, but will give an increasingly less accurate description as regions further into the center of space--time are examined. In that region, the formation of a single final Schwarzschild black hole should take place, with associated further emission of gravitational radiation.

The model then leads to  implications about the emission of radiation, which would be interesting to investigate and compare with other models on the subject, including ACV's metric. Moreover, the news function at angles fairly close to the collision axis agrees with the form found in \cite{dea}.

	On the same road of a classical and semiclassical results comparison, like the one we investigated in this work, we consider particularly interesting the program to take in account D'Eath and Payne's expansion. The metric deriving from ACV's action should be coherent with it. A key step is a deeper comprehension of the future light--cone in DP's space--time. There, the classical metric expansion should be consistent with ACV's metric.

DP metric involves a coordinate transformation that  somehow ``amends'' AS--like shifts and deflections by making geodesics' curves and their tangents continuous. That can only happen in the point of view of just one shock, so space--time can be divided in three well--defined regions, $\mathcal{P}$, $\hat{\mathcal{L}}$ and $\hat{\mathcal{R}}$, plus a fourth $\hat{\mathcal{F}}$  about which some more discussion is needed. Here well--defined means that  the metric is known, because each one of $\mathcal{L}$ and $\mathcal{R}$ is casually connected with $\mathcal{P}$ in a known way, while $\mathcal{F}$ in fact is a superposition of two regions, say $\mathcal{F}_L$ and $\mathcal{F}_R$, that represent the future of $\mathcal{L}$ and $\mathcal{R}$ each. The question is, how are $\mathcal{F}_L$ and $\mathcal{F}_R$ related? Intuitively  they should be superposable, but the technical treatment is to be investigated.

The key could be studying the fate in $\hat{\mathcal{F}}$ and of two geodesics passing ``very close'' to each other in $\hat{\mathcal{P}}$ and to the origin of coordinates, and entering one in $\hat{\mathcal{R}}$ and the other in $\hat{\mathcal{L}}$
.
The relation between their parametrizations in $\hat{\mathcal{F}}_L$ an $\hat{\mathcal{F}}_L$ respectively should be suggestive.

To visualize that situation, let us picture two test particles at rest in different spatial points near the collision, before it happens. Imagine the position of each particle be so that the first is hit by shock wave $1$ and then by shock $2$ (that means shock $2$ would already have collided with shock $1$ and is consequently deformed when hitting the test particle). Imagine the other test particle suffering the opposite fate, being hit first by shock $2$ and then by (what remains of) shock $1$. What will happen to those 
two test particles in the future? What would be their mutual fate, and with which differences? Answering those questions should lead to a test comparison of DP's result with ACV's metric, which substantially yields a semiclassical description of analogous phenomena.


\begin{figure}[t]
\begin{center}
	\begin{picture}(40,50)(0,0)
\put(5,5){\line(1,1){20}}
\put(25,25){\vector(-1,1){20}}
\put(25,5){\line(-1,1){10}}
\put(15,35){\vector(1,1){10}}
\put(15,5){$\mathcal{P}$}
\put(10,25){$\mathcal{L}$}
\put(12,40){$\mathcal{F}_L$}
	\end{picture}
	\put(-8,24){$\oplus$}
		\begin{picture}(40,50)(5,0)
\put(5,5){\line(1,1){10}}
\put(25,5){\line(-1,1){20}}
\put(15,35){\vector(-1,1){10}}
\put(5,25){\vector(1,1){20}}
\put(15,5){$\mathcal{P}$}
\put(20,25){$\mathcal{R}$}
\put(12,40){$\mathcal{F}_R$}
	\end{picture}
	\put(-15,24){$\longrightarrow$}
			\begin{picture}(40,50)(0,0)
\put(35,5){\vector(-1,1){40}}
\put(-5,5){\vector(1,1){40}}
\put(15,5){$\mathcal{P}$}
\put(0,25){$\hat{\mathcal{L}}$}\put(25,25){$\hat{\mathcal{R}}$}
\put(2,40){$\hat{\mathcal{F}}=\hat{\mathcal{F}}_R \oplus \hat{\mathcal{F}}_L$ ?}
	\end{picture}
\end{center}
\caption{Two maps of D'Eath and Payne space--time near $(u,v)=(0,0)$, where each axis is approximately straight, combined in an atlas of opportunely shifted and transformed regions. Here we denoted that peculiar combination by the $\oplus$ symbol. While $\mathcal{P}$ from each map is congruent with the other one, for $\hat{\mathcal{F}}$ the $\oplus$ symbol should stand for a particular superposition that  is to be carefully determined.}
\end{figure}
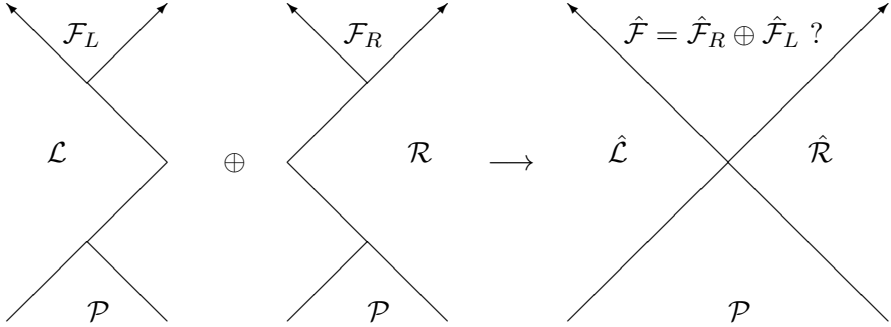


	\section{Metric of two colliding planes}\label{planes}
	
	Before introducing semiclassical approaches to gravity we would like to briefly show another classical result, due to Dray and 't Hooft \cite{d-t}, about the collision of two infinite planes of matter which can be thought of as ``smeared out'' colliding particles.
	
	In fact,   Dray and 't Hooft's approach is substantially the modification of AS' result for a different (extended) energy distribution. As in the previous cases, one of the relevant point is the fact that it can be an additional candidate for a comparison between classical and quantum (or at least semiclassical) gravity results. In fact axisymmetric scattering can be treated with the ACV's semiclassical description to be introduced in Chapter \ref{satqg}. Moreover, the resulting space--time has a resemblance with DP's one, because of its division in three flat regions and a curved one in the future of the collision. That is obviously a consequence of the geometric analogy between the scattering of two parallel shock--waves and that of two parallel energy planes.

	First of all, a single planar shell in flat space  is considered. It can be obtained by substituting the transverse Dirac's $\delta$--s in AS' source term (\ref{assource}) 
	with a $1$ factor, since the pointlike AS source must be replaced by a uniform distribution of null--matter on a plane,
	as
	\begin{eqnarray*}
	2R~ \delta(x)\delta(y) &\to& 2R
\qquad.
	\end{eqnarray*}
	We obtain the metric
	\[
	\diff{s}^2= -\diff{U}~\diff{V}+\diff{X}^2+\diff{Y}^2
	\]
	where
	\begin{eqnarray*}
	U 	&\equiv&	u-	\frac{2}{R}\\
	\\
	V	&\equiv&	v-\frac{1}{2}R~(1-\frac{1}{2}R u)~(x^2+y^2)\\
	\\
	\transverse{X}	&\equiv&	(1-\frac{1}{2}R u)~\transverse{x}\qquad.
	\end{eqnarray*}
	
As in D'Eath and Payne's description,
we identify three flat regions 
	that can be likewise labeled $\mathcal{P}$, $\mathcal{L}$ and $\mathcal{R}$, plus a fourth $\mathcal{F}$ region which 
	corresponds to the future of the collision. In that region the metric is (sometimes referred to as \emph{Robinson's nullicle}) \cite{d-t,rob}
	\begin{equation}\label{nullicle}
	\diff{s}^2 = -R~\diff{R}^2+\frac{1}{R}\diff{T}^2+R^2\left( \diff{p}^2+\diff{q}^2 \right)
	\end{equation}
with 
\[
\left(
\begin{array}{c}T\\R\\p\\q
\end{array}\right)
=
\left(
\begin{array}{c}\frac{1}{2}(A-B)~\omega^{-2/3}\\(1+A+B)^{1/2}~\omega^{-1/3}\\x~\omega^{+1/3}\\y~\omega^{+1/3}
\end{array}
\right)
\]
and
$\omega \equiv |\kappa \lambda |$, $A(u)=\kappa u (\frac{1}{4}\kappa u -1)\theta (u)$, $B(v)=\lambda v (\frac{1}{4}\lambda v -1)\theta (v)$, $\kappa$ and $\lambda$ constants.
				
				It is remarkable that this system can be analytically solved in closed form. However the $\mathcal{L}$, $\mathcal{R}$ and $\mathcal{F}$ regions turn out to limited in space and time, because the infinite amount of energy carried by the plane wave makes space--time to ``collapse''. In fact, the global structure of the solution suggests that each colliding planes focuses each other in a collapsing spherical shell that eventually become a singularity. That global structure has a similarity to an independent result by Matzner and Tipler \cite{m-t}.
				
However, locally the result shares physical sense and could be compared with DP's expansion.

\chapter{Semiclassical approaches to quantum gravity}\label{satqg}\label{chaptersemiclassical}

	Armed with the classical result from AS, our intention is now to use it as an ingredient to develop a semiclassical approach to gravitational scattering following the steps of 't Hooft, who derived the effect of an AS shock wave on a plane wave that crosses it.
	
	Plane waves, as solutions of the free particle Schroedinger equation, form a basis for the Fourier expansion of vectors representing physical states. As such, they  are a base ingredient of quantum mechanics, thus knowing the effect of AS metric on such a physical object surely is a fundamental step in our approach; for the same reasons we will then extend 't Hooft's result to wave packets.
	
	After that, it is natural to interpret the interaction between two of such wave packets using the methods of standard scattering theory, that is, with a description in terms of an $S$ interaction matrix that links initial and final states.
	Such a description has been developed by Amati, Ciafaloni and Veneziano \cite{acv} starting from
	 string theory. That is an effective field theory in which some of the fields are related to AS
\ profiles.

		\section{Scattering theory}

	Before illustrating what 
	anticipated, we would like to resume the basics of standard quantum scattering theory. For additional details, see \emph{e.g.} \cite{sak,b-a,iac,flu}.

	In scattering theory, $S$ matrix links initial and final states $\ket{i}_{in}$, $\ket{f}_{out}$ as
\[
{}_{out}\langle f | i \rangle_{in} = {}_{in}\bra{f} S \ket{i}_{in}\qquad.
\]
In that description, an \emph{elastic scattering} is an interaction process in which initial and final states are equal in number of particles and total momentum, differing 
by the individual momenta of each particle, that gets partially exchanged between them during the interaction. In that case, $S$ can be expressed as a function $S=S(s,t)$ of Mandelstam variables 
\begin{eqnarray*}
s&=&(p_1+p_2)^2=E^2\\t&=&(p_1-p'_1)^2=q^2
\end{eqnarray*}
 where $p$-s are ingoing momenta, $p'$-s outgoing ones, $E$ is the center-of-mass energy and $q$ the exchanged momentum. For an arbitrary variable $\transverse{b}$, which can take the physical role of impact parameter, we can then express a two--dimensional Fourier transform
\[
S(s,b)\simeq \int\ddiff{2}{\mathbf{q}}~e^{-i\mathbf{q}\cdot \mathbf{b}}S(s,q^2)
\]
and, again from scattering theory, it is known that
\[
S(s,b)=\exp\left\{2i\delta_0(s,b)\right\}=1+2i\delta_0+\frac{(2i\delta_0)^2}{2}+\ldots
\]
where $\delta_0(s,b)$ is the phase shift.

We recall  
 that, for a potential $V$ acting on kets $\ket{\psi^{\pm}}$ representing ingoing and outgoing scattered waves, which are the result of the interaction of $V$ with a plane wave $\ket{\phi}$, it holds the \emph{Lippmann--Schwinger equation}:
\[
\ket{\psi^{(\pm)}} = \ket{\phi} + \frac{1}{E-H_0\pm i\epsilon_0} V\ket{\psi^{(\pm)}}
\]
which is related to the Green function
\[
G_{\pm} (\vec x , \vec x' ) \equiv \frac{\hbar^2}{2m}\left\langle \vec x \left| \frac{1}{E-H_0\pm i\epsilon_0} \right| \vec x' \right\rangle
\]
of the Helmholtz equation
\[
\left(\nabla^2+k^2\right)~G_{\pm}(\vec x,\vec x') = \delta^{(3)}(\vec x-\vec x')\qquad.
\]

When $V$ is a \emph{local} potential, diagonal in the $\vec x$ representation,
\[
\bra{\vec x'}V\ket{\vec x''} = V(\vec x')\delta^{(3)}(\vec x'-\vec x'')
\]
we can write, at large distances $r$ w.r.t.\ the range of the potential, supposed to be finite,\footnote{That is worth a particular treatment when confronting with gravity, because its infinite--range character---see Section \ref{cutoff}.}
\[
\langle x | \psi^+ \rangle \longrightarrow \frac{1}{(2\pi)^{3/2}} \left[ e^{i\vec k  \cdot \vec x} + \frac{e^{ik r}}{r} f(\vec k,\vec k')\right]
\]
with $\vec k = \vec p_i$,
representing a sum of a plane wave deriving from the incident one and a spherical scattered wave, while the \emph{scattering amplitude} $f(\vec k,\vec k')$ yields by squaring the differential cross section of the process:
\[
\frac{\diff{\sigma}}{\diff{\Omega}} = |f(\vec k,\vec k')|^2\qquad.
\]

In the particular case of a weak $V$ potential, the following approach is convenient. The term \emph{weak} is to be interpreted as follows. It is reasonable to consider  a potential such if it  does not  ``largely change'' the ket $\ket{\phi}$ during the scattering process. More precisely, a potential is weak if $\ket{\psi^+}$ can be expressed by means of the \emph{Born approximation}
\[
\langle x' | \psi^+ \rangle \longrightarrow \langle x' | \phi \rangle
\]
which at first order reads (expliciting $\hbar$)
\[
f^{(1)}(\vec k,\vec k' ) = -\frac{1}{4\pi}\frac{2m}{\hbar^2}\int~\ddiff{3}{\vec x'}~e^{i\vec q\cdot \vec x'}~V(\vec x')
\]
with $\vec q=\vec k - \vec k'$ transferred three--momentum.

We also recall the \emph{optical theorem}\label{opticalthm} which links the total cross section with the forward scattering amplitude $f(\theta=0)$:
	\[
	\Im{f(\theta=0)} = \frac{k~\sigma_{total}}{4\pi} \qquad.
	\]

With a spherically symmetric potential  $V$, it is convenient (again at large $r$) to introduce the \emph{partial--wave expansion}
\[
\left[ e^{i\vec k  \cdot \vec x} + \frac{e^{ikr}}{r} f(\vec k,\vec k')\right]
=
\sum_l ~(2l+1)~\frac{P_l (\cos \theta )}{2ik}
\left[ s_l(k) \frac{e^{i k r}}{r} - \frac{e^{-i(kr-l\pi)}}{r^2}\right]
\]
 where $P_l (\cos \theta )$ are the \emph{Legendre polynomials} and a factor $1$ in the outgoing wave has been replaced by the 
combination
\[
1 \to  s_l(k) \equiv \left(1+2ikf_l(k)\right) 
\]
while the incident wave remains totally unaltered.
The term $f_l(k)$ is the \emph{partial wave scattering amplitude}.
It holds the unitarity relation for the $l$-th partial wave:
\[
|s_l(k)| = 1\qquad,
\]
hence
\[
s_l (k) = e^{2i\delta_l (k)}
\]
with real $\delta_l(k)$, which has the meaning of 
a
phase shift, and, for a potential $V$  vanishing for $r>R$, where $R$ is its range, can be determined by
\[
\tan \delta_l = \frac{k R j'_l(kR) -\beta_lj_l(kR)}{k R n'_l(kR) -\beta_l n_l(kR)}
\]
where  $j_l$ and $n_l$ are the spherical Bessel and Neumann functions (for details see \emph{e.g.} \cite{sak}), 
$\beta_l$ is
\[
\beta_l \equiv \left( \frac{r}{A_l}\frac{\diff{A_l}}{\diff{r}} \right)
=
kR \left[ \frac{ j'_l(kR) \cos \delta_l - n'_l(kR) \sin \delta_l}{ j_l(kR) \cos \delta_l - n_l(kR) \sin \delta_l} \right]
\]
and $A_l$ is the $l$-th \emph{radial--wave function}
\[
A_l(r) = e^{i\delta_l}\left[ \cos \delta_l j_l(kr) - \sin \delta_l n_l(kr) \right]\qquad.
\]
For a low energy process, $\delta_l \sim k^{2l+1}$ in $k\simeq 0$. 


			\subsubsection{Eikonal regime}
		
		For the most part of this work we consider the case of \emph{eikonal approximation regime}, that corresponds to a small scattering angle 
		where the ingoing particles are (almost) undeflected.
		The meaning of the \emph{small} adjective is to be intended as follows. First of all note that, as long as $E \gg |V|$, that does not mean that the scattering potential $V(x)$ has to be \emph{weak}---in fact we are studying scattering processes at transplanckian energies; conversely, it means that $V(x)$ varies little over a distance of order of wavelength $\lambda$, which can be then regarded as small. Under these conditions, the semiclassical path concept becomes applicable \cite{sak}.

In the eikonal approximation regime, the scattering amplitude is
\[
f(\vec k,\vec k') = -ik\int_0^{\infty}\diff{b}~b~J_0(kb\theta)[e^{2i\Delta(b)}-1]
\]
where $J_0$ is the usual Bessel function of the first kind\[
2\pi J_0(z) = \int_0^{2\pi} \diff{\phi_b} ~e^{-i z \cos \phi_b}
\]
and $\Delta (b)$ expresses a phase shift
\[
\Delta(b)\equiv\frac{-m}{2k\hbar^2}\int_{-\infty}^{+\infty}V(\sqrt{b^2+z^2})~\diff{z}
\]
which is the form assumed by phase shifts $\delta_l(k)$ in the case of eikonal regime, with $l=bk$.

	\section{Deflection of a plane wave by a gravity shock wave}
	
	Having determined the geodesics equations in Section \ref{asgeo}, we are going to obtain the deflection   of a test plane wave according to 't~Hooft's analysis \cite{tho} and then study the deflection of a localized wave packet, in Sec \ref{wavepacketdefl}.

Consider two electrically neutral particles with rest masses \[m_1, m_2 \ll m_{P}=\sqrt{\frac{\hbar c}{G}}\qquad.\]

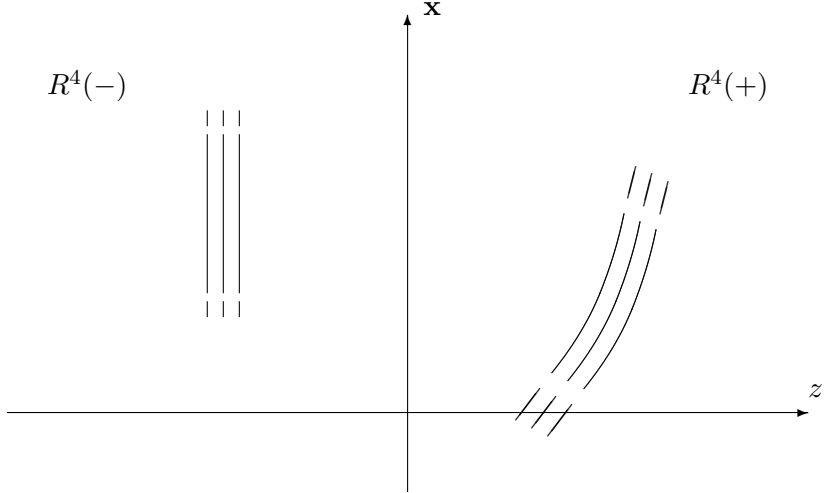
\begin{figure}
\begin{center}
	\begin{picture}(120,69)(0,10)
		\put(60,10){\vector(0,1){60}}\put(62,70){$\transverse{x}$}		
		\put(10,20){\vector(1,0){100}}\put(110,22){$z$}		
	\multiput(0,0)(2,0){3}{
	\put(35, 32){\line(0,1){2}}
	\put(35,35){\line(0,1){20}}\put(35, 56){\line(0,1){2}}
}	
				\newsavebox{\wavefrontdefp}\savebox{\wavefrontdefp}(0,0)[bl]{\qbezier(36,25)(40,30)(42,35)\qbezier(42,35)(44,40)(45,45)}
		\multiput(42,0)(2,-1){3}{\usebox{\wavefrontdefp}}
			\multiput(40,-10)(2,-1){3}{\put(33.5, 29){\line(3,4){3}}
			\put(47.5, 57){\line(1,4){1}}}	
	\put(15,60){$R^4(-)$}\put(95,60){$R^4(+)$}
	\end{picture}
\end{center}
\caption{Scattering of a plane wave through AS metric. A few wave fronts 
are shown to highlight deflection.\label{thooft}}
\end{figure}

Now take a reference system in which ingoing particle (1) is at rest, or moves slowly.
Second particle arrives from the right, along a trajectory $(x_2,y_2)\equiv \transverse{x}(r)=0$, $z_2=-t_2$, with energy 
\[
p_{0}^{(2)} = -p_{3}^{(2)} 
\qquad.
\]
Since $m_1 \ll m_{P}$ then $\beta_2\simeq 1$.
Energy $E_2$ is such that we can no longer ignore gravitational field of particle (2), that is a shock wave of AS form, determining two flat regions of space--time, glued together by a non--linear coordinate transformation at their boundaries---see Fig.~\ref{thooft}
. 

Here and thereafter we will use $(-)$ and $(+)$ 
symbols
to denote quantities in regions respectively before and after the shock, and bold symbols denote transverse quantities.

On the dividing plane,
\begin{eqnarray*}
\transverse{x}^{(+)} &\simeq& \transverse{x}^{(-)}\\
z^{(+)} &=& z^{(-)} + 2Gp_{02}\ln \frac{\transverse{x}^2}{C}\\
t^{(+)} &=& t^{(-)} - 2Gp_{02}\ln \frac{\transverse{x}^2}{C}
\end{eqnarray*}
where $C$ is an  constant which is irrelevant in this derivation, and later to be identified with the $L^2$ term deriving from the 
cut--off that will be discussed in Section \ref{cutoff}
. In short,
\[
x_{\mu (+)} = x_{\mu (-)}-2Gp_{\mu}^{(2)}\ln\frac{\transverse{x}^2}{C}\qquad.
\]
Taking particle (1) spinless for simplicity, in $R^4(-)$, that is, space--time before shock, its wave function is
\[
\Psi_{1(-)} = \exp \left\lbrace i\transverse{p}_1 \transverse{x} - i p_{(1)}^{+} u -i p_{(1)}^{-} v \right\rbrace
\]
with $u=(t-z)/2$, $v=(t+z)/2$ being light--cone coordinates.

In $R^4(+)$, \emph{i.e. }immediately after the shock wave, that is at $v=0^{(+)}$,
\[
\psi_{1(+)}=\exp\left\lbrace i\transverse{p}_{(1)} \cdot \transverse{x} - i p_{(1)}^{+}\left[ u+2Gp_{(2)}^{0}\ln\frac{\transverse{x}^2}{C}\right] \right\rbrace 
\]
Expanded in plane waves:
\[
\psi_{1(+)}=\int A(k_+,\transverse{k})~\diff{k_+}~\ddiff{2}{\transverse{k}}\exp\left\lbrace i\transverse{k}\transverse{x}-ik^+u-ik^-v\right\rbrace \qquad.
\]
with 
\begin{itemize}
\item $k_-=(\transverse{k}^2+m_1^2)/k_+$,
\item $2~Gp_{1,+}p_{2,0} = - 2G\left( p_1\cdot p_2\right) = Gs$,
\item $\diff{k_+}~\ddiff{2}{\transverse{k}} = (k_+/k_0)~\diff{k_3}~\ddiff{2}{\transverse{k}}$ and
\item $A(k_+,\transverse{k}) = \delta\left(k_+-p_{1,+}\right)\frac{1}{(2\pi)^2}\int\ddiff{2}{\transverse{x}}\exp\left\lbrace i \transverse{q}\cdot \transverse{x}-2iGp_{(1)}^{+}p_{(2)}^{0}\ln\frac{\transverse{x}^2}{C}\right\rbrace$,
\end{itemize}
with  an exchanged momentum $q=k_1-p_1$.
We then get
\begin{equation}\label{thooftsmatrix}
{}_{out}\langle k_1 | p_1 \rangle_{in} = \frac{k_+}{4\pi~k_0}\delta(k_+-p_+)\frac{\Gamma(1-iGs)}{\Gamma(iGs)}\left(\frac{4}{(\transverse{p}-\transverse{k})^2}\right)^{1-iGS}\qquad.
\end{equation}
No particle production or \emph{Bremsstrahlung} is seen 
 as long as (1) is electrically neutral and $m_1\ll m_{P}$, hence this  is the case of elastic scattering; as usual
$\delta\left(k_+-p_{+}\right)$ expresses energy conservation.
Considering the Mandelstam variable $t=-q^2$, we get an elastic scattering amplitude, apart from a canonical factor $(k_+/k_0)~\delta\left(\Sigma k - \Sigma p\right)$,
\[
U(s,t)=\frac{\Gamma(1-iGs)}{4\pi\Gamma(iGs)}\left(\frac{4}{-t}\right)^{1-iGS}
\]
and a cross section
\[
\diff{
\sigma 
}
(\transverse{p}_1\to\transverse{k}_1)
= \frac{4}{t^2}\left| \frac{\Gamma(1-iGs)}{\Gamma(iGs)} \right|^2~\ddiff{2}{\transverse{k}} = 4G^2\frac{s^2}{t^2}~\ddiff{2}{\transverse{k}}\qquad.
\]

		\subsection{Deflection of a wave packet}\label{wavepacketdefl}
		
		Using 't Hooft's result for a plane wave, we can now determine the scattering angle of a wave packet in AS metric, that is, passing through an AS shock wave.
		
		Consider a wave packet
		\[
		\ket{f}_{in} = \int \ddiff{}{\widetilde p}~D(\vec p)~\ket{\vec p}_{in}
		\]
		where $\{\ket{p}_{in}\}$ is a complete set of ingoing plane waves and $D(p)$ is the wave packet's distribution, that we can take gaussian:
		\[
		D(\vec p) = \mathcal{N} \exp\left\lbrace -\frac{1}{2} \left(\frac{\vec p - \vec{p}_0}{\sigma}\right)^2 \right\rbrace\qquad.
		\]
		Formally, the ingoing wave packet $\ket{f}_{in}$ can be written in terms of a complete set of outgoing plane waves $\{\ket{k}_{out}\}$ as
		
\begin{eqnarray*}
\tilde{f}_{out}(k) &=& {}_{out}\langle k | f \rangle_{in} =
\int \ddiff{4}{p}~\delta_+(p^2)~{}_{out}\langle k | p \rangle_{in}~{}_{in}\langle p | f \rangle_{in}\\
&=& \int \ddiff{4}{p}~\delta_+(p^2)~{}_{out}\langle k | p \rangle_{in}~ \int \ddiff{4}{p'}~\delta_+(p'^2)~D(p')~\langle p | p' \rangle\\
&=& \int \ddiff{4}{p}~\delta_+(p^2)~{}_{out}\langle k | p \rangle_{in}~D(p)\\
&=& \int \frac{\diff{p^+}}{p^+}\ddiff{2}{\tilde{p}}~\frac{k^+}{4\pi~k_0}~\delta(k^+-p^+)~\frac{\Gamma(1-iGs)}{\Gamma(iGs)}\left(\frac{4}{(\tilde{p}-\tilde{k})^2}\right)^{1-iGs}D(p)
\end{eqnarray*}
where we used the completeness of plane waves sets, so that the outgoing wave function is given by
 \[
f_{out}(x) = \int \ddiff{4}{k}~\delta(k^2)~e^{-ikx}~\tilde{f}_{out}(k)\qquad.
\]
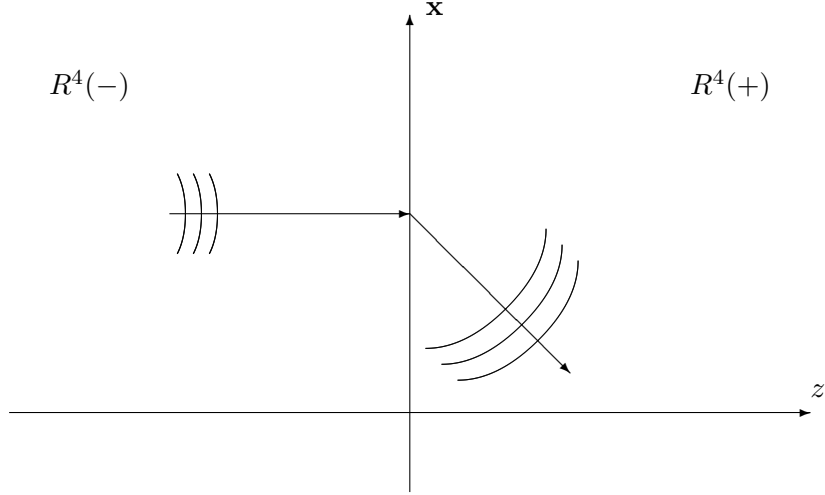
\begin{figure}
\begin{center}
	\begin{picture}(120,69)(0,10)
		\put(60,10){\vector(0,1){60}}		\put(10,20){\vector(1,0){100}}
		\put(30,45){\vector(1,0){30}} 		\put(60,45){\vector(1,-1){20}}
		\newsavebox{\wavefront}\savebox{\wavefront}(0,0)[bl]{\qbezier(31,50)(32,48)(32,45)\qbezier(31,40)(32,42)(32,45)}
		\multiput(0,0)(2,0){3}{\usebox{\wavefront}}
				\newsavebox{\wavefrontdef}\savebox{\wavefrontdef}(0,0)[bl]{\qbezier(0,0)(5,0)(10,5)\qbezier(10,5)(15,10)(15,15)}
		\multiput(62,28)(2,-2){3}{\usebox{\wavefrontdef}}
		\put(62,70){$\transverse{x}$}		
		\put(110,22){$z$}		
			\put(15,60){$R^4(-)$}\put(95,60){$R^4(+)$}
	\end{picture}
\end{center}
\caption{Scattering of a wave packet, represented by a few of its front waves. Group speed is shown.}
\end{figure}
Let us then consider a particle moving along $z$ in negative direction with ultrarelativistic energy $E$, source of a shock wave in the plane $x^+=t+z=0$, and a test particle passing through it.

If $x^{\mu}_A\to 0^-$ denotes the entering point (space--time event) of the test particle into the shock--wave, then the particle would emerge in $x_B^{\mu}$ with
\mbox{$x_B^{+}\to 0^+$},
$\transverse{x}_B = \transverse{x}_A$ and
$x_B^{-}=x_{A}^- -4GE\ln\frac{\transverse{x}^2}{L^2}$.

In other words, it would be subject to a space--time translation along $z$ and $t$ coordinates, emerging in the same state it entered the shock--wave, in particular with the same phase $\varphi_B(x^{(+)})=\varphi_A(x^{(-)})$,
and
\begin{eqnarray*}
x^{\mu(-)} = \left(\begin{array}{c}x^+\to 0^-\\x^-\\\transverse{x}\end{array}\right), \qquad
x^{\mu(+)} = \left(\begin{array}{c}x^+\to 0^+\\x^-+4GE\ln \frac{\transverse{x}^2}{L^2}\\\transverse{x}\end{array}\right)\qquad.
\end{eqnarray*}
Then\[
 \varphi_B(x^{(+)}) = \exp \left\{-\frac{i}{2}p^+\left(x^-+4GE\ln \frac{\transverse{x}^2}{L^2} \right)+i \transverse{p}\cdot\transverse{x}\right\}\qquad.
\]

In case of an on--shell test particle with mass $m\geq 0$, the emerging wave could be expanded as a superposition of on--shell plane waves:
\begin{eqnarray*}
\varphi_B(x) &=& 2\pi\int \frac{\ddiff{4}{k}}{(2\pi)^4}~e^{-ikx}~\delta_+(k^2-m^2)~\mathcal{B}(k)\\
&=&
\int \frac{\diff{k^+}}{k^+} \frac{\ddiff{2}{\transverse{k}}}{(2\pi)^3}~e^{-ikx}~\delta_+(k^2-m^2)~\mathcal{B}(k)
\end{eqnarray*}
with 
$\mathcal{B}(k)$ opportune factors. In the limit $x^+\to 0^+$, $\varphi_B(x)$ amounts to
\[
\varphi_B(0^+,x^--4GE\ln\frac{\transverse{x}^2}{L^2},\transverse{x})
=
\frac{1}{(2\pi)^3}\int\diff{k^+}~\ddiff{2}{\transverse{k}}~e^{-i\frac{k^+x^-}{2}+i\transverse{k}\cdot\transverse{x}}~\frac{\mathcal{B}(k)}{k^+}\qquad.
\]
The result is that, passing through the shock--wave,
\begin{equation}\label{deflectedpacket}
e^{-ipx} \mapsto \int \diff{k^+}~\ddiff{2}{\transverse{k}}~e^{-ikx}~\frac{i\alpha}{\pi}~(\transverse{q}^2)^{i\alpha-1}\delta (q^+) ~\left( \frac{L}{2} \right)^{2i\alpha}\frac{\Gamma (1-i\alpha)}{\Gamma(1+i\alpha)}\qquad.
\end{equation}

In fact, in non--relativistic 
normalization, the $S$--matrix is defined by
\[
e^{-ipx} \mapsto \int\ddiff{3}{\vec k}~e^{-ikx}~S_{\vec k, \vec p}
\]
so that, given $\diff{k^+}/k^+ = \diff{k^z}/k^0$ and (\ref{thooftsmatrix}), we find 
\[
S_{\vec k, \vec p} = \frac{k^+}{k^0}~\frac{i\alpha}{\pi}~(\transverse{q}^2)^{i\alpha-1}~\delta(q^+)\left(\frac{L}{2}\right)^2\frac{\Gamma (1-i\alpha)}{\Gamma(1+i\alpha)}
\]
with
$q^{\mu}\equiv k^{\mu}-p^{\mu}$ and $\alpha \equiv 2GEp^+=Gs$.
Anti--transforming $\varphi_B(x)$:
\begin{eqnarray*}
\frac{\mathcal{B}(k)}{k^+} &=& \int \diff{\left(\frac{x^-}{2}\right)}~\ddiff{2}{\transverse{x}}~e^{ik^+\frac{x^-}{2}-i\transverse{k}\cdot\transverse{x}}~\varphi_B(0,x^-,\transverse{x})\\
&=&
8\pi^2\left(\frac{L}{2}\right)^{2i\alpha}\frac{\Gamma (1-i\alpha)}{\Gamma (1+i\alpha)}~i\alpha~(\transverse{q}^2)^{i\alpha-1}~\delta(q^+)
\end{eqnarray*}
we find at last that upon crossing the shock, wave packets change as
for Eq.~\ref{deflectedpacket}.
We can check that in the no--interaction limit, that is $G\to 0$, the pure phase factor limit value
\[
\lim_{\alpha\to 0}\left(\frac{L}{2}\right)^{2i\alpha}\frac{\Gamma (1-i\alpha)}{\Gamma (1+i\alpha)} = 1
\]
leads to the natural result
\[
\lim_{G \to 0} \frac{i\alpha}{\pi}\left(\transverse{q}^2\right)^{i\alpha-1} = \delta^2 \left( \transverse{q} \right)\qquad.
\]

We would like now to derive the packet's deflection upon crossing the wave front. We consider, without loss of generality, a packet passing by the event $(t_0,\vec x_0)$ with average momentum $\vec p_0$ and momentum spread $\sigma$ such that $|\sigma |\ll |\vec p_0|$:
\[
\widetilde\psi(t_0,\vec p) = \widetilde N ~\exp\left\{ -\frac{1}{2}\left(\frac{\vec p - \vec p_0}{\vec \sigma}\right)^2-i\vec p \cdot \vec x_0 \right\}\qquad.
\]
We also require $\sigma$ to be much smaller than the impact parameter $b=|\transverse{x}_0|$.
We find natural, having to do with wave packets, to opt for a saddle--point method approximation.
In order to manage $\exp\{\Gamma(1-iGs)\}$, we approximate it by means of Stirling's formula,
\ using Euler's \emph{digamma function}
\[
\psi (x) \equiv \frac{\diff{}}{\diff{x}}\ln \Gamma (x) = \frac{\Gamma'(x)}{\Gamma (x)}
\]
and ($\alpha\equiv 2GEp^+=Gs \gg 1$)
\[
\psi (1-i\alpha) \stackrel{\alpha \gg 1}{\sim} \ln (i\alpha) + \ln (-i\alpha) = \ln (2\alpha)\qquad.
\]
We
recall
\ that the factor $\Gamma(1-iGs)$  in 't~Hooft's derivation derives from the integral
\[
\int \ddiff{2}{\transverse{x}}~\exp\left\lbrace i\transverse{k}\cdot \transverse{x} - i Gs\ln \frac{\transverse{x}^2}{C} \right\rbrace\qquad.
\]

After the passage through the shock wave, the wave packets in coordinate space has changed to
\begin{eqnarray*}
\psi_B(x) &\simeq & \tilde{\mathcal{N}} \frac{i\alpha}{\pi} \int \frac{\ddiff{3}{\vec p}}{(2\pi)^3}\ddiff{2}{\transverse{k}}~\exp\left\lbrace -\frac{1}{2}\left(\frac{\vec p - \vec p_0}{\sigma}\right)^2 +i~p\cdot x_0 \right\rbrace\\
&&\times \exp\left\lbrace-ikx\right\rbrace\left(\transverse{q}^2\right)^{i\alpha-1}\\
&=& \tilde{\mathcal{N}} \frac{i\alpha}{\pi} \int \frac{\diff{p^+}\ddiff{2}{\transverse{p}}}{(2\pi)^3}~\ddiff{2}{\transverse{q}}~\frac{p^0}{p^+}~\frac{1}{\transverse{q}^2}~\exp\left\lbrace -\frac{1}{2}\left(\frac{\vec p - \vec p_0}{\sigma}\right)^2\right\rbrace \\
&&\times \exp\left[ i~\mathcal{F}\left(p^+,\transverse{p},\transverse{q}\right)\right]
\end{eqnarray*}
where:
\begin{itemize}
\item we treated a dependence on $p^+\simeq p^+_0\pm \sigma^+$ of the phase factor\[
\left(\frac{L}{2}\right)^{2i\alpha}\frac{\Gamma (1-i\alpha)}{\Gamma (1+i\alpha)}
\]
by recalling that $\sigma^+ \ll p^+$ and we used Stirling's formula as stated;
\item we made a variable transformation $\diff{p^z} = \diff{p^+}~p^0/p^+$;
\item we introduced a phase function
\begin{eqnarray*}
\mathcal{F} &\equiv& -k\cdot x +p \cdot x_0 + \alpha ~\ln \transverse{q}^2\\
&=&
-\frac{1}{2}p^+x^--\frac{1}{2}\frac{(\transverse{p}+\transverse{q})^2+m^2}{p^+}x^+ +
(\transverse{p}+\transverse{q})\cdot \transverse{x} \\
&&+ \frac{1}{2} p^+x_0^- + \frac{1}{2}\frac{\transverse{p}^2+m^2}{p^+}x_0^+ - \transverse{p}\cdot \transverse{x}_0+\alpha ~\ln \transverse{q}^2\qquad.
\end{eqnarray*}
\end{itemize}
We now approximate the integral appearing in scattered wave packet's form by steepest descent method, thus finding stationary point of $\mathcal{F}$ as
\[
\nabla_{(p^+,\transverse{p},\transverse{q})}\mathcal{F}=0\qquad.
\]

In the case under consideration, the coordinate values $x^{\mu}$ 
 are much larger than the packet's width $\sigma$, because we consider observers far away from the scattering region. 
 The derivative w.r.t\ $p^+$ yields the wave packet's delay, while
informations about the deflection are related to the gradient with respect to $\transverse{p}$ and $\transverse{q}$. We choose to focus on the latter, to show that the resulting deflection of the wave packet is consistent with that of a particle.
From the difference 
\[
\pardev{\mathcal{F}}{\transverse{p}}-\pardev{\mathcal{F}}{\transverse{q}}=0
\]
we find
\[
\frac{\transverse{q}}{\transverse{q}^2} = \frac{1}{2\alpha}\left( \transverse{p}\frac{x_0^+}{p^+}-\transverse{x}_0\right)
\]
which is manifestly dependent on $x_0^+$; $\transverse{x}(x^+=0)$ is the transverse coordinate of the front crossing event, for which we can choose $t_0=0$.

Then (exactly)
\[
\left\lbrace
\begin{array}{lcr}
\hat q &=& \hat x_0 \\
\\
|\transverse{q}|&=& - \frac{2\alpha}{b}
\end{array}\qquad.
\right.
\]

The scattering angle in dependence of the crossing event distance on the field--generating particle is then
\[
\frac{\diff{\transverse{x}}}{\diff{x^+}}\bigg|_{\text{after crossing}} = \frac{\transverse{p}+\transverse{q}}{p^+} =
\frac{\diff{\transverse{x}}}{\diff{x^+}}\bigg|_{\text{before crossing}} +  \frac{\transverse{q}}{p^+} \qquad.
\] 
Those quantities are related to the difference between the incident and emerging angles measured as $\arctan (\diff{|\transverse{x}}|/\diff{x^+})$, that is the deflection we would like to evaluate. 
In the case of vanishing mass, 
\[
\tan \theta = \frac{\diff{|\transverse{x}|}}{\diff{z}}, \qquad \tan \frac{\theta}{2} = \frac{\diff{|\transverse{x}}}{\diff{x^+}}
\]
where $\theta$ is the ordinary scattering angle, and
the deflection is
\[
\tan \frac{\theta_i}{2}-\tan \frac{\theta_e}{2}=
\frac{\transverse{q}}{p^+}  = - \frac{2\alpha}{b~p^+}~\hat x_0 = - \frac{4GE}{b}~\hat x_0
\]
\emph{i.e.} the same result for a classical single particle's geodesic in AS metric. 


	\section{An $S$--matrix description of gravity
	}\label{sectionacv}
	
	Amati, Ciafaloni and Veneziano developed an $S$--matrix approach as a model of gravitational scattering, one of whose goals is to implement a gravitational collapse interpretation. Looking for quantum effects relative to that phenomena could help to remove some problematic behavior in regard to classical physics. Classical collapse in fact would give rise to a space--time singularity which seems incoherent with 
	a quantum interpretation of the world. In fact, it 
	clashes with the ordinary interpretation of particles in terms of quantum field theory. A first result which can be interpreted as a quantum symptom is the lack of  real  field solutions  in the case of a transplanckian scattering with impact parameter under a critical value $b_c$. The complex solutions are everywhere regular and suggest a quantum tunneling rather than a singular geometry \cite{acv,fal}.
	
	ACV's approach lives in a quantum string--gravity framework, where the fundamental scale is the \emph{string length} $\lambda_s=\sqrt{\alpha' \hbar}$
.

There are three distinct regimes related to the ratios among the three length scales, the string length $\lambda_s$, the impact parameter $b$, and the gravitational radius $R=2 G\sqrt{s}$. The three regimes are distinguished by which one of the three parameters in larger with respect to the other two. 

In the case of  $b,R\ll \lambda_s$ string effects would be dominant, and that is the most speculative regime.
In that regime, string effects soften gravity according to the generalized uncertainty relation \cite{acv}
\[	\Delta x > \frac{\hbar}{\Delta p} + \alpha'\Delta p > \lambda_s	\]
and string scale itself is the minimal observable size of the system; classical gravitational collapse are never met because it exceeds $R$.

 The case of a large $b$ is the one of small deflection angles scattering, thus is is referred to as the eikonal regime. That region is well described, and has been analyzed by ACV themselves and more recently by Giddings, Gross and Maharana \cite{ggm}, by a leading eikonal approximation with small string--size and classical corrections corresponding to the expansion parameters $(\lambda_s/b)^2$ and $(R/b)^2$ respectively.
 

The third regime concerns $R\gg b, \lambda_s$. In the transition from the second to the third regime, that is, $b\sim R\gg \lambda_s$, a classical gravitational collapse is expected to take place. Also, a semiclassical treatment is expected to be valid, and has been analyzed by ACV \cite{acv}. In this work we are mainly concerned with that situation. 
The effective Lagrangian which is used to describe the second regime and, hopefully, that transition, is motivated by a string--gravity expansion but does not contain explicit string corrections. In that particular regime
	string--size effects can be small while gravitational interaction can still be strong; however
	ACV showed that eikonal regime includes both string-- and strong--gravity effects \cite{c-c}.
	
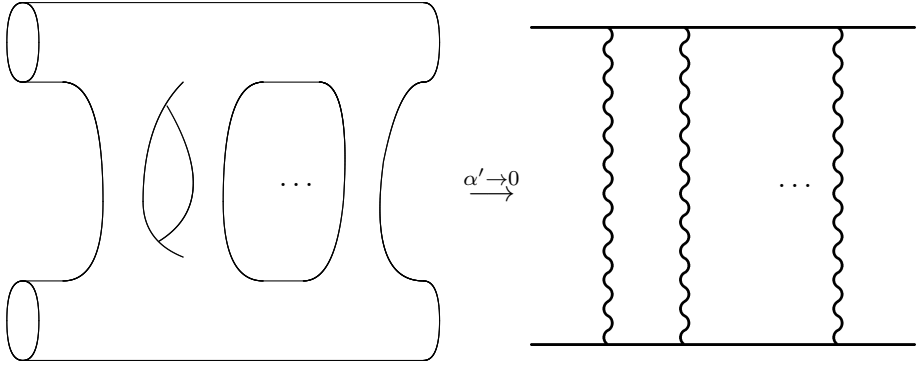
\begin{figure}
\begin{center}
	\begin{picture}(60,40)(-5,0)
		\put(0,45){\line(1,0){50}}
		\qbezier(0,45)(-2,45)(-2,40)\qbezier(0,45)(2,45)(2,40)
			\qbezier(0,35)(-2,35)(-2,40)\qbezier(0,35)(2,35)(2,40)
						\put(0,-35){\qbezier(0,45)(-2,45)(-2,40)\qbezier(0,45)(2,45)(2,40)
			\qbezier(0,35)(-2,35)(-2,40)\qbezier(0,35)(2,35)(2,40)}
				\put(0,0){\line(1,0){50}}
\put(5,0){					\qbezier(0,35)(5,35)(5,20)
								\qbezier(0,10)(5,10)(5,20)					}
\put(0,0){				\qbezier(20,13)(15,15)(15,20)	\qbezier(20,35)(15,30)(15,20)
					\qbezier(17,15)(25,20)(18,32)	}	
\put(10,0){				\qbezier(20,10)(15,10)(15,20)	\qbezier(20,35)(15,35)(15,20)	}	
\put(17,0){				\qbezier(18,10)(22,10)(23,20)	\qbezier(20,35)(24,35)(23,20)	}	
\put(32,22){\ldots}
\put(30,0){				\qbezier(20,10)(13,10)(15,25)	\qbezier(20,35)(17,35)(15,25)	}
\put(50,-35){\qbezier(0,45)(2,45)(2,40)\qbezier(0,35)(2,35)(2,40)}
\put(50,0){\qbezier(0,45)(2,45)(2,40)\qbezier(0,35)(2,35)(2,40)}
\put(0,35){\line(1,0){5}}\put(30,35){\line(1,0){7}}
\put(0,10){\line(1,0){5}}\put(30,10){\line(1,0){5}}
	\end{picture}
\put(0,20){$\stackrel{\alpha'\to 0}{\longrightarrow}$}
	\begin{picture}(60,40)(0,-2)
	\begin{fmffile}{ladderdiagramstring}
\begin{fmfgraph*}(60,40)
\fmfleft{i1,i2}
\fmfright{o1,o2}
\fmf{plain}{i1,v1,v3,v5,v7,o1}
\fmf{plain}{i2,v2,v4,v6,v8,o2}
\fmffreeze
\fmf{wiggly}{v1,v2}
\fmf{wiggly}{v3,v4}
\fmf{phantom,label=\ldots}{v5,v6}
\fmf{wiggly}{v7,v8}
\end{fmfgraph*}
\end{fmffile}
	\end{picture}
\end{center}
\caption{String limit to eikonal Feynman diagrams.\label{strings}}
\end{figure}

	Considering a small string length $\lambda_s$ is analogous to taking the string parameter $\alpha'$ in the limit $\alpha' \to 0$, and shifting the interest  from string  diagrams to Feynman diagrams---see Fig.~\ref{strings}.

		\subsection{Amati, Ciafaloni and Veneziano's results}

In string--gravity, ACV found that $S$--matrix in impact parameter representation has an eikonal form that can be expanded in terms of $R^2/b^2$, ($R\equiv 2G\sqrt{s}$). The expansion terms  are in correspondence with connected tree diagrams interacting with colliding strings via the exchange of $2n$ gravitons at tree level \cite{acv}.

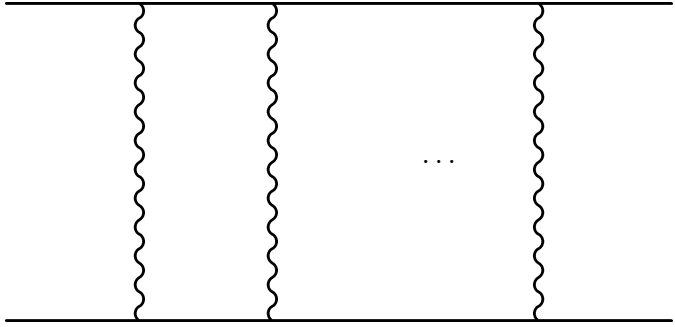
\begin{figure}
\begin{center}
	\begin{picture}(120,50)(0,0)
	\begin{fmffile}{ladderdiagram}
\begin{fmfgraph*}(104,40)
\fmfleft{i1,i2}
\fmfright{o1,o2}
\fmf{plain}{i1,v1,v3,v5,v7,o1}
\fmf{plain}{i2,v2,v4,v6,v8,o2}
\fmffreeze
\fmf{wiggly}{v1,v2}
\fmf{wiggly}{v3,v4}
\fmf{phantom,label=\ldots}{v5,v6}
\fmf{wiggly}{v7,v8}
\end{fmfgraph*}
\end{fmffile}
	\end{picture}
\end{center}
\caption{Ladder diagrams representing the leading eikonal contributions to gravitational scattering, as the eikonal exchange of $n$ particles.\label{diagramsladder}}
\end{figure}

At leading eikonal order,  contributions to the tree are all the $n$--order diagrams depicted in Fig.~\ref{diagramsladder}, representing the eikonal exchange of $n$ gravitons between the two scattering particles, and which due to their appearance can be labeled as ``ladder'' diagrams, where each ``rung'' depicts a particle exchange.

Besides one--loop correction $\delta_1(b,s)$, ACV showed that the lowest term in the series is the Regge--Gribov ``H--diagram''. At high energies, the graviton emission amplitude of the H--diagram takes the form \cite{acv}
\begin{equation}\label{hdiagramemission}
A^{\mu\nu} = \frac{\kappa^3s^2}{\transverse{k}^2} \left( \sin^2 \theta_{12}\epsilon^{\mu\nu}_{TT}-\sin\theta_{12}\cos\theta_{12} \epsilon^{\mu\nu}_{LT}\right)
\end{equation}
where $\epsilon^{\mu\nu}_{ij}$ represent polarization along $i$, $j$ components, denoted by $L$ and $T$ respectively for longitudinal and transverse. 

Taking in account the H--diagram,
we have that
\[
S(s,b)=\exp\left\{i\mathcal{A}(b,s)\right\}
\]
with an effective action 
\[
\mathcal{A}(b,s)=\int\ddiff{2}{\transverse{x}}~\mathcal{L}(a,\bar a, \phi)
\]
reminder of Lipatov's action \cite{lip}
\begin{equation}\label{lipatovaction}
A=\int\ddiff{4}x ~L(h_{++},h_{--},h_{TT})
\end{equation}
where fields $h_{\mu\nu}$ represents gravitons. At high energies, only combinations of the three components appearing in Lipatov's lagrangian $L$ contribute, with $h_{\pm\pm}=h_{00}\pm h_{03} \pm h_{03} + h_{33}$ and $T$ standing for \emph{transverse}.  Light--cone components are related as $4h_{--} = h^{++}$. These are the reasons to focus on the field component $h^{++}$ in leading--orders computations like ours, detailed in Chapter \ref{theresult}.

\begin{figure}
\begin{center}
	\begin{picture}(120,50)(-5,0)
	\begin{fmffile}{hdiagram}
\begin{fmfgraph*}(104,40)
\fmfleft{i1,i2}
\fmfright{o1,o2}
\fmf{plain}{i1,v1,v3,v5,o1}
\fmf{plain}{i2,v2,v4,v6,o2}
\fmffreeze
\fmf{wiggly,label=$\bar a$}{v1,vl}\fmf{wiggly,label=$a$}{vl,v2}
\fmf{wiggly}{v5,vr,v6}
\fmfdot{vl}
\fmfv{label=$\mathcal{J}$}{vl}
\fmffreeze
\fmf{wiggly,label=$\phi$}{vl,vr}
\end{fmfgraph*}
\end{fmffile}
	\end{picture}
	\begin{picture}(120,50)(-5,0)
	\begin{fmffile}{multihdiagram}
\begin{fmfgraph*}(104,40)
\fmfleft{i1,i2}
\fmfright{o1,o2}
\fmf{plain}{i1,v1,v3,v5,o1}
\fmf{plain}{i2,v2,v4,v6,o2}
\fmffreeze
\fmf{wiggly}{v1,vlb,vlt,v2}
\fmf{wiggly}{v3,vcb,vct,v4}
\fmf{wiggly}{v5,vrb,vrt,v6}
\fmffreeze
\fmf{wiggly}{vlb,vcb}
\fmf{wiggly}{vct,vrt}
\end{fmfgraph*}
\end{fmffile}
	\end{picture}
	\begin{picture}(120,50)(-5,0)
	\begin{fmffile}{rescatteringdiagram}
\begin{fmfgraph*}(104,40)
\fmfleft{i1,i2}
\fmfright{o1,o2}
\fmf{plain}{i1,v1,v3,v5,o1}
\fmf{plain}{i2,v2,v4,v6,o2}
\fmffreeze
\fmf{wiggly}{v1,vl,v2}
\fmf{wiggly}{v6,vr,v5}
\fmffreeze
\fmf{wiggly}{vl,vc,vr}
\fmffreeze
\fmf{wiggly}{vc,v3}
\end{fmfgraph*}
\end{fmffile}
	\end{picture}
\end{center}
\caption{The H--diagram, a multi--H--diagram and a Regge--Gribov rescattering diagram. A Lipatov's vertex $\mathcal{J}$ is highlighted.\label{diagramscorrections}}
\end{figure}
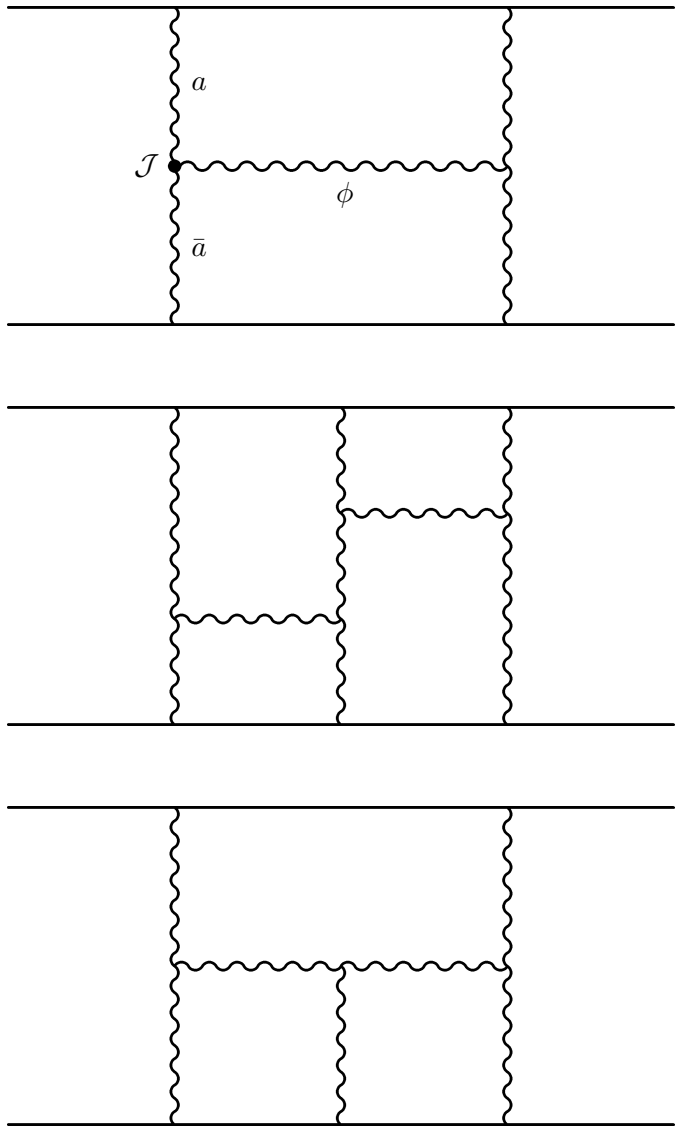

Beyond the leading eikonal order and H--diagrams, tree diagrams contain corrections represented by multi--H--diagrams and rescattering diagrams (see Fig.~\ref{diagramscorrections});
further terms in the $R^2/b^2$--expansion are obtained by considering both kind of diagrams.


Limiting ourselves to multi--H diagrams, ACV find that they are described by a reduced two--dimensional effective action \cite{acv}
\begin{equation}\label{acvaction}
\frac{\mathcal{A}}{2\pi Gs} = a(\transverse{b}) + \bar a (0) -\frac{1}{2}\int\ddiff{2}{\transverse{x}}~\nabla a \nabla \bar a + \frac{(\pi R)^2}{2} \int \ddiff{2}{\transverse{x}}~\left(-(\left(\nabla^2\phi\right)^2+2\mathcal{H}\nabla^2\phi\right)
\end{equation}
with\[
-\nabla^2\mathcal{H} \equiv \nabla^2 a \nabla^2 \bar a - \nabla_i \nabla_j a \nabla_j \nabla_i \bar a 
\]
representing an internal interaction vertex between a so--called Regge graviton and Lipatov's current. In Feynman diagrams' nomenclature that is referred to as \emph{Lipatov's vertex} $\mathcal{J}$.

It would be interesting to evaluate the physical consequences of the metric derived from ACV approach and to compare that effective metric with analogous classical GR results, like D'Eath and Payne's expansion in terms of $\gamma^{-2}=1-\beta^2$ \cite{dep} we introduced in Section \ref{death}, and to  Lipatov metric \cite{lip};
the latter should be done considering also a work by Kirschner and Szymanowski \cite{k-s} which derive the reduced action due to Lipatov starting from  Einstein's equations in high--energies approximation.

Once fields $a(\transverse{x})$, $\bar a (\transverse{x})$ and $\phi(\transverse{x})$ are known, or supposed so, the corresponding metric can be derived. The shock--wave interpretation of the reduced action framework allows to provide the effective metric produced by the solutions of the lagrangian equations.  

We recall that the metric is expressed by $\diff{s}^2=g_{\mu\nu}\diff{x}^{\mu}\diff{x}^{\nu}$ where $g_{\mu\nu}$ is the metric tensor. The metric tensor can be expressed as metric perturbations  $h_{\mu\nu}$ of a background metric $\eta_{\mu\nu}$:
\[
g_{\mu\nu}=\eta_{\mu\nu}+h_{\mu\nu}
\]
and of the components $h_{\mu\nu}$, as stated, only three appear in Lipatov's reduced action (\ref{lipatovaction}). Those are related to $a$, $\bar a$ and $\phi$ by
\begin{eqnarray*}
{h}^{++} &=& \kappa\sqrt{s} \delta(x^-) a (\transverse{x}) \\
\\
{h}^{--} &=& \kappa\sqrt{s} \delta(x^+) \bar a (\transverse{x}) \\
\\
h_{TT} &=& \frac{\kappa^3s}{4}\Theta(x^+x^-)\phi(\transverse{x})
\end{eqnarray*}
and the explicit dependence allows to calculate the longitudinal components of the metric induced by the $h_{TT}$ field. That is done by generalizing Eq.~\ref{hdiagramemission}
to $x^{\pm}$ space and by using the longitudinal components of the $TT$ polarization that, in coordinate space, read
\[
\epsilon^{TT}_{++}=-\frac{\partial_+}{4\partial_-},\qquad
\epsilon^{TT}_{--}=-\frac{\partial_-}{4\partial_+},\qquad
\epsilon^{TT}_{+-}=-\frac{1}{4}.
\]

		\subsection{Axisymmetry case and particle \emph{vs.} ring scattering}\label{particlevsring}

A particularly simple and suggestive situation in which to explore ACV's results is the case of axisymmetric scattering, where every quantity is a function of the radial $r$ variable
\[
r\equiv \left( x^2+y^2 \right)^{\frac{1}{2}}\qquad.
\]
 In that particular case an exact solution can be found. 
 
 While it has been previously considered the reduced action (\ref{acvaction}) in a perturbative region where the relation \mbox{$\lambda_s \ll b \ll R$} holds, here ACV treat the case $b\sim R \gg \lambda_s$ by considering the following. In that regime, dynamics is described by the effective action (\ref{acvaction}). In that action, there any dependence on $\lambda_s$ disappears; that is, string corrections are not considered explicitly.  In the practical situation, we can formally take $\lambda_s\to0$. $R$ is the only explicit coupling left, and that allows to consider smaller impact parameters, including the case of a head--on collision, taking $b=0$.

In this case we can look for axisymmetric solutions for the fields $a$, $\bar a = a$ and $\phi$ which are functions of $r^2 = \transverse{x}^2$ only, and obey a set of equations which turn out to be simple enough to allow a complete treatment
\ In fact, the lack of a dependence on azimuthal angle yields an ODE: 
\begin{eqnarray*}
	\pardev{}{r^2} \left[ r^2\dot a (1-(2\pi R)^2\dot \phi) \right]&=& 0\\
	\pardev{}{r^2} \left[ r^2 \frac{\partial^2}{(\partial r^2)^2}{(r^2 \dot{\phi})} \right] + \frac{1}{2} \pardev{}{r^2}(r^2\dot{a}^2)&=& 0
\end{eqnarray*}
where the dot $(\dot{})$ stands for a $r^2$--derivative
\ \cite{acv}. Such a notation looks natural because of the role of $r^2$ as a unique variable to describe the process, like it would be done with a time variable $\tau\equiv r^2$.

		The case   introduced above can be described as a central collision of two homogeneous beam by an action term \cite{fal}
		\[
		T_{\mu\nu} h^{\mu\nu} = T^{++}h_{++}+T^{--}h_{--}
		\]
		with
\begin{eqnarray*}
		T^{++} &=& \delta (\vec x -\vec b) \delta(x^-)~\frac{E}{2} \\
		T^{--} &=& \delta (\vec x) \delta(x^+)~\frac{E}{2} 
		\end{eqnarray*}
		and
\begin{eqnarray*}
		h_{++}
		&=& (2\pi R)\bar a (\vec x) \delta (x^+)\\
		h_{--}
		&=& (2\pi R) a (\vec x) \delta (x^-)\qquad.
		\end{eqnarray*}
		

		In the effective action (\ref{acvaction}), $R$ plays the role of the coupling constant between $\mathcal{H}$ and $\phi$.
		From $\phi$'s equations of motion, we derive
		\[
		\mathcal{H} = \nabla^2\phi \Rightarrow \nabla^2 \mathcal{H} = \nabla^4\phi
		\]
		and taking axisymmetry into account,
		\[
		\frac{\diff{\mathcal{H}}}{\diff{r}^2} \equiv \dot{\mathcal{H}} = -2\dot a \dot{\bar a}\qquad.
		\]
		In this way, the r.h.s.\ of Eq.~\ref{acvaction} becomes an unidimensional integral and one can express the lagrangian in terms of a single field describing the dynamics of transverse gravitons. We then introduce the function
\[
\rho (r^2) \equiv r^2\left( 1-(2\pi R)^2\dot{\phi} \right)
\]
which can be interpreted as an effective radius $r^2$
. Then
		\[
		(\nabla^2 \phi)^2 =  \frac{(1-\dot{\rho})^2}{2(\pi R)^2}
				\]
assumes the role of 		the kinetic term and the resulting
action is 
		\[
		\frac{\mathcal{A}}{2\pi^2 Gs} = \int \left[ a\bar s + \bar a s -2\dot{\bar a}\dot{a} \rho - \frac{(1-\dot{\rho})^2}{2(\pi R)^2} \right] ~\diff{r^2}\qquad.
		\]
As previously stated, in the case of $b=0$ we have $\bar a = a$, then 
\[
\dot{\rho}^2 + \frac{R^2}{\rho} = 1,	\qquad 	\dot a = \dot{\bar a}= -\frac{1}{2\pi\rho}\qquad.
\]
Explicit solutions are obtained in terms of a hyperbolic angle $\chi(r^2)$ as
\begin{eqnarray*}
\rho(r^2) &=& R^2\cosh^2\chi (r^2)\\
a(r^2) &=& \frac{1}{2\pi}\int_{r^2}^{L^2}\frac{\diff{r^2}}{\rho(r^2)}=\frac{1}{\pi}\left( \chi(L^2)-\chi(r^2) \right)\\
\frac{r^2}{R^2} &=& \chi + \cosh \chi \sinh \chi -\chi_0 -\cosh \chi_0\sinh\chi_0
\end{eqnarray*}
where 
 $L^2$ is the IR cut--off needed to regularize the logarithmic Coulomb phase (see Section \ref{cutoff}), 
 and $\chi_0=\chi(0)$ is an arbitrary value of $\chi$ at the origin. 
 An explicit form of the function $r(\chi)$ is not  derivable, but the asymptotic behavior at large $r$ can be found by iteration:
\begin{eqnarray*}
	\rho(r^2) &\simeq& r^2 - \frac{R^2}{2}\ln \frac{4r^2}{\bar{r}^2(\chi_0)}\\
	\phi(r^2)&\simeq& \frac{1}{16\pi^2}\ln^2\frac{4r^2}{\bar{r}^2(\chi_0)}\\
	a(r^2) &\simeq& \frac{1}{2\pi}\left( \ln \frac{L^2}{r^2} + \frac{R^2}{2r^2}\ln\frac{4r^2}{\bar{r}^2}\right)
\end{eqnarray*}
with $
\bar{r}^2(\chi_0) = R^2 \exp \left\lbrace 1+2\chi_0+\sinh \chi_0 \right\rbrace
$.
That solution is actually valid strictly for $b=0$ but can be extended to $b>0$ as
\[
\dot{a}\rho(r^2) = -\frac{1}{2\pi},\quad \ddot{\rho} = 2(\pi R)^2 \dot{a}\dot{\bar{a}}
\]
which substantially acts as a replacement by a factor $\dot a$ by $\dot{\bar{a}}$.
		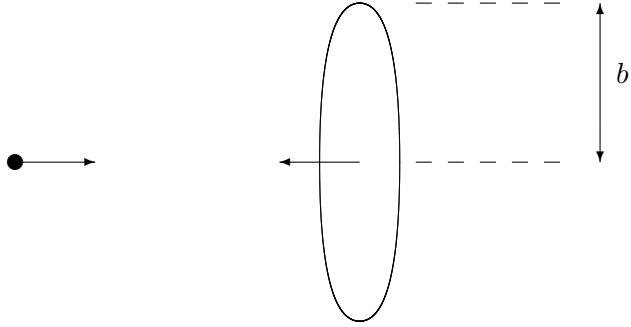
\begin{figure}
\begin{center}
	\begin{picture}(120,50)(20,10)
		\put(42,35){\circle*{2}}
		\qbezier(85,55)(90,55)(90,35)
		\qbezier(85,15)(90,15)(90,35)
		\qbezier(85,15)(80,15)(80,35)
		\qbezier(80,35)(80,55)(85,55)
		\put(42,35){\vector(1,0){10}}
				\put(85,35){\vector(-1,0){10}}
					\put(115,40){\vector(0,1){15}\vector(0,-1){5}}
					\put(117,45){$b$}
					\multiput(92,35)(4,0){5}{\line(1,0){2}}
										\multiput(92,55)(4,0){5}{\line(1,0){2}}
	\end{picture}
\end{center}
\caption{Axisymmetric scattering of a particle and a ring of matter. Ring's speed is shown in its center of matter.}\label{ring}
\end{figure}
		Field $a$ shows two different regimes and reads 
		\[
\bar{a}(r^2) = a(r^2)\Theta(r^2-b^2)+a(b^2)\Theta(b^2-r^2)
\]
 for a ring--shaped source as in Fig.~\ref{ring}. In the limit $r\to\infty$, $a(r^2)$ goes to infinity as $\ln \frac{r^2}{L^2}$, that is, as an AS profile.
 For the particle \emph{vs.}\ ring scattering, ingoing states can be defined as
		\[
		R_1=R, \quad R_2=R~\Theta (r^2-b^2)
		\]
yielding
		\[
			\ddot{\rho} = \frac{1}{2}\frac{R^2}{\rho^2}\Theta (r^2-b^2)
		\]
which can be resolved analytically.		Veneziano and Wosiek \cite{vw} extended  the approach to the  case of extended axisymmetric sources, solving the equations numerically. 
The extended sources are expressed by
\[
R_i(\tau)=\int_0^{\tau}s_{i}(\tau ')~\diff{\tau '}\qquad.
\]

		Once $\rho(\tau)$ is known, metric can be derived by $a(\tau)$, $\bar a (\tau)$, $\phi(\tau)$ \cite{acv}.
The results of those derivations could be compared with known classical GR behaviors,
\ like the collision of two planes of matter resumed in Section \ref{planes}.


		\subsection{Impact parameter and unitarity defect}\label{impactparameter}
		The analytic solution of the axisymmetric scattering of a particle and a ring described above yields the introduction of a critical impact parameter. In terms of $\chi(r^2)$ 
		we find, for $r^2\geq b^2$ (in units of $R=1$),
		$
		\rho = \cosh^2 \chi (r^2) 
		$.
		
		Two boundary conditions are introduced. The condition $\rho(0)=0$ is required  for self--consistency of the reduced action (\ref{acvaction})  \cite{acv}. We also note that $\rho(r^2)$ is linear for $r^2<b^2$, and that suggests an enforcement of the first condition and 
		to write 
$\rho(r^2) \sim t_b r^2$ for $r^2<b^2$
\ with
$t_b = \tanh \chi_b$
and 
$\chi_b = \chi (r^2=b^2)$.

		That yields the following \emph{criticity equation}:
		\[
		t_b (1-t_b^2) = \frac{1}{b}
		\]
		that cannot be satisfied when
		\[
		b<b_c=\sqrt{\frac{3\sqrt{3}}{2}}~R
		\]
		
Real--valued field solutions with $\rho(0) = 0$ exist only for $b \geq b_c$.
In the $b>0$ case, $b_c$ separates, in the real--valued domain, the class of ``weak--field'' solutions having $\rho(0) = 0$ (for $b > b_c$) from that of ``strong--field'' solutions with $\rho(0) > 0$ for $b<b_c$ \cite{acv}. 
		
		When $b>b_c$ there are two distinct solutions;\footnote{Strictly speaking, there are three solutions, but one is discarded because it yields a negative impact parameter \cite{fal}.} when $b=b_c$ two coincident solutions; when $b<b_c$ keeping $\rho(0)=0$, two complex solutions that could be symptom of gravitational collapse \cite{acv,fal}.
		In fact the appearance of non--real values can be  associated with the passage form classical to quantum behavior, that is expected to play an important role in the collapse physics. That is because the action becomes a complex quantity, and consequently the elastic $S$--matrix is no longer unitary 
.
\ Thus, the critical parameter is 
related to 
that unitarity defect
. 
That is related to the topic of information loss, and the incongruences between classical gravitational collapse, black hole GR solutions and quantum theories.
It can be conjectured that the semiclassical unitarity defect could be recovered by some quantum process to be clarified; possibly a quantum tunneling effect could replace the classical collapse singularity.
 To address the subject it clearly would  be interesting to further investigate the meaning of the critical impact parameter $b_c$, along the lines of \cite{fal}.

		\subsection{Cut--off}\label{cutoff}

We now explain the procedure to treat an IR divergence in terms of the cut--off  we cited earlier
.
An algebraic cut--off is necessary because of the peculiar infinite--range nature of gravitational interaction, for which an $S$--matrix 
requires a particular treatment%
. The interaction picture in which the $S$--matrix lives requires that the interaction is treated as vanishing apart from a limited region, where the whole scattering interaction is considered to happen. In fact, $S$--matrix links states which are free at infinity. But, in the case of gravitation,  initial states cannot really be prepared in a region where gravitational interaction is somehow ``turned off'', because of its infinite range.
That can be pictured by considering that AS profiles have logarithmic dependence on distance. That substantially derives from an integration of the Newton potential $\sim 1/r$ 
giving rise to a $\ln r$ which diverges in the limit $r\to\infty$.

Technically, as shown in Chapter \ref{chapterresult}, we face the problem in the integration of a $\transverse{Q}^2$ propagator
\[
\int 	\frac{\ddiff{2}{\transverse{Q}}}{\transverse{Q}^2} ~\exp\left\lbrace i\transverse{Q}\cdot \transverse{b}
 \right\rbrace
\] 
for which adding  a $\Theta \left( |\transverse{Q}|-q_0\right)$ cut--off is needed \cite{cas}.
This leads to the appearance of a Bessel function of the first kind $J_0(bQ)$ represented by the last integral in
\[
\int	\frac{\ddiff{2}{Q}}{Q^2} ~\exp\left\lbrace i\transverse{Q}\cdot \transverse{b}
 \right\rbrace
\Theta \left( |\transverse{Q}|-q_0\right)
=
\int_{q_0}^{\infty}  \frac{\diff{Q}}{Q}
\int_0^{2\pi} \diff{\theta}~e^{ iQb\cos\theta}
\]
Then, the problem is technically solved with 
	\[
2
\lim_{q_0\to 0} \int_{q_0}^{+\infty} \diff{Q}~\frac{J_0(bQ)}{Q}
= \lim_{q_0\to 0} \left[ 2~\ln\left(\frac{2}{q_0 b}\right) + \gamma_E+\mathcal{O}(q_0b) \right]
	\]
 Then, 
to be coherent with our notation, we define 
\[
L = 2~e^{-\gamma_E} \frac{1}{q_0}
\]
and we can write   \cite{c-c}
\begin{equation}\label{cutoffeq}
\int 	\frac{\ddiff{2}{Q}}{Q^2} ~
~\exp\left\lbrace i\transverse{Q}\cdot \transverse{b}\right\rbrace
\to
\ln \left( \frac{L^2}{b^2} \right) 
\qquad .
\end{equation}

\chapter{Gravitational fields calculations}

\label{theresult}\label{chapterresult}

In this chapter we use the technique of Feynman diagrams to calculate 
the gravitational fields occurring in regions ``far away'' from, but sensitive to, a gravitational scattering.
	The meaning of ``far away'' will be detailed and will be clearer in the following.
	The measured field of interest is the effective $h^{++}$ field perturbation at lowest--level order.
	The aim is  to compute $h^{++}$ field on  states $\Psi$ of two ingoing particles. First, we describe how to perform the calculation, then we turn on the calculation details themselves.
	The result of the resummation of all leading eikonal diagrams with $n$ graviton exchanges is a field proportional to a shifted $\delta$--function, that corresponds to AS' classical shifts. The ACV semiclassical procedure is then able to reproduce, in terms of eikonal exchanges (ladder--diagrams) at every order of perturbation, the classical interaction predicted by AS.


	\section{Gravitational scattering 
	 diagrams}
		We  perform the calculation using the method and formalism of Feynman diagrams, which supplies rules for  computing  scattering amplitudes starting from a graphical representation of the concerning scattering process.

	In order to perform such a calculation  for gravitational interaction scattering between two particles, we should clarify how to depict the process and how to derive the appropriate rules to calculate it.
		In other words, we should derive 
Feynman rules and establish them for the present case. Some of them will result analogous to well--known rules of the Standard Model, while others will be peculiar of this treatment of gravitational interaction.

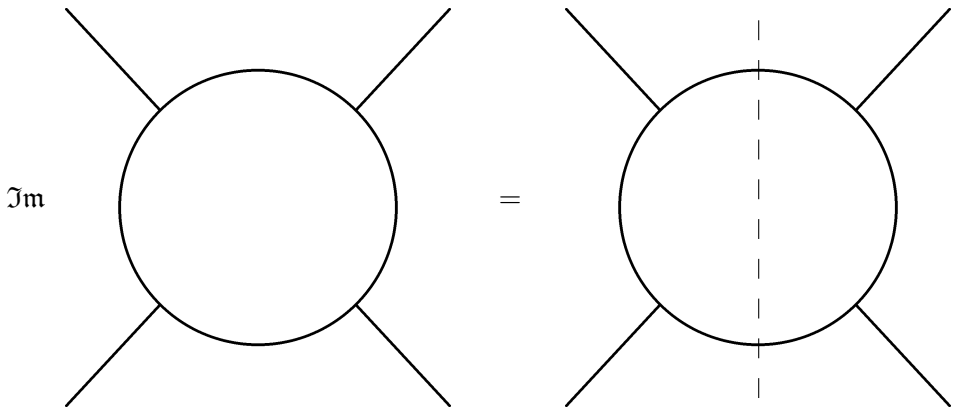
\begin{figure}
\begin{center}
\makebox[\textwidth]{
	\put(,25){$\mathfrak{Im}$}
	\begin{fmffile}{opticalL}
		\begin{fmfgraph*}(60,50)
		\fmfleftn{l}{2}\fmfrightn{r}{2}
\fmfrpolyn{empty,smooth, tension=0.5}{E}{4} \fmf{plain}{l1,E1}\fmf{plain}{l2,E2} \fmf{plain}{E4,r1}\fmf{plain}{E3,r2}	
		\end{fmfgraph*}
	\end{fmffile}
	\put(0,25){$=$}
	\begin{fmffile}{opticalR}
		\begin{fmfgraph*}(60,50)
		\fmfleftn{l}{2}\fmfrightn{r}{2}
\fmfrpolyn{empty,smooth, tension=0.5}{E}{4} \fmf{plain}{l1,E1}\fmf{plain}{l2,E2} \fmf{plain}{E4,r1}\fmf{plain}{E3,r2}	
		\end{fmfgraph*}
	\end{fmffile}
	\multiput(-30,1)(0,5){10}{\line(0,1){2.5}}
	}
\end{center}
\caption{General Feynman diagram with outgoing states cut by dashed lines. The rightmost part of the diagram is to be taken in its hermitian conjugate form, so that the whole diagrams results in a squared amplitude.\label{optical}}
\end{figure}

	As stated, our aim is  to evaluate the $h^{++}$ field generated by two particles which we represent as a state $\Psi$ of two localized wave packets
	.
		In standard scattering diagrams, each line cut by dashed lines corresponds to an outgoing state, while dashed lines themselves stand for a sum over all those states.  That is related to the optical theorem (see Section \ref{opticalthm})
	 as for Fig.~\ref{optical}. 
To have the diagrams yield directly a squared amplitude,
\ in our derivations,
every contribution deriving from diagram parts standing on the right of dashed lines has to be taken in its hermitian conjugate form   \cite{pes,rom2}.

We would like to recall that we are interested in the value of an $h^{++}$ field as it would result from a hypothetical measure of it.
To derive it, we consider diagrams with a leg attached on a particle line and ending in an $x$--coordinate event denoted by a cross. That feature is referred to as a \emph{field insertion}, for which we will supply a rule in the following.

To obtain $h^{++}(x)$ as a function of coordinates, we also Fourier--transform the resulting field from momentum space to coordinate space.

The expectation value of such a field is given by
\[
 \langle h^{++}(x) \rangle_{\Psi} \equiv \langle\Psi | h^{++}(x) | \Psi \rangle
\]
and we work in Heisenberg picture.  The two-particles wave packet is expanded in terms of $\ket{in}$ states according to
\[
\ket{\Psi} = \int \diff{\tilde p_1}~\diff{\tilde p_2}~ \Psi (p_1,p_2) |p_1, p_2, in\rangle
\]
where $\{ \ket{p_1,p_2;in} \}$ is a complete basis of the two--particles states and 
$\diff{\tilde p}$  denotes the Lorentz--invariant phase space (LIPS) measure defined as
\begin{equation}\label{lips}
\ddiff{}{\tilde p}=\frac{\ddiff{4}{p}}{(2\pi)^3}~ \delta(p^2) = (2\pi)^{-3} \frac{\ddiff{3}{\vec p}}{2E_p} \qquad.
\end{equation}
Inserting a two--particles identity decomposition
\[
\mathbb{I}_{(2~particles)} = \int\diff{\tilde k_1}~\diff{\tilde  k_2}~  \ket{k_1,k_2,in}\bra{k_1,k_2,in}
\]
we can evaluate
\begin{IEEEeqnarray*}{rcl}\label{hpsi}
 \langle h^{++}(x)\rangle_{\Psi} = &
\int \diff{\tilde  p_1} \diff{\tilde p_2} \diff{\tilde p'_1} \diff{\tilde p'_2} \diff{\tilde k_1} \diff{\tilde k_2} \Psi (p_1, p_2) \Psi^* (p'_1, p'_2) 
\\
&\phantom{\int}\times
{}_{out}\bra{k_1k_2}h(x)\ket{p_1p_2}_{in}~
{}_{out}\bra{k_1k_2}p'_1p'_2\rangle_{in}^* \qquad. \IEEEyesnumber
\end{IEEEeqnarray*}
In Eq.~\ref{hpsi} two brackets appear. The first one is just an (adjoint) $S$--matrix element
\[
{}_{out}\bra{k_1k_2}p'_1p'_2\rangle_{in}^*=
\left( {}_{in}\bra{k_1k_2}S\ket{p'_1p'_2}_{in} \right)^*
\equiv
S^*_{f(k_1k_2)\leftarrow i(p'_1p'_2)}
\]
while the second
\[
{}_{out}\bra{k_1k_2}h(x)\ket{p_1p_2}_{in} \equiv S_{h,f(k_1k_2)\leftarrow i(p_1p_2)}
\]
 contains the field insertion to be evaluated in the following, with the method of Feynman diagrams.

			\subsubsection{Localized wave packets}
	
	
In order to study the scattering of particles at given impact parameter, we ought to localize wave packets in coordinates space. We should handle that procedure with care because of Heisenberg principle effects which could cause an unwanted total uncertainty in momenta space.

 We start considering one--particle spinless states of mass $m$, for which \mbox{$p^2=m^2$}, leaving only $\vec p$ independent.
We assume the set of one--particle states $\{\ket{\vec p}, \vec p \in \mathbbm{R}^3\}$ to be complete:
\[
\ket{\psi}_{1~particle} = \int\ddiff{}{\tilde p}~\psi(\vec p)\ket{\vec p}\qquad.
\]
 We also ask for kets leading to covariant integration measures, that is
\[
\langle \psi | \psi \rangle = 1 = \int \frac{\ddiff{3}{\vec p}}{2E_p(2\pi)^3}|\psi(\vec p)|^2
\]
and
\[
||\psi||=\langle \psi | \psi \rangle = \int \ddiff{3}{\vec x}~|\psi(\vec x)|^2\qquad.
\]
Then wave packet kets should 
obey 
the covariant rule
\[
\langle \vec{p} | \vec{k} \rangle = (2\pi)^3~2E_p~\delta^3(\vec p - \vec k)
\]
and the LIPS defined in Eq.~\ref{lips}.
The Fourier transform reads as
\[
\psi(\vec x) \equiv \int \frac{\ddiff{3}{\vec p}}{(2\pi)^3\sqrt{2E_p}}\widetilde{\psi}(\vec p)e^{-i\vec p \cdot \vec x}\qquad.
\]

One-dimensional localized wave packets would be
\[
\widetilde{\psi}(\vec p)\propto \widetilde{\mathcal{N}}\exp\left\{\frac{1}{2}\left(\frac{\vec p-\vec{\bar{p}}}{\sigma}\right)^2+ip\bar{x}\right\}
\]

\[
\psi(\vec x)\propto \mathcal{N}\exp\left\{\frac{1}{2}\left(\frac{\vec x-\vec{\bar{x}}}{\lambda}\right)^2-i\bar{p}x\right\}
\]
where $\lambda=\frac{1}{\sigma}$ is the width in coordinate space. Phase factors have the effect of localizing Fourier transforms in the barred quantities $\bar x$ and $\bar p=\sqrt{s}/2$.

 Introducing the vector $\vec b_0\equiv(
(
b_{0x},b_{0y},z_0)$
, a three--dimensional packet passing by the space--time event $(t_0,\vec b_0)$ is then expressed by
\begin{equation}\label{wavepacket}
\frac{\widetilde{\psi}(\vec p)}{\sqrt{2E_p}} = \widetilde{\mathcal{N}}_z \widetilde{\mathcal{N}}_{x,y}
\exp\left\{\frac{1}{2}\left(\frac{\vec{p}-\vec{\bar{p}}}{\sigma}\right)^2-i\left[(\vec p-\vec{\bar{p}})\cdot \vec b_0 + \left(E_p - \sqrt{s}\right)t_0 \right]\right\}\qquad.
\end{equation}
				

			\subsection{Gravity vertices}\label{vertexsec}

We now estimate each vertex contribution 
 in the special case of gravity. While in eikonal high-energy ($p\simeq p'$) electro--magnetic scattering we get a vertex contribution
\[
i \lambda_e \equiv ie(p^{\mu}+p'^{\mu}) \simeq ie2p^{\mu}
\]
where $e$ is the electron charge, 
in (eikonal) gravity 
the interaction charge 
is proportional to the particle's
energy $\sqrt{s}$ itself. 
Actually, we deal with spin--2 gravitons whose contribution will be proportional to the combination
\[
p^{(\mu}p'^{\nu)} \equiv
p^{\mu}p'^{\nu}+p^{\nu}p'^{\mu}
\]
as follows:
\[
i \Gamma^{\mu\nu} \equiv -i\kappa~p^{(\mu}p'^{\nu)}  \qquad,
\]
$G$ being the Newton constant  $G=\kappa^2/8\pi$.

Consider a particle  $i$ moving along the positive light--cone direction, having $p^+=\sqrt{s}$ as the only non--vanishing component of momentum, and emitting the field $h^{++}$ as in the only vertex of Fig.~\ref{insertion0gravitons}. That vertex contribution amounts to
\[
i\Gamma^{\mu\nu} \to i\Gamma^{++} =-i\kappa~p_i^+p_i'^+ \equiv i\lambda_G 
\]
that is a form which will be used in the following. An analogous form holds for negative light--cone direction particles, involving $p(')_i^-$.


\subsection{Gravitational Feynman rules}\label{frules}

We can now summarize Feynman rules used in our derivation, apart from standard ones:

\begin{itemize}
\item for each eikonal vertex involving momenta $k_i$, $k_j$ and a (small) transferred momentum $k_l=k_i-k_j$ there is a factor $-i\kappa k_ik_j$ and an explicit factor $(2\pi)^4\delta^4(\sum_i k_i)$ of momentum conservation;
\item for each internal particle line with momentum $k$ there is a propagator\footnote{For particles with mass $m$,  energies in transplanckian regime are such that $k^0 \gg m$ and $D(k)=(k^2-m^2+i\epsilon)^{-1}\simeq(k^2+i\epsilon)^{-1}$.} \mbox{$D(k)=(k^2+i\epsilon)^{-1}$} and an integration with measure $\ddiff{4}{k}/(2\pi)^4$;
\item for each ingoing or outgoing external particle line there is its  wave function $\psi(k)$ and an integration over its momentum $k$;
\item the field insertion  at point $x$ (which we depict in diagrams using a cross) contains a factor $\kappa \exp\{-iqx\}$ where $q$ is the momentum of the attached exchanged particle.
\end{itemize}

Propagators and vertices can be derived from their exact versions \cite{ssv}.

Other general Feynman rules are obtained in standard ways for a $\phi^3$ theory by a perturbative procedure involving sum over all field contractions accounted for by Wick's theorem; see \emph{e.g.} \cite{cas,pes,rom1}.

We are now ready to perform explicit field calculations.

	\section{Field insertion without graviton exchanges}
	
	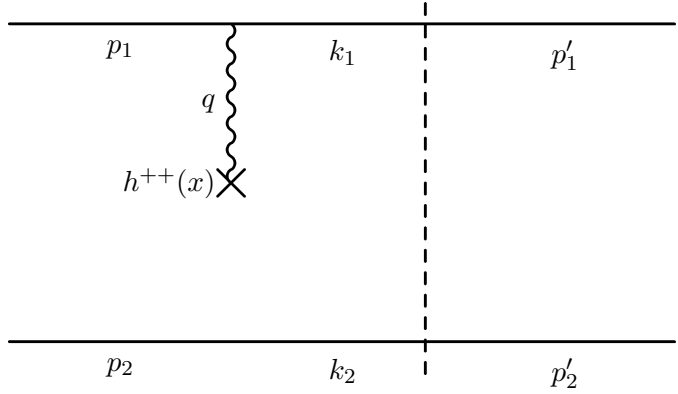
\begin{figure}
\begin{center}
	\begin{fmffile}{insertiononly1}
\begin{fmfgraph*}(104,40)
\fmfleft{i1,i2}
\fmfright{i3,i4}
\fmf{plain,label=$p_2$}{i1,v1}\fmf{plain,label=$k_2$}{v1,o1}\fmf{plain,label=$p'_2$}{o1,i3}
\fmf{plain,label=$p_1$}{i2,v2}\fmf{plain,label=$k_1$}{v2,o2}\fmf{plain,label=$p'_1$}{o2,i4}
\fmffreeze
\fmf{boson,label=$q$
}{v2,vx}
\fmf{phantom}{v1,vx}
\fmfv{decor.shape=cross,label=$h^{++}(x)$}{vx}
\fmfforce{(.6w,-0.1h)}{o3}
\fmfforce{(.6w,1.1h)}{o4}
\fmf{dashes}{o3
,o4}
\end{fmfgraph*}
	\end{fmffile}
\end{center}
\caption{Diagram with a field insertion without graviton exchanges, resulting in field  $\langle h^{++}\rangle^{(0)}$.\label{insertion0gravitons}}.
\end{figure}

	Calculation of a field insertion in a high--energy scattering at zeroth order, that is without interaction via gravitons exchange, will result in the known AS--like logarithmic field profile function with a delta behavior on the generating particle $x^-$ surface:
\begin{equation}\label{insertiononly}
\langle h^{++}\rangle^{(0)} = R~\ln \frac{L^2}{x^2}~\delta(x^-)\qquad.
\end{equation}

The derivation of that result is performed as follows, in reference to Fig.~\ref{insertion0gravitons}.
The rightmost part of the diagram yields two momentum Dirac's deltas, which select $p'_2=k_2$ and $p'_1=k_1$. From the leftmost part, another delta selects $p_2=k_2$. The unique vertex accounts to $-i\kappa p_1^+k_1^+$. There is a propagator $(q^2-i\epsilon)^{-1}$ of the unique exchanged particle and a factor $\kappa\exp\{-iqx\}$ for the presence of the insertion. Then, there are the four particle wave packets $\psi$, expressed in the form (\ref{wavepacket}). Momentum conservation at the vertex allows to change integration variable from $k_1=p'_1$ to $q$.
Summing up,
\begin{eqnarray*}
\langle h^{++}\rangle^{(0)} &=&
(-i)\kappa^2\widetilde{\mathcal{N}}^4\int \ddiff{}{ p^+_1}~\ddiff{2}{\transverse{p}_1}~\ddiff{}{q^+}\ddiff{2}{\transverse{q}}~\frac{1}{q^+q^--\transverse{q}^2+i\epsilon}
\\
&\times& 
p_1^+
 \exp\left\lbrace -\frac{1}{2\sigma^2}\left[ (p_1^+-\sqrt{s})^2 + (p_2^--\sqrt{s})^2 +(q^+)^2+(q^-)^2+\right. \right.\\
&\times&\left.\left.+2(\transverse{p}_1)^2 +2(\transverse{p}_2)^2+(\transverse{q})^2\right] -\frac{i}{2}\left[q^-x^++q^+x^--2\transverse{q}\cdot(\transverse{x}-\transverse{b})\right]\right\}.
\end{eqnarray*}
where $p_2^-$, $\transverse{p}_2$, $q^-$ are functions of integration variables.
	
	Integration over $p_1^+$ yields the factor $R=2G\sqrt{s}$, due to the presence of $\kappa^2$ and the momentum gaussian with selects $p_1^+=\sqrt{s}$.

	Integrating over $\transverse{q}$ the factor 
	\[
	\frac{1}{\transverse{q}^2}\exp\left\lbrace -\frac{\transverse{q}^2}{2\sigma^2} +i \transverse{q}\cdot(\transverse{x}-\transverse{b}) \right\rbrace
	\]
	requires the IR cut--off described in Section \ref{cutoff} and yields the AS--like function profile $\ln \left(L^2/(\transverse{x}-\transverse{b})^2\right)$.
	
	It is the integration over $q^+$ which yields the $\delta$ support to the profile function, because\footnote{In the last step we consider the limit $\sigma \to 0$ because in the scattering process we describe we do not resolve wave packet's widths.}
\[
\int \diff{q^+} \exp\left\{-\frac{\sigma^2}{16}(q^+)^2-\frac{i}{2}q^+x^-\right\}\\
\stackrel{\sigma \to 0}{=}
4 \pi  \delta(x^-) \qquad.
\]

	\section{Field insertion before a graviton exchange}
	


\begin{figure}
		
\begin{center}
\begin{fmffile}{insertionbefore1rung1}
\begin{fmfgraph*}(104,40)
\fmfleft{i1,i2}
\fmfright{i3,i4}
\fmf{plain,label=$p_2$}{i1,v1}
\fmf{plain}{v1,v3}
\fmf{plain,label=$k_2$}{v3,o1}
\fmf{plain,label=$p'_2$}{o1,i3}
%
\fmf{plain,label=$p_1$,label.side=left}{i2,v2}
\fmf{plain,label=$k$,label.side=left}{v2,v4}
\fmf{plain,label=$k_1$,label.side=left}{v4,o2}
\fmf{plain,label=$p'_1$,label.side=left}{o2,i4}
\fmffreeze
\fmf{wiggly,label=$Q$}{v3,v4}
\fmf{boson,label=$q$
}{v2,vx}
\fmf{phantom}{vx,v1}
\fmfv{decor.shape=cross,label=$h^{++}(x)$,label.side=left}{vx}
\fmfforce{(.7w,-0.1h)}{o3}
\fmfforce{(.7w,1.1h)}{o4}
\fmf{dashes}{o3,o4}
\end{fmfgraph*}
\end{fmffile}
\end{center}
\caption{ Diagram $\mathcal{A}$, representing a field insertion before one graviton exchange.\label{beforeonerung}}
\end{figure}
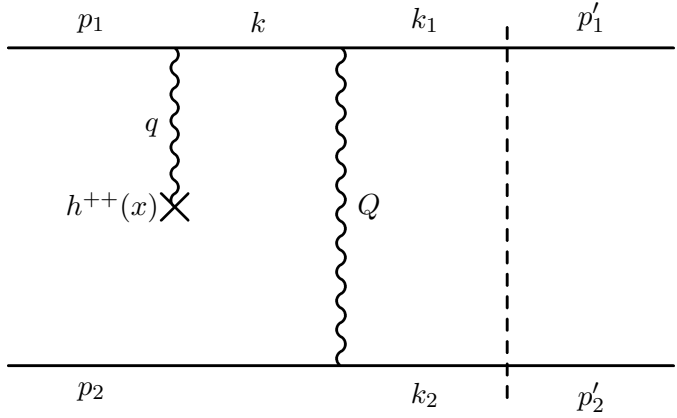

Here we find the contribution $ \langle h^{++}(x) \rangle^{\mathcal{A}}$ to
$
\bra{\Psi}h^{++}(x)\ket{\Psi}$
given by diagram $\mathcal{A}$ depicted in Fig.~\ref{beforeonerung}.

The rightmost part of that diagram, for which $S$ is the identity $\mathbb{I}$, yields
\[
{}_{in}\bra{k_1k_2}\mathbb{I}\ket{p'_1p'_2}_{in}=
\widetilde\delta(k_1-p'_1)~\widetilde\delta(k_2-p'_2)
\]
while the leftmost part sums up to
\[
i\lambda_A ~i\lambda_B ~i\lambda_C ~iD(q)~iD(Q)~i\Delta(k)~e^{-iqx}
\]
where $A,B,C$ label each one of the three vertices. For each vertex we have a $ i \lambda_j (p_{in},p_{out}) = - i \kappa p_{in} p_{out}$ factor where $p_{in},p_{out}$ are the relevant light--cone components of the involved momenta at that vertex, as explained in Section \ref{vertexsec}. 
Given
\[
\Delta(k)=\frac{1}{k^2-m^2+i\epsilon},\qquad
D(q)=\frac{1}{q^2+i\epsilon}
\]
with $i\epsilon \equiv i0^+$, we obtain
\begin{eqnarray*}
\langle h^{++}(x) \rangle^{\mathcal{A}} &=& \int \diff{\tilde{p}_1}~\diff{\tilde{p}_2}~\diff{\tilde{p}'_1}~\diff{\tilde{p}'_2}~\diff{\tilde{k}_1}~\diff{\tilde{k}_2}~
\tilde\psi_1(p_1)~\tilde\psi_2(p_2)~\tilde\psi^*_1(p'_1)~\tilde\psi^*_2(p'_2)~\\
&&
\times (-\lambda_A\lambda_B\lambda_C)
~D(q)~D(Q)~\Delta(k)~\tilde{\delta}(k_1-p'_1)~\tilde{\delta}(k_2-p'_2)~e^{-iqx}.
\end{eqnarray*}

Thanks to momenta conservations at vertices,
\[
\left\lbrace
\begin{array}{rcl}
Q&=&k_2-p_2=p'_2-p_2\\
k&=&k_1+Q=p'_1+p'_2-p_2\\
q&=&p_1-k=p_1-p'_1+p_2-p'_2
\end{array}
\right.
\]
 we can change integration variables to
\begin{eqnarray*}
\diff{p'_2}{}^- &=& \diff{(p_2^-+q^-)}=\diff{Q^-}\\
\ddiff{2}{\transverse{p}'_2}&=&\ddiff{2}{\transverse{Q}}\\
\diff{p'_1}{}^+ &=& \diff{(p_1^+-q^+-Q^+)}=\diff{q^+}\\
\ddiff{2}{\transverse{p}'_1}&=&\ddiff{2}{(\transverse{p}_1-\transverse{Q}-\transverse{q})}
=\ddiff{2}{\transverse{q}}\qquad.
\end{eqnarray*}
We choose $p_1$, $p_2$, $q$ and $Q$ as integration variables, and the usual (explained above) relativistic measure.

\begin{figure}
\begin{center}
	\begin{picture}(120,69)(-5,0)
		\put(0,10){\vector(1,1){55}}\put(55,60){$x^+$}
		\put(25,35){\vector(1,1){25}}\put(45,50){$p^+$}
				\put(50,10){\vector(-1,1){55}}\put(0,65){$x^-$}
				\put(30,50){$p'{}^+$}
				 \qbezier(0,60)(0,57)(25,57)\qbezier(25,57)(50,57)(50,60)
				\qbezier(0,60)(5,62)(25,62)\qbezier(25,62)(50,62)(50,60)
		\put(25,35){\vector(3,4){17}}
\put(-5,0){
		\put(65,35){\vector(1,0){50}}\put(100,40){$p^+$}
				\put(65,35){\vector(4,-1){48}}\put(90,20){$p'{}^+$}
						\put(115,35){\vector(0,-1){12}}\put(117,27){$\transverse{q}$}
						\put(115,23){\vector(-1,0){2}}\put(113,20){$q^-$}
		}
	\end{picture}
\end{center}
\caption{Momenta on their light--cone. The angle between ingoing and outgoing momenta has been exaggerated for clarity purposes.\label{lightconemomenta}}
\end{figure}
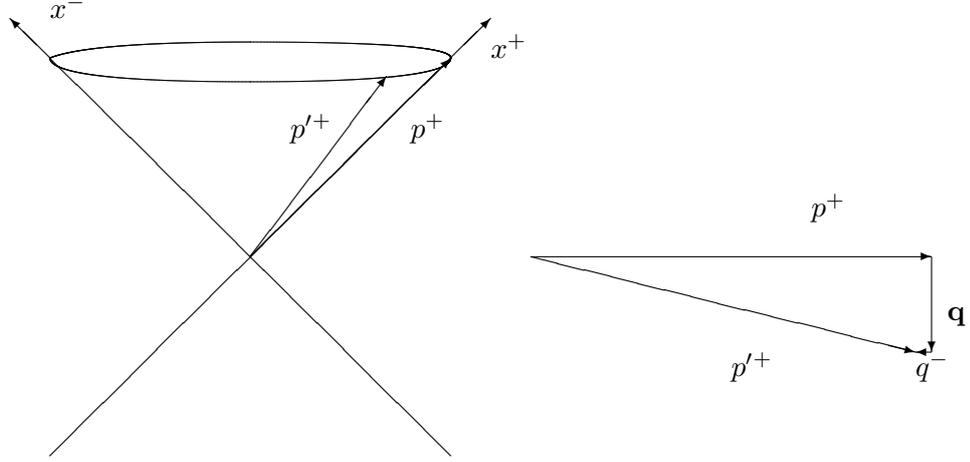

Momenta's order of magnitude hierarchy is
\[
p_1^+, p_2^-, q^- \gg \transverse{p}_1, \transverse{q}_1, \transverse{p}_2 \gg p_1^-, q^-, p_2^+
\]
given that leftmost momenta are of order $\sqrt{s}$ (energy in the center--of--mass), middle ones of order $\eta \sqrt{s}$ and rightmost ones of order $\eta^2 \sqrt{s}$, where 
\[
\eta \sim \sqrt{\frac{t}{s}} \sim \theta_s \sim \frac{R}{b}
\]
 is a measure of validity of eikonal regime condition for this scattering and the sharpness of the wave packets gaussian. 
That can be easily pictured by considering a light--cone on which momenta lay---see Fig.~\ref{lightconemomenta}.
Then we find the  useful relations
\begin{itemize}

\item $\transverse{p}^2\ll p_2^-$ and $q^-\ll p_1^+$
\item $q^-q^+ \ll \transverse{q}^2$.
\end{itemize}
Those relations let neglect $p_2^+$, $p_1^-$ and $(q^-+Q^-)$ so that $q^-\simeq - Q^-$ and $\diff{q^-}=\diff{Q^-}$. 
That also affects each wave packet; let us take $\widetilde{\psi}_1(p)$ as an example. The covariant normalization factor is approximated as
\[
\sqrt{p^++p^-}\simeq \sqrt{p^+}
\]
and longitudinal part as
\[
\widetilde{\psi}_{1L}\simeq \exp\left\lbrace -\frac{1}{2}\left(\frac{p^+ - \bar{p}^+}{2\sigma_z}\right)^2 -\frac{i}{2}\left[(p^+-\bar{p}^+)z_0-p^+t_0\right]\right\rbrace\qquad.
\]
Transverse part in unchanged; we recall that it contains the transverse  parameter $\transverse{b}_1$.
Summing up, diagram $\mathcal{A}$ yields
\begin{eqnarray*}
\langle h^{++}(x) \rangle^{\mathcal{A}} &\simeq& 	-\frac{i\kappa^3}{2^{4}(2\pi)^{12}} \int \frac{\diff{p_1^+}}{p_1^+}~\frac{\diff{p_1'^+}}{p_1'^+}~\frac{\diff{p_2^-}}{p_2^-}~\frac{\diff{p_2'^-}}{p_2'^-}~\ddiff{2}{\transverse{p}_1}~\ddiff{2}{\transverse{p}'_1}~\ddiff{2}{\transverse{p}_2}~\ddiff{2}{\transverse{p}'_2}
\\
	&&\times \left( \widetilde{\mathcal{N}}_z~\widetilde{\mathcal{N}}_{x,y} \right)^4\sqrt{p_1^+ p_1'^+ p_2^- p_2'^-} ~p_1^+ ~k^+{}^2 ~p'^+_1 ~p_2^-~p'^-_2 \\
	&&\times \frac{1}{q^+q^--\transverse{q}^2+i\epsilon} ~ \frac{1}{Q^+Q^--\transverse{Q}^2+i\epsilon}  ~\frac{1}{k^+k^--\transverse{k}^2+i\epsilon} \\
	&&\times \exp \left\lbrace -\frac{1}{8\sigma^2} \left[ (p_1^+-\sqrt{s})^2 
	+(p'^+_1-\sqrt{s})^2 + (p_2^+-\sqrt{s})^2 + \right.\right.\\ 
	&& \phantom{\exp \lbrace }\left.\left. + (p'^+_2-\sqrt{s})^2 \right] + \frac{i}{2} \left[ (p^+_1-p'^+_1)x^- + (p^-_1-p'^-_1)x^+  \right] \right\rbrace \\
	&&\times \exp \left\lbrace -\frac{1}{2\sigma^2}\left( \transverse{p}^2_1 +  \transverse{p}'^2_1 +  \transverse{p}^2_2+ \transverse{p}'^2_1\right) -i \bigg[\left( \transverse{p}_1 -\transverse{p}'_1\right) \cdot \transverse{b}_1 \right. +\\
	&&	 \phantom{\exp \lbrace }\left. + \left( \transverse{p}_2 -\transverse{p}'_2\right)\cdot \transverse{b}_2 +\frac{q^+x^-}{2}+\frac{q^-x^+}{2}-\transverse{q}\cdot\transverse{x}\bigg] \right\rbrace
\end{eqnarray*}
where we ought to change integration variables to components of $q$ and $Q$ as stated above.

The quadratic forms of longitudinal components can be diagonalized by using new integration variables to $P$ and $p$, defined by\footnote{We remark that $P$ and $p$ are opportune light--cone components of \mbox{4--vectors}.}
\[
p_1^+ \equiv P+\frac{q^+}{2}, \quad p_2^-\equiv p+\frac{q^-}{2}\qquad.
\]
We also note that
\begin{equation*}
\sqrt{p_1^+-q^+} = \sqrt{P-\frac{q^+}{2}} \simeq \sqrt{P}\left( 1-\frac{1}{4}\frac{q^+}{P} \right)
\end{equation*}
causes the appearance of a delta derivative as detailed below.

First we perform integration on longitudinal components of $q$, \emph{i.e.} $q^+$ and $q^-$, that are expected to result in the wave front, while those in $p_1^+$ and $p_2^-$ in a normalizing factor.

The integrand has two poles in $q^-$:
\[
 q^-_{1P} = \frac{-(\transverse{p}-\transverse{q})^2+i\epsilon}{p_1^+-q^+}
, \quad q^-_{2P} = \frac{\transverse{q}^2-i\epsilon}{q^+}
\]
whose the first is responsible of the appearance of $\Theta (x^+_0-x^+)$; that is a general result of the presence of the leg between field insertion and graviton exchange, which causes, by $q^-$ integration, the appearance of a $\Theta(x^+)$.
\begin{eqnarray*}
&&\int \diff{q^-}  \frac{\sqrt{p}\left(1-\frac{1}{4}\frac{q^-}{p}\right)e^{-\frac{i}{2}q^-(x^+-x_0^+)}}{(q^+q^--\transverse{q}^2+i\epsilon)\left(P-\frac{q^+}{2}\right)} \frac{1}{q^-+\frac{(\transverse{P}-\frac{\transverse{q}}{2})^2}{P-\frac{q^+}{2}}-i\epsilon}
=\\
&=&\Theta (x_0^+ - x^+) ~2\pi i \frac{\sqrt{p}-\frac{1}{4}\frac{q^-}{\sqrt{p}}}{(q^+q^--\transverse{q}^2)(P-\frac{q^+}{2})}e^{-\frac{i}{2}q^-(x^+-x_0^+)}\\
&\simeq & 2\pi i \frac{\sqrt{p}}{-\transverse{q}^2(P-\frac{q^+}{2})}\Theta(x_0^+-x^+)\qquad.
\end{eqnarray*}
 In the numerators of the integrand, pole values themselves, being of order
$
q^-_{P}\sim \frac{
\transverse{q}}{\sqrt{s}},
$
 can be neglected
.

While $q^-$ integration causes the appearance of a $\Theta$ as a function of $x^+$, now we show how $q^+$ integration causes the appearance of a  derivative of a $\delta$ as a function of $x^-$.
We can expand a denominator factor in $q^+/P$ as
\[
\left( 1-\frac{1}{4}\frac{q^+}{P} \right)^{-1} \simeq
1+\frac{1}{4}\frac{q^+}{P}\qquad.
\]
Because of the presence of $\exp\{q^+(x^--x_0^-)\}$ in integrand, a factor $q^+$ can be extracted from the integral in the form of a $\partial_-\equiv \partial_{x^-}$ derivative:
\begin{eqnarray*}
&&\int \diff{q^+} \frac{\exp\{-\frac{\sigma^2}{16}(q^+)^2-\frac{i}{2}q^+(x^--x_0^-)\}}{\left( 1-\frac{1}{4}\frac{q^+}{P} \right)}\\
&\simeq&
\left( 1-\frac{1}{4}\frac{i\partial_-}{2P} \right) \int \diff{q^+} \exp\left\{-\frac{\sigma^2}{16}(q^+)^2-\frac{i}{2}q^+(x^--x_0^-)\right\}\\
&\stackrel{\sigma \to 0}{=}&
4 \pi \left[ \delta(x^--x_0^-) + \frac{i}{2P}\delta' (x^--x_0^-)\right]
\end{eqnarray*}
where in the last step we consider the limit $\sigma \to 0$ because in the scattering process we describe we do not resolve wave packet's widths. That is also  used in the following.

Then 
 we perform integration on transverse components $\transverse{q}$. The cut--off procedure detailed in Section \ref{cutoff} is needed. The result is a term proportional to the profile function $a(x)$:
 \[
 \int\ddiff{2}{\transverse{q}}\frac{e^{i\transverse{q}\cdot\left(\transverse{x}-\transverse{b}_1\right)-\frac{\transverse{q}^2}{2\sigma^2}}}{\transverse{q}^2}
 \stackrel{cut-off}{=}\ln \frac{L^2}{|\transverse{x}-\transverse{b}_1|^2}\qquad.
 \]
 
 The same procedure is performed w.r.t.\ integration over $\transverse{Q}$, resulting in a term proportional to the first--order amplitude $\mathcal{A}(s,b)$:
\[
 \int\ddiff{2}{\transverse{Q}}\frac{e^{i\transverse{Q}\cdot\left(\transverse{b}_2-\transverse{b}_1\right)-\frac{\transverse{Q}^2}{2\sigma^2}}}{\transverse{Q}^2}
 \stackrel{cut-off}{=}\ln \frac{L^2}{|\transverse{b}_2-\transverse{b}_1|^2}\qquad.
 \]

Then we find that the eikonal exchange of one graviton on the right of a field insertion, 
 at eikonal first--order,
yields a contribution
\[
\langle h^{++}(x) \rangle^{\mathcal{A}} = -R~\ln \frac{L^2}{|\transverse{x}-\transverse{b}_1|^2}~\delta'(x^--x^-_0)~R~\ln \frac{L^2}{|\transverse{b}_2-\transverse{b}_1|^2}~\Theta(x^+_0-x^+)\qquad.
\]

				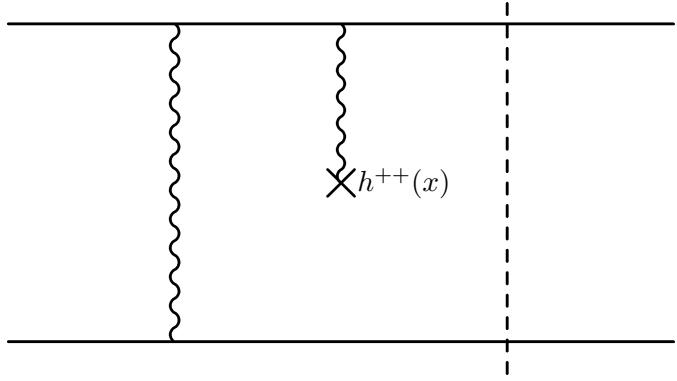
\begin{figure}[!t]
\begin{center}
\begin{fmffile}{insertionafter1rung1}
\begin{fmfgraph*}(104,40)
\fmfleft{i1,i2}
\fmfright{i3,i4}
\fmf{plain}{i1,v1,v3,o1,i3}
\fmf{plain}{i2,v2,v4,o2,i4}
\fmffreeze
\fmf{wiggly}{v1,v2}
\fmf{boson$
}{v4,vx}
\fmf{phantom}{vx,v3}
\fmfv{decor.shape=cross,label=$h^{++}(x)$}{vx}
\fmfforce{(.7w,-0.1h)}{o3}
\fmfforce{(.7w,1.1h)}{o4}
\fmf{dashes}{o3,o4}
\end{fmfgraph*}
\end{fmffile}
\end{center}
\caption{Diagram $\mathcal{A}'$, representing a field insertion after a graviton exchange.\label{afteronerung}}
\end{figure}

	\section{Field insertion after a graviton exchange}

The contribution yielded by diagram $\mathcal{A}'$ of Fig.~\ref{afteronerung}  is analogous to the previous one, with the notable difference of the presence of $\Theta (x^+-x^+_0)$ in place of $\Theta (x^+_0-x^+)$. Choosing, in order to simplify notation, $x^+_0=x^-_0=0$ and defining $\transverse{x}_b\equiv \transverse{x}-\transverse{b}_1$ and $\transverse{b}\equiv\transverse{b}_2-\transverse{b}_1$, we have
\[
\langle h^{++}(x) \rangle^{\mathcal{A}'} = R~\ln \frac{L^2}{\transverse{x}_b^2}~\delta'(x^-)~R~\ln \frac{L^2}{\transverse{b}^2}~\Theta(x^+)\qquad.
\]



In both computations of one graviton exchange we find a typical field structure. Let us focus on the latter. The results shows the following structure: a wave front appearing in the future of the scattering (expressed by the $\Theta$ function), multiplied by the usual profile function and the scattering amplitude. In this case, that is the first--order amplitude, resulting proportional to
$\delta'(x^-)$. 

 The delta--derivative is indeed interpreted as a first--order expansion of the $\delta$ selecting the plane of motion $x^-$ of the field--generating particle. We conjecture that all the $n$--orders eikonal exchanged would be resummed in a shifted $\delta$ which expresses AS' shifts. 
 Then, we turn to the computation  in the case of the exchange of an infinite number of gravitons. 

	\section{Field insertion in infinite gravitons exchanges}\label{sectionresult}
	
	We present here the result that is the original part of our work.

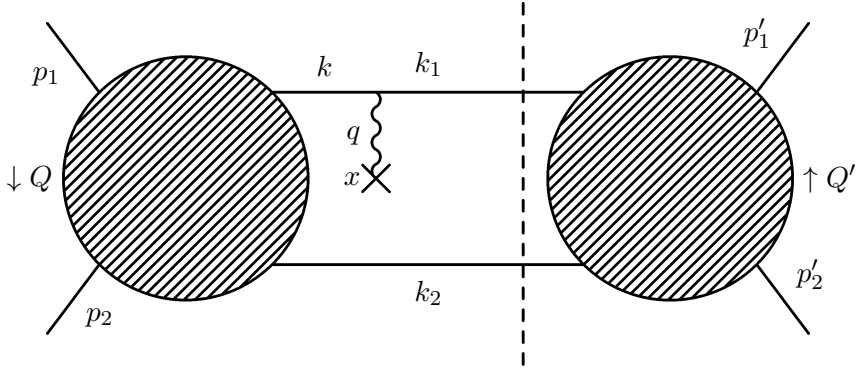
\begin{figure}
\begin{center}
\begin{fmffile}{insertionbetween2blobs1}
\begin{fmfgraph*}(119,39)
\fmfleft{i1,i2}
\fmfright{i3,i4}
\fmfrpolyn{smooth,tension=.5,filled=shaded,label=$\downarrow Q$\phantom{this to shift momentum 
}
}{v}{4}
\fmfrpolyn{smooth,tension=.5,filled=shaded,label=\phantom{this to shift momentum label a little more}$\uparrow Q'$
}{b}{4}
\fmf{plain,tension=2,label=$p_2$}{i1,v1}
\fmf{plain,tension=2,label=$p_1$}{i2,v2}
\fmf{plain,tension=1}{v4,v5}
\fmf{plain,tension=1,label=$k_2$,label.side=right}{v5,o1}
\fmf{plain,tension=1}{o1,b1}
\fmf{plain,tension=1,label=$k$,label.side=left}{v3,v6}
\fmf{plain,tension=1,label=$k_1$,label.side=left}{v6,o2}
\fmf{plain,tension=1}{o2,b2}
\fmf{plain,tension=2,label=$p'_2$}{i3,b4}
\fmf{plain,tension=2,label=$p'_1$}{i4,b3}
\fmffreeze
\fmf{boson,label=$q$,label.side=left
}{v6,vx}
\fmf{phantom}{vx,v5}
\fmfv{decor.shape=cross,label=$x$}{vx}
\fmfforce{(.6w,-0.1h)}{o3}
\fmfforce{(.6w,1.1h)}{o4}
\fmf{dashes}{o3,o4}
\end{fmfgraph*}
\end{fmffile}
\end{center}
\caption{Field insertion between two blobs. Each blob represent an infinite ladder of eikonal exchanges, transferring from topmost to bottom lines total momenta $Q$ and $-Q'$ respectively.\label{insertionbetween2blobs1}}
\end{figure}

We consider the diagram of Fig.~\ref{insertionbetween2blobs1}, which represents a process involving all orders of gravitons exchanges of simple ``ladder'' type, that is, without rescattering or H--diagrams contributions. 

Each blob represents a ladder made of an infinite number of rungs, where each rung stands for a single graviton exchange, summing up in a total exchanged momentum $Q$ and $Q'$. Let us consider leftmost blob; the other one is analogous.  The blob computation, performed by iteration of the one--graviton diagram,\footnote{See the derivation by Ciafaloni and Colferai in \cite{c-c}.}  yields
\[
\bra{k_1k_2}S\ket{p_1p_2} =
2~s~(2\pi)^4~\delta^4(p_1+p_2-k_1-k_2)
\int \ddiff{2}{\mathbf{b}}~e^{i\mathbf{Q}\cdot\mathbf{b}+i\mathcal{A}}
\]
where $s=(p_1+p_2)^2$, $\mathcal{A}=\mathcal{A}(s,b)=Gs\ln (L^2/b^2)$ and $\mathbf{Q}$ is a fictitious \mbox{2--dimensional} vector with
\[
\mathbf{Q}^2 = -(p_1-k_1)^2 > 0
\]
which can be identified with the transverse component of $Q$.
Then, apart from momentum conservations deltas, the product of the two blob  accounts to 
\[
4~s_1 s_2
\int \ddiff{2}{\mathbf{b}}~ \ddiff{2}{\mathbf{b'}} \exp \left\{ i \left[ \mathbf{Q}\cdot\,\mathbf{b}+\mathcal{A}(s_1,b) - \mathbf{Q'}\cdot\,\mathbf{b'}-\mathcal{A}(s_2,b') \right] \right\}
\]
where $ s_1\equiv s(p_1,p_2) $ and $s_2\equiv s(p'_1,p'_2)$.

As usual we will calculate that diagram contribution in momentum space and integrate it over external momenta with wave packets to obtain $\langle h \rangle$.

Apart from blobs, we have two vertices and two internal momenta $k$ and $q$. Vertex $A$ accounts to $i \kappa \lambda_A = -i \kappa k k_1$ and in $x$ we have another $\kappa$ factor:\footnote{See the rule for field insertions in Section \ref{frules}.}
\[
-i\kappa^2 \int \ddiff{}{\tilde{k}}~ \ddiff{}{\tilde{q}} ~\Delta(k) ~D(q) ~e^{-iqx}\qquad.
\]
Here the last exponential factor is the unique ``survivor'' from coordinate integrations because it links the crossed vertex $x$ where it is not associated with any particle.
As said we integrate over all external momenta using relativistic measure.
Momentum conservation at each vertex leads to a factor
\[
(2\pi)^{12}\delta^4(k-k_1-q)\delta^4(p_1+p_2-k-k_2)\delta^4(p'_1+p'_2-k_1-k_2)
\]
and we weight $b_1$, $b_2$ and ingoing momenta integrations on wave packets
\[
\Psi(p_1,p_2,b_1,b_2)\Psi^*(p'_1,p'_2,b_1,b_2) = \widetilde{\psi}_1(p_1,b_1) ~ \widetilde{\psi}_2(p_2,b_2) ~ \widetilde{\psi}^*_1(p'_1,b_1)  ~ \widetilde{\psi}^*_2(p'_2,b_2) \ .
\]

All that said, diagram in Fig.~\ref{insertionbetween2blobs1} accounts to

\begin{eqnarray*}
\langle h^{++}(x) \rangle
 &=& 
(2\pi)^{4\cdot 3}\int \ddiff{4}{\tilde{p}_1}~\ddiff{4}{\tilde{p}_2}~\ddiff{4}{\tilde{p}'_1}~\ddiff{4}{\tilde{p}'_2}~
\ddiff{4}{\tilde{k}_1}~\ddiff{4}{\tilde{k}_2}~\ddiff{4}{\tilde{k}}~\ddiff{4}{\tilde{q}}~
\ddiff{2}{\mathbf{b_1}}~\ddiff{2}{\mathbf{b_2}}~
\\ 
&\times &
\delta^4(k-k_1-q)~\delta^4(p_1+p_2-k-k_2)~\delta^4(p'_1+p'_2-k_1-k_2)~
\\
&\times& i\Delta(k)~iD(q)~e^{-iqx}~\kappa~i\lambda_G (k,k_1)~2s(p_1,p_2)~2s(p'_1,p'_2)\\
&\times&\exp\left\lbrace i\left[ \mathbf{Q}\cdot \mathbf{b_1} + \mathcal{A}\left( s(p_1,p_2)\right) 
- \mathbf{Q'}\cdot \mathbf{b_2} - \mathcal{A}\left( s(p'_1,p'_2)\right) \right] \right\rbrace \\
&\times &
\widetilde{\psi}_1(p_1)~ \widetilde{\psi}_2(p_2)~\widetilde{\psi}^*_1(p'_1)~ \widetilde{\psi}^*_2(p'_2)~
\end{eqnarray*}
where
\begin{eqnarray*}
\widetilde{\psi}_1(p_1,b_1) &=& \sqrt{p_1^++p_1^-}\widetilde{\mathcal{N}}_1\widetilde{\mathcal{M}}_1
\exp\left\lbrace -\frac{1}{2}\left( \frac{p_1^+-\bar{p_1^+}}{2\sigma_z}\right) 
+\frac{1}{2\sigma_x}\transverse{p}_1^2 \right. +\\
&&- \left. \frac{i}{2} \left[ \left(p_1^+-\bar{p_1^+}\right) z_{01} - \left( p_1^+-\bar{p_1^+}\right) t_0
+2\transverse{p}_1\cdot \transverse{b}_{01}\right]  \right\rbrace \qquad,
\end{eqnarray*}
$\transverse{p}_i$ denotes 2-dimensional transverse momenta components. An analogous expression holds for $\widetilde{\psi}_2(p_2,b_2)$. We used $p_i^+=p_i^0+p_{iz}=2p_{iz}$  and we take:
\begin{eqnarray*}
\widetilde{\mathcal{N}}\widetilde{\mathcal{M}} = \widetilde{\mathcal{N}}_i\widetilde{\mathcal{M}}_i &=& 2^{1/2}\pi^{1/4}\lambda^{1/2}~2~\pi^{1/2}\lambda=(2\lambda)^{3/2}\pi^{3/4}\\
\lambda^{-1} &=& \sigma_z =\sigma_x=\sigma_y=\sigma\\
\bar{p_1^+} = \bar{p_2^-}  &=& \sqrt{s}\qquad.
\end{eqnarray*}
We also consider ingoing particles along light-cones coordinates; for mono\-chromatic plane waves that would mean
\[
p(')_1 = (p_1^+,0,0,0),\qquad p(')_2 = (0,p_2^-,0,0)
\]
however we take highly--peaked wave packets which are centered around those values, so that in extremely good\footnote{The approximation is as good as neglecting a correction laying at several standard deviations from the $p$--s gaussian's peak.} approximation\footnote{For ease of visualization of every approximation considered, we will denote each with a bullet in the following.}
\begin{itemize}
\item $p(')_1^+\gg p(')_1^-$ and $p(')_2^-\gg p(')_2^+$
\item $p(')_1\simeq (p(')_1^+,0,0,0)$ and $p(')_2\simeq (0,p(')_2^-,0,0)$.
\end{itemize}
Because of the eikonal scattering condition, the same holds for outgoing particles:
\begin{itemize}
\item $k_1^+\gg k_1^-$ and $k_2^-\gg k_2^+$
\item $k_1\simeq (k_1^+,0,0,0)$ and $k_2\simeq (0,k_2^-,0,0)$
\end{itemize}
so that, to sum up, $p_1^+ +p_1^-\simeq p_1^+$, $p_2^+ +p_2^-\simeq p_2^-$ and $k_1^2\simeq (k_1^+)^2,$ etc.
Each one of the 6 external momenta has a measure analogous to (\emph{e.g.} for $p_1$)
\[
\ddiff{4}{\tilde{p_1}}~\delta(p_1^2) = 
\frac{1}{(2\pi)^3}\frac{\diff{p_1^+}\diff{p_1^-}}{2}\ddiff{2}{\transverse{p}_1}~\delta(p_1^+p_1^--\transverse{p}^2)=
\frac{\diff{p_1^+}}{2(2\pi)^3p_1^+}\ddiff{2}{\transverse{p}_1}
\]
while, for the 2 internal momenta $k$ and $q$, to 
\[
\ddiff{4}{\tilde{k}} = \frac{1}{(2\pi)^4}\frac{\diff{k^+}\diff{k^-}}{2}\ddiff{2}{\transverse{k}}
\]
resulting in an overall factor $2^{-6-2}(2\pi)^{-3\cdot 6-4\cdot 2}$.

The unique vertex accounts to $\sigma(k,k_1)=-\kappa k k_1$ with $\kappa^2\equiv 8\pi G$.
Then (using a abbreviated notation for vector variables differentials)
\begin{eqnarray*}
\langle h^{++}(x) \rangle &=& -i^3~2^{-8}~(2\pi)^{-26}
~(2\pi)^{12}~4\kappa^2\\
&\times &\int 
\frac{\diff{p_1^+}}{p_1^+} \frac{\diff{p_2^-}}{p_2^-} \frac{\diff{p_1'^+}}{p_1'^+} \frac{\diff{p_2'^-}}{p_2'^-}
\frac{\diff{k_1^+}}{k_1^+} \frac{\diff{k_2^-}}{k_2^-}
\ddiff{12}(\transverse{p}_1\transverse{p}_2\transverse{p}'_1\transverse{p}'_2\transverse{k}_1\transverse{k}_2 \mathbf{b_1} \mathbf{b_2})
\ddiff{4}{k}\ddiff{4}{q}\\
&\times &\delta^4(k-k_1-q)~\delta^4(p_1+p_2-k-k_2)~\delta^4(p'_1+p'_2-k_1-k_2)\\
&\times &\tilde{\psi}_1(p_1) \tilde{\psi}_2(p_2)\tilde{\psi}^*_1(p'_1) \tilde{\psi}^*_2(p'_2)~
\frac{1}{k^2+i\epsilon}\frac{1}{q^2+i\epsilon}~e^{-iqx}~kk_1p_1^+p_2^-{p'}_1^+{p'}_2^-\\
&\times &
\exp\left\lbrace i\left[ \mathbf{Q}\cdot \mathbf{b_1} + \mathcal{A}\left( s(p_1,p_2)\right) 
- \mathbf{Q'}\cdot \mathbf{b_2} - \mathcal{A}\left( s(p'_1,p'_2)\right) \right] \right\rbrace \qquad.
\end{eqnarray*}

For analogous reasons as in the computation of Section \ref{beforeonerung}, we define
\[
p_1\equiv P+\frac{Q+q-Q'}{2},\quad p_2 \equiv p-\frac{Q-Q'}{2}
\]
and consequently find, because of momenta conservation deltas, 
\[
p'_1= P-\frac{Q+q-Q'}{2},\quad p'_2= p+\frac{Q-Q'}{2}
\]\[
k= P+\frac{-Q+q-Q'}{2},\quad k_1= P-\frac{Q+q+Q'}{2},\quad k_2= p+\frac{Q+Q'}{2}\qquad.
\]
We also have to pay attention to a change in form of wave functions
\begin{eqnarray*}
\psi_{1}(p_1)&\to&\psi_P(P)\\
\psi_{1}(p'_1)&\to&\psi'_P(P)\\
\psi_{2}(p_2)&\to&\psi_p(p)\\
\psi_{2}(p'_2)&\to&\psi'_p(p)
\end{eqnarray*}
and the fact that in new coordinates a 2 factor appears for each one of the three conservation $\delta^4$-s.
We also consider that in eikonal regime  a low transferred momentum is involved. From momenta conservations  and subsequent integrations  on saddle--point values, we find the conditions
\begin{itemize}
\item $Q^+\simeq Q'{}^+\simeq Q^-\simeq 0$ and $Q'{}^-\simeq -q^-$
\end{itemize}
so that
$
p_1^+ \simeq k^+\simeq P^++\frac{q^+}{2},\quad p_2^- \simeq k_2^-\simeq p^--\frac{q^-}{2}
$
are to be expected.
We integrate over $k$, $k_1$, $k_2$ 
 using conservation $\delta$-s, obtaining
\begin{eqnarray*}
\langle h^{++}(x) \rangle &=& \frac{i\kappa^2~4}{(2\pi)^{14}~2^{(8-3)}}
\int 
\diff{P^+}~\diff{p^-}~\ddiff{2}{\transverse{P}}~\ddiff{2}{\transverse{p}}~\diff{q^+}~\diff{q^-}~\ddiff{2}{\transverse{q}}~
\ddiff{2}{\transverse{Q}}~\ddiff{2}{\transverse{Q}'}~\\
&\times &\tilde{\psi}_P(P) \tilde{\psi}_p(p)\tilde{\psi}'{}^*_P(P) \tilde{\psi}'{}^*_p(p)~
\frac{k}{k_2^-}\frac{\exp\left\{-\frac{i}{2}\left(q^+x^- + q^-x^+ - 2 \transverse{q}\cdot\transverse{x}\right)\right\}}{(k^2+i\epsilon)(q^2+i\epsilon)}\\
&\times &\ddiff{2}{\mathbf{b_1}}~\ddiff{2}{\mathbf{b_2}}~
\exp\left\lbrace i\left[ \mathbf{Q}\cdot \mathbf{b_1} + \mathcal{A}\left( s(p_1,p_2)\right) 
- \mathbf{Q'}\cdot \mathbf{b_2} - \mathcal{A}\left( s(p_1,p_2)\right) \right] \right\rbrace 
\end{eqnarray*}
with
\begin{eqnarray*}
k&=&k_1+q=P+\frac{q-Q-Q'}{2}\\
k_2&=&p+\frac{Q+{Q'}}{2}\\
\widetilde{\psi}_P(P)&\simeq&\sqrt{P^++\frac{q^+}{2}}\widetilde{\mathcal{N}}\widetilde{\mathcal{M}}
\exp\left\lbrace -\frac{1}{8\sigma^2}\left( P^+-\sqrt{s}+\frac{q^+}{2}\right) \right. \\
&&\left. -\frac{i}{2}\left[ z_0\left( P^+-\sqrt{s}+\frac{q^+}{2}\right) -t_0\left( P^++\frac{q^+}{2}\right)\right] \right\rbrace 
\end{eqnarray*}
with an analogous form for $\widetilde{\psi}_p(p)$. 

Consequently, $\Psi(P,p)\equiv \widetilde{\psi}_P(P)\widetilde{\psi}'{}^*_P(P)\widetilde{\psi}_p(p)\widetilde{\psi}'{}^*_p(p)$ is
\begin{eqnarray*}
\Psi(P,p) &\simeq&
\sqrt{\left( (P^+)^2-\frac{(q^+)^2}{4}\right) \left( (p^-)^2-\frac{(q^-)^2}{4}\right)} \left( \widetilde{\mathcal{N}}\widetilde{\mathcal{M}}\right)^4\\
&\times &\exp\bigg\lbrace -\frac{1}{4\sigma^2}\bigg[ (P^+-\sqrt{s})^2+(p^--\sqrt{s})^2+\frac{1}{4}(q^{+2}+q^{-2})+ \\
&&\phantom{\exp\lbrace}\left. +4(\transverse{P}^2+\transverse{p}^2)+2
\left( (\transverse{Q}-\transverse{Q}'+\transverse{q})^2+(\transverse{Q}-\transverse{Q}')^2\right) \right] +\\
&&\phantom{\exp\lbrace} -\frac{i}{2}\left[ q^+(z_{01}-t_0)-q^-(z_{02}+t_0)+\right.\\
&&\phantom{\exp\lbrace}+2(\transverse{Q}-\transverse{Q}'+\transverse{q})\cdot\transverse{b}_{01}- 2(\transverse{Q}-\transverse{Q}')\cdot\transverse{b}_{02}\bigg] \bigg\rbrace\qquad.
\end{eqnarray*}
It is evidently useful to introduce a translation
\[
\transverse{Q}''\equiv \transverse{Q}+\frac{\transverse{q}}{2}
, \qquad \ddiff{2}{\transverse{Q}}''~\ddiff{2}{\transverse{q}} = \ddiff{2}{\transverse{Q}}~\ddiff{2}{\transverse{q}}\]
so that
\begin{eqnarray*}
\frac{1}{2}\left[ (\transverse{Q}-\transverse{Q}'+\transverse{q})^2+(\transverse{Q}-\transverse{Q}')^2\right]
&=& (\transverse{Q}''-\transverse{Q}')^2+\frac{\transverse{q}^2}{4}\\
\left(\transverse{Q}-\transverse{Q}'+\transverse{q}\right)\cdot\transverse{b}_{01}- \left(\transverse{Q}-\transverse{Q}'\right)\cdot\transverse{b}_{02}
&=&
\left( \transverse{Q}''-\transverse{Q}'\right)\cdot\left(\transverse{b}_{01}-\transverse{b}_{02}\right)
+\frac{\transverse{q}}{2} \cdot\left(\transverse{b}_{01}+\transverse{b}_{02}\right).
\end{eqnarray*}
Also we take for
\begin{eqnarray*}
&&\sqrt{\left( (P^+)^2-\frac{(q^+)^2}{4}\right) \left( (p^-)^2-\frac{(q^-)^2}{4}\right)}
\simeq \\
&&\simeq P^+ p^- \left\lbrace 1-\frac{1}{8}\left[\left(\frac{q^+}{P^+}\right)^2 + \left(\frac{q^-}{p^-}\right)^2\right] + \left(\frac{q^+ q^-}{8 P^+ p^-}\right)^2\right\rbrace
\end{eqnarray*}
a first--order approximation
\begin{itemize}
\item $\sqrt{\left( (P^+)^2-\frac{(q^+)^2}{4}\right) \left( (p^-)^2-\frac{(q^-)^2}{4}\right)} \simeq P^+ p^-$.
\end{itemize}
Now we proceed with a
\begin{itemize}
\item saddle--point approximation. 
\end{itemize}
We integrate over transverse components of $P$, $p$, $Q$, $Q$, $b_1$ and $b_2$ using steepest descent method (see Appendix \ref{steepest}, 
for a more detailed calculation) on the exponent part
\begin{eqnarray*}
\Phi &\equiv& -\frac{1}{4\sigma^2}\left[ 4\left( \transverse{P}^2+\transverse{p}^2\right) + 4\left( \transverse{Q}''-\transverse{Q}'\right)^2\right]
-\frac{i}{2}\bigg\{\left. 2\left( \transverse{Q}''-\transverse{Q}'\right)\cdot\left( \transverse{b}_{01}-\transverse{b}_{02}\right)+\right. \\
&&\left. -2\left( \transverse{Q}''\cdot\transverse{b}_{1}-\transverse{Q}'\cdot\transverse{b}_{2}\right)
-2Gp^-\left[ \left( P^+ +\frac{q^+}{2}\right) \ln \frac{L^2}{\transverse{b}_1^2} -\left( P^+ -\frac{q^+}{2}\right) \ln \frac{L^2}{\transverse{b}_2^2}\right] \right\} 
\end{eqnarray*}
obtaining the following saddle--point values:
\begin{eqnarray*}
\transverse{P}_c=\transverse{p}_c&=&0\\
\transverse{Q}'_c &=& \frac{\left(\transverse{b}_{01}-\transverse{b}_{02}\right)}{\left|\transverse{b}_{01}-\transverse{b}_{02}\right|^2} ~Gp^- ~(2P^+-q^+)\left[1+o\left(\Xi\right)\right]  \\ 
\transverse{Q}''_c &=& \frac{\left(\transverse{b}_{01}-\transverse{b}_{02}\right)}{\left|\transverse{b}_{01}-\transverse{b}_{02}\right|^2} ~Gp^- ~(2P^++q^+)\left[1+o\left(\Xi\right)\right]  \\ 
\left( \transverse{Q}''-\transverse{Q}'\right)_c&=&\frac{\left(\transverse{b}_{01}-\transverse{b}_{02}\right)}{\left|\transverse{b}_{01}-\transverse{b}_{02}\right|^2}~2Gp^-q^+\left[1+o\left(\Xi\right)\right]\\
\left( \transverse{Q}''+\transverse{Q}'\right)_c&=&\frac{\left(\transverse{b}_{01}-\transverse{b}_{02}\right)}{\left|\transverse{b}_{01}-\transverse{b}_{02}\right|^2}~2Gp^-P^+\left[1+o\left(\Xi\right)\right]\\
\transverse{b}_{1c}=\transverse{b}_{2c}&=&\frac{1}{2}\left(\transverse{b}_{01}-\transverse{b}_{02}\right)\left[1+o\left(\Xi\right)\right]\\
\Phi_c&=&iGp^-q^+\ln\frac{L^2}{\left|\transverse{b}_{01}-\transverse{b}_{02}\right|^2}+o \left(\Xi\right)
\end{eqnarray*}
with
\[
\Xi \equiv \frac{Gp^-q^+}{\sigma^2 \left(\transverse{b}_{01}-\transverse{b}_{02}\right)^2}
\]
and an Hessian determinant
\begin{equation}\label{hessian}
H 
= \frac{2^4}{\sigma^8}\left[ 1+8\left(\frac{Gp^-q^+}{\sigma^2 \left(\transverse{b}_{01}-\transverse{b}_{02}\right)^2}\right)^2+o \left(\Xi\right)^2
\right]\qquad.
\end{equation}
We consider large impact parameters:
\begin{itemize}
\item $|\transverse{b}_{02}-\transverse{b}_{01}| \gg R$
\end{itemize} 
and consequently peaked wave packets:
\begin{itemize}
\item $\sigma^2 \left( \transverse{b}_{02}-\transverse{b}_{01}\right)^2 \ll 1$
\end{itemize}
and we can take $H\simeq {2^4}/{\sigma^8}$
if
\begin{itemize}
\item $Gp^-q^+ \ll \sigma^2 \left(\transverse{b}_{01}-\transverse{b}_{02}\right)^2$.
\end{itemize}
Then we obtain
\begin{eqnarray*}
\langle h^{++}(x) \rangle &\simeq& \frac{i\kappa^2}{(2\pi)^{14}~2^{3}}
\left( \widetilde{\mathcal{N}}\widetilde{\mathcal{M}}\right)^4
\frac{(2\pi)^6}{2^2\lambda^4}
\int 
\diff{P^+}\diff{p^-}\diff{q^+}\diff{q^-}\ddiff{2}{\transverse{q}}~\\
&\times &
P^+p^-
\frac{P^+ +\frac{q^+}{2}}{p^- -\frac{q^-}{2}}\frac{1}{k_c^2+i\epsilon}\frac{1}{q^2+i\epsilon}\\
&\times &\exp \left\lbrace -\frac{1}{4\sigma^2}\left[ (P^+-\sqrt{s})^2+(p^--\sqrt{s})^2+\frac{1}{4}\left[(q^+)^2+(q^-)^2\right]-\transverse{q}^2\right]+ \right. \\
&&\phantom{\exp \left\lbrace\right.}\left. -\frac{i}{2}\left[ q^+\left(x^- -t_0+z_{01} -2Gp^-\ln\frac{L^2}{\left| \transverse{b}_{02}-\transverse{b}_{01}\right|^2}\right)+\right.\right.\\
&&\phantom{\exp \left\lbrace -\frac{i}{2}\right.}\left.\left. +q^-(x^+ - t_0 - z_{02})+ 2\transverse{q}\cdot(\transverse{x}-\transverse{b}_{01})\bigg]\bigg\rbrace \right.\right.
\end{eqnarray*}
where
\[
{k_c^2}={\left(P^+ +\frac{q^+}{2}\right)\left(\frac{\transverse{P}_c}{P^+}+\frac{q^-}{2}\right) -\left(\transverse{P}_c+\frac{\transverse{q}}{2}\right)^2}={\left(P^+ +\frac{q^+}{2}\right)\frac{q^-}{2} -\left(\frac{\transverse{q}}{2}\right)^2}\qquad.
\]
Then we perform $P^+$ and $p^-$ integrations, again with the aid of steepest descent method, which yields
\begin{eqnarray*}
P^+_c = p^-_c &=&
\sqrt{s}\\
H&=&\frac{1}{4\sigma^4}
\end{eqnarray*}
\begin{IEEEeqnarray*}{rcl}
\langle h^{++}(x) \rangle \simeq& & \frac{i\kappa^2}{(2\pi)^{8}~2^{5}\sigma^4}
\left( \widetilde{\mathcal{N}}\widetilde{\mathcal{M}}\right)^4
\frac{2(2\pi)}{\sigma^2}
\int 
\diff{q^+}\diff{q^-}\ddiff{2}{\transverse{q}}~s
\frac{\sqrt{s} +\frac{q^+}{2}}{\sqrt{s} -\frac{q^-}{2}}
\\
&&\times \frac{1}{q^+ q^- - \transverse{q}^2 +i\epsilon}\frac{1}{\left(\sqrt{s}+\frac{q^+}{2}\right)\frac{q^-}{2} -\frac{\transverse{q}^2}{4}+i\epsilon}\\
&&\times \exp \left\lbrace -\frac{1}{4\sigma^2}\left[ \frac{(q^+)^2}{4}+\frac{(q^-)^2}{4}-\transverse{q}^2\right]+ \right. \\
&&\left. -\frac{i}{2}\left[ q^+\left(x^- -t_0 + z_{01} -2G\sqrt{s}\ln\frac{L^2}{\left| \transverse{b}_{02}-\transverse{b}_{01}\right|^2}\right)+\right.\right.\\
&&+q^-(x^+ - t_0 -z_{02})+ 2\transverse{q}\cdot(\transverse{x}-\transverse{b}_{01})\bigg]\bigg\rbrace\qquad. \IEEEyesnumber \label{poleseq}
\end{IEEEeqnarray*}
Apart from  pole at $q^-=2\sqrt{s}$ whose contribution can be neglected, because it is strongly suppressed by the zero--centered $\frac{q^-}{\sigma}$ gaussian,\footnote{The wave packets under consideration have large width $\lambda$ with respect to Compton length $\lambda_C$.} the integrand has two poles in $q^-$:
\begin{eqnarray*}
q^-_{P1}&=&
\frac{\transverse{q}^2-i4\epsilon}{2\left(P^+ + \frac{q^+}{2}\right)} \simeq \frac{\transverse{q}^2}{2\sqrt{s}+q^+} - i \frac{2\sqrt{s}+q^+}{|2\sqrt{s}+q^+|}\varepsilon'\\
q^-_{P2}&=&\frac{\transverse{q}^2-i\epsilon}{q^+} = \frac{\transverse{q}^2}{q^+} - i \frac{q^+}{|q^+|}\varepsilon'
\end{eqnarray*}
\begin{figure}
\begin{center}
	\begin{picture}(120,69)(5,0)
		\put(5,35){\vector(1,0){50}}		\put(30,10){\vector(0,1){50}}
			\put(50,37){$\Re{q^-}$}			\put(31,60){$\Im{q^-}$}
			\put(35,30){\circle*{1}}\put(35,25){$q^-_{P1}$}
			\put(15,40){\circle*{1}}\put(15,45){$q^-_{P2}$} 
			\put(25,0){($q^+<0$)}							
		\put(70,35){\vector(1,0){50}}		\put(95,10){\vector(0,1){50}}
			\put(115,37){$\Re{q^-}$}			\put(96,60){$\Im{q^-}$}
			\put(100,30){\circle*{1}}\put(100,25){$q^-_{P1}$}
			\put(110,30){\circle*{1}}\put(110,25){$q^-_{P2}$}
						\put(90,0){($q^+>0$)}
	\end{picture}
\end{center}
\caption{Poles in $q^-$ complex plane.\label{poles}}
\end{figure}
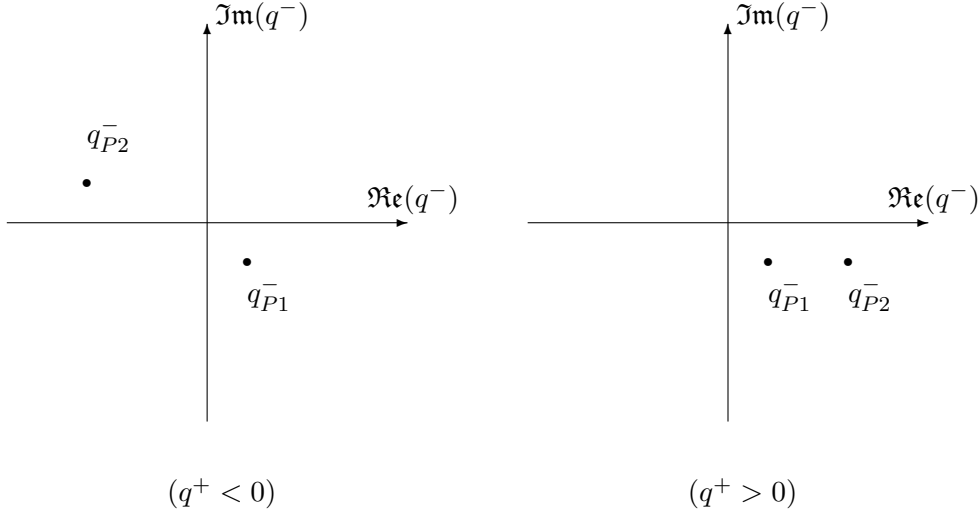
and for the moment we consider just the first pole, assuming%
\begin{itemize}
\item $\Res \left(q^-_{P2}\right) \ll \Res \left(q^-_{P1}\right)$.
\end{itemize}
For an analogous reason we can take $2\sqrt{s}+q^+ >0$ in $\Res \left(q^-_{P1}\right)$.
The second pole, whose position is dependent on $q^+$'s sign (Fig.~\ref{poles}), will be considered as a further approximation
\ in Section \ref{conditions}.
We also take, coherently with 
saddle--point values,
\begin{itemize}
\item $
\frac{\transverse{q}^2}{\sqrt{s}} \simeq 0$
\end{itemize}
so that if
\begin{itemize}
\item $\sqrt{s} + \frac{q^+}{2} > 0$
\end{itemize}
then
\[
q^-_{P1} \simeq 0 -i \varepsilon'
\]
and integration in the $q^-$ complex plane along a path closed in negative imaginary region requires the appearance of an (expected) Heaviside theta:
\begin{eqnarray*}
\langle h^{++}(x) \rangle &\simeq& \frac{i\kappa^2}{(2\pi)^{7}~2^{4}\sigma^6}
\left( \widetilde{\mathcal{N}}\widetilde{\mathcal{M}}\right)^4
2\pi~i
\int 
\diff{q^+}\ddiff{2}{\transverse{q}}~
\Theta\left(x^+ - \left(t_0 +z_{02}\right)\right)~
\frac{s}{\sqrt{s}}
\\
&\times &\frac{1}{-\transverse{q}^2}\exp \left\lbrace -\frac{1}{4\sigma^2}\left( \frac{(q^+)^2}{4}-\transverse{q}^2\right)+\right.\\
&&\left. -\frac{i}{2}\left[ q^+\left(x^- - (t_0-z_{01}) -2G\sqrt{s}\ln\frac{L^2}{\left| \transverse{b}_{02}-\transverse{b}_{01}\right|^2}\right)+\right.\right.\\ &&+2\transverse{q}\cdot(\transverse{x}-\transverse{b}_{01})\bigg]\bigg\rbrace \qquad.
\end{eqnarray*}
Then we perform $q^+$ integration which leads to
\begin{eqnarray*}
\langle h^{++}(x) \rangle &\simeq& \frac{-\kappa^2}{(2\pi)^{6}~2^{4}\sigma^6}\sqrt{s}~
\left( \widetilde{\mathcal{N}}\widetilde{\mathcal{M}}\right)^4~
\Theta\left(x^+ - \left(t_0 +z_{02}\right)\right)\\
&\times & 2\pi~2
\delta\left(x^- - (t_0-z_{01}) -2G\sqrt{s}\ln\frac{L^2}{\left| \transverse{b}_{02}-\transverse{b}_{01}\right|^2}\right)\\
&\times &\int 
\ddiff{2}{\transverse{q}}~
\frac{1}{-\transverse{q}^2}
\exp \left\lbrace \frac{\transverse{q}^2}{4\sigma^2}  - i ~ \transverse{q}\cdot(\transverse{x}-\transverse{b}_{01})\right\rbrace \qquad.
\end{eqnarray*}
Considering a high distance between the observer and the scattering region,
 \begin{itemize}
\item $|\transverse{x}-\transverse{b}_{01}| \gg R$
\end{itemize}
will lead to  a high phase even in sharp--peaked zero--centered $\transverse{q}$ gaussians. That practically makes integration endpoints equivalent to infinity.
Given that, using the usual cut--off $Q_0 \propto L^{-1}$ and $\sigma \to 0$,  we get
\[
\int 
\ddiff{2}{\transverse{q}}~
\frac{1}{\transverse{q}^2}
\exp \left\lbrace \frac{\transverse{q}^2}{4\sigma^2}  - i ~ \transverse{q}\cdot(\transverse{x}-\transverse{b}_{01})\right\rbrace 
=
 2\pi ~\ln \frac{L}{|\transverse{x}-\transverse{b}_{01}|}
=
 \pi ~\ln \frac{L^2}{|\transverse{x}-\transverse{b}_{01}|^2}.
\]
The leading factor accounts to
\[
 \frac{8\pi G\sqrt{s}}{(2\pi)^{5}~2^{3}\sigma^6}~
(2\sigma^2)^{3}(2\pi)^{3}~\pi
= 2G\sqrt{s}\equiv
R\qquad.
\]
The final result is
\begin{IEEEeqnarray*}{rcl}\label{ourresult}
\langle h^{++}(x) \rangle & ~\simeq~ &
R~\ln \frac{L^2}{|\transverse{x}-\transverse{b}_{01}|^2}~\Theta\left(x^+ - \left(t_0 +z_{02}\right)\right)\\
&&\times ~
\delta\left(x^- - (t_0-z_{01}) -R\ln\frac{L^2}{\left| \transverse{b}_{02}-\transverse{b}_{01}\right|^2}\right)\IEEEyesnumber
\end{IEEEeqnarray*}
that is, as usual, a shift in time and space, occurring in the future, of the shock profile function. Here however we find that the infinite--rungs ladder of eikonal exchanges gets resummed in a delta whose argument is shifted. The shift is an amplitude depending on impact parameter, which has the form of Aichelburg--Sexl's shifts (\ref{asshifts}).

 As noted earlier, although striking, that is not surprising. It could be expected on the basis of the following conjecture. In the perturbation field theory point of view, the blobs contain contributions from every orders of eikonal exchange. The conjecture is that those orders contribute with derivatives of increasing orders of the delta distribution, and get resummed in the shifted delta.
 
We resume  all the approximations and conditions used in this derivation
in Table \ref{conditionstable}.
As we will discuss in Section \ref{conditions}, these approximations and conditions can be divided in three main categories
.
Depending on which categories of conditions one would choose to relax, quantifiable effects would take place as next--order approximations of our calculation.
				

		\subsection{Approximations and conditions relaxation}\label{conditions}

\begin{table}[t]
\begin{center}
\begin{tabular}{lll}
\hline
measure procedures	&	packets preparation	&	eikonal regime	\\
\hline
\\
$q^+$, $q^-$, $|\transverse{q}| \ll \sqrt{s}$	&	$\frac{1}{\sigma^2}=\lambda^2 \ll |\transverse{b}_{02}-\transverse{b}_{01}|^2$	&	$k_1\simeq (k_1^+,0,0,0)$, etc.	\\
\\
$2\sqrt{s}+q^+>0$	&	$p_1\simeq (p_1^+,0,0,0)$, etc.	&	$\theta\simeq 0$	\\
\\
$\transverse{q}^2/\sqrt{s}\simeq 0$	&	$p_1^+ \gg p_1^-$, etc.		&	$Q'^-\simeq -q^-, Q^+\simeq 0$	\\
\\
$|\transverse{x}-\transverse{b}_{01}|\gg R$	&	&	$|\transverse{b}_{02}-\transverse{b}_{01}|\gg R$ \\
\\
$\Res(q^-_{P2}) \ll \Res(q^-_{P1})$	&	&	ladder--diagrams only\\
\\
\hline
\\
\multicolumn{3}{c}{$
\sigma~|\transverse{b}_{02}-\transverse{b}_{01}|^2
\gg 2Gq^+ = Rq^+/\sqrt{s}$}\\
\\
\hline
\end{tabular}
\end{center}
\caption{Approximations and conditions categories. Condition in last line  involves all three categories, because it states that a measure exchanging a momentum $q^+$ should not substantially influence the scattering with impact parameter $b_0$ of two wave packets prepared with a given width of order $\sigma$.
\label{conditionstable}}
\end{table}
		
					As stated, in spite of being an all--orders calculation, our result is limited to eikonal exchanges represented by ladder diagrams and also is subject to a number of approximations deriving from physical conditions we chose. We now would like to briefly discuss the possible effects of the relaxation of some of such conditions, which we hope will be useful for a program of future investigations on the subject.
		
Those conditions can be divided into three main categories (see Table \ref{conditionstable}): 
measure procedures, wave packets preparation, and eikonal regime scattering.
Roughly speaking, the former two are somewhat more of experimental nature and subject to larger observation arbitrariness, while the latter expresses a parameter regarding more intrinsically physical aspects of the studied interaction.



			\subsubsection{Additional $q^-$ poles}
			
If we consider now the second $q^-$ pole in Eq.~\ref{poleseq}, 
\[
q_{2P}^- \simeq \frac{\transverse{q}^2}{q^+}-i\frac{|q^+|}{q^+}\epsilon
\]
we find the following quantity, to be added in $\langle h \rangle$:
\begin{eqnarray*}
\langle h \rangle^{(2)} &=& -\left[\Theta(q^+)\Theta(x^+-x^+_0)-\Theta(-q^+)\Theta(x^+_0-x^+)\right] \\
&&\times \int \ddiff{2}{\transverse{q}}~\diff{q^+}~\frac{\sqrt{s}+\frac{q^+}{2}}{\sqrt{s}-\frac{\transverse{q}^2}{2q^+}}\frac{1}{q^+}\frac{1}{\sqrt{s}\frac{\transverse{q}^2}{q^+}+\frac{\transverse{q}^2}{4}} \exp\left\lbrace\Phi\right\rbrace
\end{eqnarray*}
with 
\begin{eqnarray*}
\Re{\Phi} &=& -\frac{1}{16\sigma^2}\left[ \left( \frac{\transverse{q^+}}{q^+} \right)^2 + \left( q^+ \right)^2 \right]\\
\Im{\Phi} &=&
-\frac{1}{2}\left[ \frac{\transverse{q}^2}{q^+} \left( x^+ - x^+_0 \right) + q^+ \left( x^- - x^-_0 -2G\sqrt{s}\ln\frac{L^2}{|\transverse{b}_{02}-\transverse{b}_{01}|^2} \right) \right].
\end{eqnarray*}
Because of the presence of Heaviside thetas, $\langle h \rangle^{(2)}$ integration domain can be treated as follows:
\[
\left[\Theta(q^+)\Theta(x^+-x^+_0)-\Theta(-q^+)\Theta(x^+_0-x^+)\right] \int \diff{q^+}
\leadsto
\int_0^{+\infty}\diff{q^+}
\]
where, although the exponent's real part is quadratic in $q^+$, one should handle  sign changes with care 
because the integrand's real part is not strictly an even function%
.

In order to treat the $\langle h \rangle^{(2)}$ correction, one could first ignore quadratic terms in the exponent. Due to the presence of $q^+$ both in numerator and denominator, very small and very large $q^+$ values in the exponent's imaginary part correspond to a highly oscillating phase  while the effect concerning the real part is that they lead to a suppressed exponential. In first approximation, the expression is an integral representation of the \emph{modified Bessel functions of the second kind} $K_1(\transverse{q}^2)$, and is to be treated accordingly.

It could be noteworthy to stress that the second pole arises from the factor $k^{-2}$, where $k$ is an internal momentum.

			\subsubsection{Second order in steepest descent method}
			
			Taking second order in $\det \mathcal{H}$ (see Eq.~\ref{hessian}),
			\[
			\frac{\det \mathcal{H}}{2^4 \sigma^{-8}} = 1 + 0 + 8 \left( \frac{Gp^-q^+}{\sigma^2 b_0^2} \right)^2 + \mathcal{O} \left( \frac{Gp^-q^+}{\sigma^2 b_0^2} \right)^3
			\]
			would mean to relax the condition $q^+ \ll 2|b_0|^2\sigma^2/R$. That evaluates how much, having previously prepared a packet with a certain $\sigma$ with respect to impact parameter (in our conditions $|b_0|^2\gg\sigma^2$), the measuring procedure would influence the physics of the system. Thus, that approximation involves all of the three conditions categories (see Table \ref{conditions}). That can be considered somehow unexpected, but on the basis of the presence of $G$ it could be conjectured as 
			 peculiar of gravitation interaction
.
			
			The effect is the following: $\langle h^{++} \rangle$ is now proportional to
			(we omit other factors to highlight just the changed ones)
\begin{equation*}
			\int \diff{q^{+}}  \frac{1}{2^2\lambda^4\sqrt{1+8\left(\frac{Gp^-q^+}{\sigma^2b_0^2}\right)^2}} 
			\exp \left\lbrace -\frac{1}{4\sigma^2} ~(q^+)^2\left( \frac{1}{4}-4\left(\frac{Gp^-}{b_0}\right)^2 \right)\right\rbrace \times \ldots
			\end{equation*}
			where the additional exponent factor has the effect of a shift of the $q^+$ gaussian by an irrelevant quantity\footnote{As long $R \ll b_0$, obviously.} $R/b_0$.
			
			The effect of the additional determinant term in square root is the appearance of a new pure imaginary $q^+$ pole
			\[
			q^+_{HP} = \frac{i}{\sqrt{8}}\frac{\sigma^2 b_0^2}{Gp^-}\qquad.
			\]
			
			As usual $p^-$ will be later integrated and evaluated as $\sqrt{s}$ with steepest descent method, making the pole position again dependent on the relation between $q^+$ and the quantity
			$
			G\sqrt{s}/(\sigma^2 b_0^2)\qquad.
			$
			
			As long as all the other conditions we considered are maintained, it should be acceptable to take
				\[
				q^+ \ll \frac{G\sqrt{s}}{(\sigma^2 b_0^2)}\qquad.
				\]			
				That question could however be worth further investigations.

			\subsubsection{Smaller impact parameter}
			
			A small impact parameter would be particularly interesting for possible indications of gravitational collapse \cite{acv, fal}.
			
			As known from general classical mechanics, quantum field theory, and as investigated by ACV's studies, it obviously is the relevant parameter when evaluating a strong gravity effect which could cause such a collapse. 
			
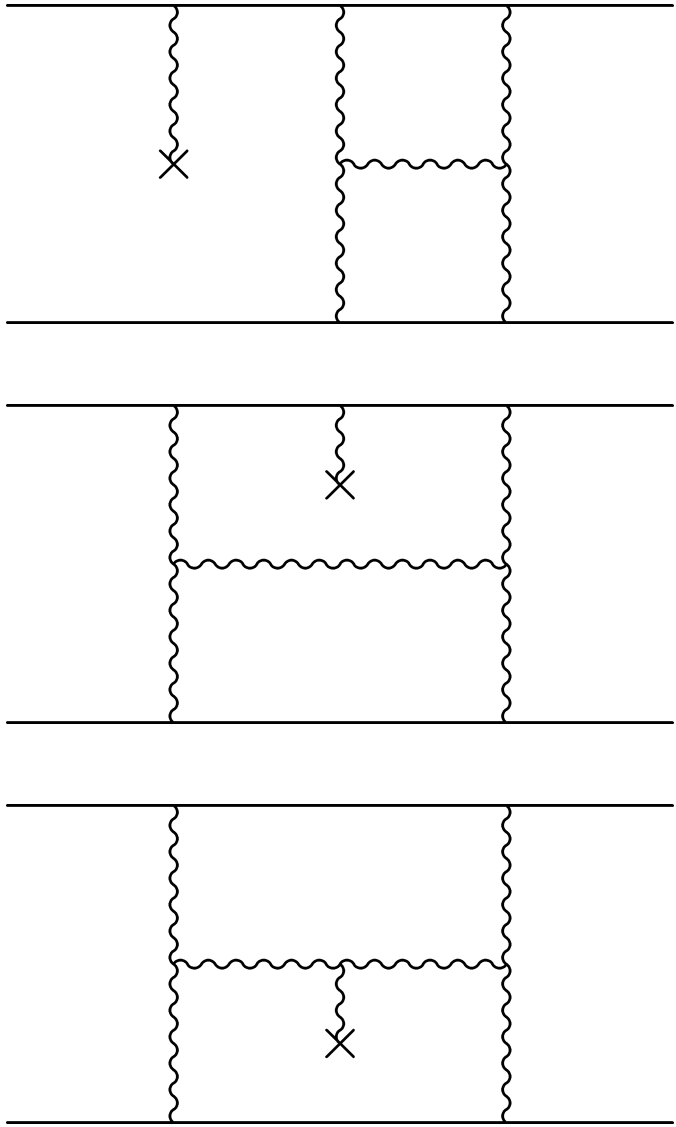
\begin{figure}
\begin{center}
	\begin{picture}(120,50)(-8,0)
	\begin{fmffile}{hdiagramsinsertion1}
\begin{fmfgraph*}(104,40)
\fmfleft{i1,i2}
\fmfright{o1,o2}
\fmf{plain}{i1,v1,v3,v5,o1}
\fmf{plain}{i2,v2,v4,v6,o2}
\fmffreeze
\fmf{wiggly}{v3,vl,v4}
\fmf{wiggly}{v5,vr,v6}
\fmffreeze
\fmf{wiggly}{vl,vr}
\fmf{wiggly}{v2,vx}
\fmf{phantom}{vx,v1}
\fmfv{decor.shape=cross}{vx}
\end{fmfgraph*}
\end{fmffile}
	\end{picture}
	\begin{picture}(120,50)(-8,0)
	\begin{fmffile}{hdiagramsinsertion2}
\begin{fmfgraph*}(104,40)
\fmfleft{i1,i2}
\fmfright{o1,o2}
\fmf{plain}{i1,v1,v3,v5,o1}
\fmf{plain}{i2,v2,v4,v6,o2}
\fmffreeze
\fmf{wiggly}{v1,vl,v2}
\fmf{wiggly}{v6,vr,v5}
\fmffreeze
\fmf{wiggly}{vl,vc,vr}
\fmffreeze
\fmf{wiggly}{v4,vx}
\fmf{phantom}{vx,vc}
\fmfv{decor.shape=cross}{vx}
\end{fmfgraph*}
\end{fmffile}
	\end{picture}
	\begin{picture}(120,50)(-8,0)
	\begin{fmffile}{hdiagramsinsertion3}
\begin{fmfgraph*}(104,40)
\fmfleft{i1,i2}
\fmfright{o1,o2}
\fmf{plain}{i1,v1,v3,v5,o1}
\fmf{plain}{i2,v2,v4,v6,o2}
\fmffreeze
\fmf{wiggly}{v1,vl,v2}
\fmf{wiggly}{v6,vr,v5}
\fmffreeze
\fmf{wiggly}{vl,vc,vr}
\fmffreeze
\fmf{wiggly}{vc,vx}
\fmf{phantom}{vx,v3}
\fmfv{decor.shape=cross}{vx}
\end{fmfgraph*}
\end{fmffile}
	\end{picture}
\end{center}
\caption{Field insertions in H--diagrams.\label{hinsertions}}
\end{figure}

			In our study, we focused on transverse component $|\transverse{b}|$ and saw it intervening in approximations as a large length when compared to other ones, which we used as expansions parameters; in final $\langle h^{++}(x) \rangle$ calculation result it appears as a logarithm argument in the scattering amplitude. Considering a relaxation on $|\transverse{b}|$'s magnitude will thus have a sensitive effect on the technical side of the computation, but one should pay attention also to the fact that, with smaller $b$--s, the eikonal approximation is less valid. One of the most important effect is that other diagrams start to give sensitive contribution. Thus, while repeating the computation of $\langle h^{++}(x)\rangle$ just removing the large impact parameter condition, while still useful for a deeper insight into the subject, would be an incomplete treatment. That is, at least as long as one doesn't take in account the contributions from field insertion in other diagrams, that is, in the next order, in H--diagrams (Fig.~\ref{hinsertions}).

			\subsubsection{Energy conservation}

			In our derivation we do not take in account the possible effects of transferred energies between the two interacting particles. When two particles interacts via exchanged particles, they exchange momenta and energies transported by those mediators. In the case of gravity, energy takes the role of the interaction charge and appears in factors at the scattering vertices. Sources which emit gravitons experience an energy loss that should influence their subsequent gravitational interactions. On the other hand, lost energies end up in gravitons and  are released to other objects, like another particle or  a detector,  in  subsequent interactions.
			Consequently it would be interesting to compute the resulting fields, and compare it with results of the theory of gravitational radiation emission.

\chapter{Conclusions}\label{conclusions}\label{chapterconclusions}

In this work we describe a semiclassical approach which can give insight for a more fundamental investigation on the path towards a reliable quantum theory of gravity. We introduce the $S$--matrix description due to Amati, Ciafaloni and Veneziano \cite{acv} which is built on string theory; however it should be stressed that its validity could be independent from string theory itself, at least how it is known today. That is, given its nature of an effective description (in this case of quantum gravitational fields), the $S$--matrix approach has chances to remain valid even if the underlying fundamental  theory undergoes substantial changes.
	Consider that 
	Kirschner and Szymanowski  were able to derive Lipatov's reduced action starting from  Einstein's equations in high--energies approximation \cite{k-s}.

In Chapter \ref{chapterclassical} we review  classical results of general relativity which are expected to be reproduced by the classical limit of any non--wrong\footnote{We recall that a scientific theory should be \emph{falsifiable}, that is, admit a confrontation with an experimental observation or any other accepted theory that in principle can happen to prove it scientifically wrong \cite{pop}.} quantum theory: Aichelburg and Sexl's metric given by a source moving at the speed of light \cite{a-s}, D'Eath and Payne's expansion of the metric deriving from the collision of two of such sources \cite{dep} and Dray and 't Hooft's metric of two colliding plane shells of matter \cite{d-t}.

We see that AS profile can be thought of as the flattening of the spherical Schwarzschild metric  on a plane, being a length contraction effect. In fact, apart from a technical difficulty in the procedure to treat the limit  $\beta \to 1$, the metric is derived with the standard method of applying a Lorentz transformation in the form of a boost. Depicting  the static newtonian potential by its field lines, and visualizing a spherical equipotential surface, then applying the boost, we would see a deformed sphere, contracted in a similar manner. Upon reaching $\beta=1$ the contracted sphere would degenerate in a plane. 
The physics is unchanged by a coordinate transformation, and the ability to derive that metric by a Lorentz boost  suggests that informations about the global field should be conserved. In other words, the total newtonian field, which in the $\beta =0$ RS permeates all the three-dimensional space,  in the $\beta=1$ RS must be in some sense all contained  on the plane. Quantitatively, that yields  the logarithmic profile of AS' metric. In fact, we recall that AS' metric shows a profile $\ln (\transverse{x}^2/L^2)$, $\transverse{x}$ being the transverse distance from the field--generating particle (we  return on $L$ later); that logarithm can be thought of as the integration on all space of the newtonian potential $\propto 1/r$. That is also visualizable in terms of the Dirac's $\delta(z)$ distribution, which selects the AS profile's plane and has a finite integral. Despite restricting the effects of the potential to $z=0$, its measure is finite. In other words, it measures as the whole space. In that sense we could think of all the Schwarzschild static solution as flattened in the plane selected by $\delta(z)$.

While suggestive, that behavior of the $1/r$ newtonian potential is all but unexpected; it is known that such kind of infinite--range potentials show peculiar effects when investigated at spatial infinity. Being its $r$-integral proportional to a logarithm of $r$, it shows, as we say, a logarithmic divergence. That is because the logarithm at infinity lacks a finite limit. We could say that even if we try to escape gravitational effects by getting away at infinity, all the previously unobserved gravitational effects cumulate in a infinite effect which is the sum of the global newtonian potential. That is the reason of our inability, shown in Section \ref{asgeosec}, to put a test particle with a finite impact parameter $b$ at rest w.r.t.\ some observer at infinity: that observer is actually not well--defined.

Pursuing a satisfactory description in that sense, we find another peculiar behavior typical of gravitation. That is an effect analogous to the often--cited twin paradox.
It two (or more) observers, equipped with synchronized clocks, cross an AS shock wave at different impact parameters, then after the crossing their clocks are no longer synchronized, because their time and space--shifts depend on their impact parameters. Naturally an observer measuring times and lengths with respect to a local RS moving with himself would not measure discontinuities in its own motion. But, as we show in Section \ref{asgeo}, in that RS other observers' motion is discontinuous. We should remark that the shift is consequently dependent on the choice of a RS, that is, of the impact parameter of the observer, which is identified with  the arbitrary cut--off $L$ described in Section \ref{cutoff}
.  

We see that, while for a single particle source the metric can be expressed exactly and in compact form, in the case of the collision of such two particles the situation is trickier. D'Eath and Payne describe the collision of ultrarelativistic sources by expanding the so--called \emph{news function} (see Section \ref{death}) in terms of $\lambda/\nu\sim\gamma^{-2}$. $\lambda/\nu$ measures the ratio of particle's energies in the RS in which one source can be considered to produce a weak shock w.r.t.\ the strong shock of the other one. First of all, we note that an expansion seems to be necessary when describing such sources. 
W%
e see that the description is much more complicate than AS metric. While that can be considered just a technical aspect, it is suggestive of a deeper physics to be explored. 
A%
nother aspect which remains to be investigated concerns some details of geodesics in DP's space--time. It is not clear how different would be the fate  of two geodesics that crosses the two shock waves in different order. 

Another situation we illustrate concerns the treatment of the collision of infinite planes of matter as described by Dray and 't~Hooft. In that case, infinities also appear in the total energy, so it is not surprising that in the future of that ideal experiment the whole space--time collapses. In fact, each plane shell focuses the other one in a collapsing sphere that closes all the Universe in a singularity. That is obviously a rather unphysical situation, but it is remarkable that despite that peculiarity, the local structure near the collision is solved in analytic form. We recall that also DP's result is valid in confined regions. That opens the possibility of a further comparison between the two results. Summing up, both results could be compared with each other and with ACV's metric.

In Chapter \ref{chaptersemiclassical}  we recall standard quantum scattering theory to introduce semiclassical results and ACV's description.

A basic result we introduce is due to 't~Hooft \cite{tho}, and describes the effect of an AS wave front on a plane wave. That opens the road to a semiclassical treatment of gravitational interaction in terms of AS shock profiles. 

't~Hooft's result is coherent with the AS' description of particle deflection. In other words, a plane wave is subject to the same space--time shift by a gravitational shock wave than a test particle. We extend 't~Hooft's result to wave packets, and confirm that their centers are subject to the same deflection of a test particle. That is completely coherent with the usual particle description in terms of wave packets, and our test confirms that the approach we follow preserves that coherence.

We then resume the ACV's approach \cite{acv} which introduces an $S$--matrix description of gravitational interaction. ACV derive their description from string theory. In the  particular regime they investigate, string effects can be neglected while strong gravity effects are sensible. That leads to the possibility of describing gravitational interaction in terms of standard scattering theory. The approach then involves the construction of $S$--matrices and makes possible the computation of expectation values of fields using Feynman diagrams. Feynman rules contain propagators deriving from the $S$--matrix description. In perturbation theory, gravitational fields $h_{\mu\nu}$ are interpreted as  perturbations of the metric with respect to a fixed Minkowski metric
$\eta_{\mu\nu}$. The metric tensor is $g_{\mu\nu}=\eta_{\mu\nu}+h_{\mu\nu}$, thus field computations  yields the metric. It is remarkable that a metric can be computed in terms of Feynman diagrams. It is even more remarkable that, as we show with our result, Feynman diagrams are able to yield a term proportional to the Dirac's delta of a shifted coordinate, and that such shift reproduces the one described by AS in a purely classical context.
Moreover, fields $a$ and $\bar a$ in ACV's treatment show the same AS--like function profiles.

We introduce a particular solvable system in ACV's treatment, that is the axisymmetric collision of a particle \emph{vs.}\ a ring of matter. That system shows a few very interesting features. Principally, it implies the presence of a critical impact parameter $b_c$ that divides the physics of that system in two distinct regimes. Those regimes differ on the reality  of the field solutions, and that suggests the presence of some  physical effect in the passage from the above to below the critical parameter. Incidentally, the critical parameter value $b_c \sim 1.6~R$ lays in the region of the gravitational radius $R$, where a classical gravitational collapse is expected to take place. Consequently it is conjectured that impact parameters below $b_c$ could be related to some quantum effect in gravitational collapse.

In Chapter \ref{chapterresult} we show how to compute effective gravitational fields expectation values on  states with two colliding particles, represented by wave packets, in the $S$--matrix semiclassical interaction description. We use Feynman diagram techniques to perform calculations for leading eikonal diagrams with $0$ and $1$ graviton exchanges.

Then we use a result due to Ciafaloni and Colferai (CC thereafter) \cite{c-c}, which takes into account the contribution of a blob diagram that stands for the eikonal exchange of an infinite number of gravitons. A direct calculation of the eikonal exchange of two or more gravitons has not been performed yet; CC inferred the result with a recursive procedure. Using that result, we directly compute a diagram with such infinite gravitons, that should be in principle equivalent to a resummation of $n$ gravitons eikonal exchanges with $n$ from $0$ to infinity.
That computation represents the original part of this work.
	
	The resummation is consistent with the interaction picture in terms of phase shifts.
In other words, for small exchanged momenta and big impact parameters $b$, that is, $b\gg R=2G\sqrt{s}$, the expansion (in terms of opportune factors)
\[
\Delta \simeq  \delta (x^--x^-_0) + A \delta'(x^--x^-_0) + \frac{A^2}{2} \delta''(x^--x^-_0) + \ldots
\]
gets resummed up in a shift
\[
\Delta =  \delta (x^- -x^-_0 + A) 
\]
where $A$ in our case is the amplitude $-R\ln(L^2/b^2)$.

The most remarkable part is that a direct calculation has been performed only for the $0$ and $1$ gravitons eikonal exchange, which yield the factors $\delta$ and $
\delta'$; the possibility that the $n$--gravitons exchange would yield a factor $\delta^{(n)}A^n/n!$ is a conjecture.  Here $(n)$ expresses the $n$--th derivative of Dirac's delta function with respect to $x^-$, that should be taken in the sense of distributions.	Our derivation confirms, and proves in a more rigorous way, that the result of the direct calculation is equivalent to the sum of the series. 

 Moreover, we successfully found the validity of a particular classical limit  for 
 ACV's $S$--matrix semiclassical approach.		The shifts appearing in our result is consistent with the AS' shift, derived in a classical description. The classical AS shift can thus be thought of, in terms of field theory, as the result of an eikonal exchange of an infinite number of gravitons with total transferred momentum $Q$, exchanged between two particles interacting with impact parameter $b$.
Thus, ACV's semiclassical approach is able to reproduce that precise shift value. That picture suggests that next orders in ACV's expansion 
 would lead to quantum effects not predicted by a classical approach. 
 
 Our result can also be seen as a consistency check of the effective description, in particular regimes, of the gravity interaction as supplied by ACV's treatment. 

We further expand our result suggesting the possible effects of the relaxation of some of the approximations we consider (see Section \ref{conditions} for further details). 
In other words, some  effects we neglected could possibly yield quantum corrections to AS' classical result.

A first option would be to take  into account the pole $q_{2P}^-$ when integrating over the exchanged momentum $q^-$. That pole derives from the presence of an internal particle, of momentum $k$, in the diagram. Moreover we see that taking into account the pole yields an additional dependence on the transverse transferred momentum $\transverse{q}$. That could mean that the pole would possibly take into account a sensible  momentum transfer from $k$ to $\transverse{q}$, maybe expressing a discrepancy from eikonal regime. Also, poles could be in general related to the emission of gravitational radiation.

Taking second order in the Hessian $\mathcal{H}$ of the steepest descent method approximation (see Appendix \ref{steepest}) seems not to yield particularly sensible effects. The approximation is related to the order of magnitude of the exchanged momentum component $q^+$ of the leg attached to the computed field. We recall that components of $q$ are related to an influence  on the scattering process by the measure of the field. It could then be conjectured that next orders in $\mathcal{H}$ would reveal the physical effects of a measuring procedure that sensibly influences the field itself.

Considering a smaller impact parameter $b=|\transverse{b}|$, in the light of the above discussion, should lead to interesting comparisons with the role of the parameter $b_c$ in ACV's  description of axisymmetric scattering. We recall that parameter $b$ is related to ACV's expansion in terms of $R^2/b^2$, which at leading order yields eikonal diagrams
. An investigation on smaller impact parameters would supposedly be incomplete without the computation of insertions in next--order diagrams, that is, the H--diagrams of Fig.~\ref{hinsertions}.

Another suggestive consideration concerns the conservation of energy. In our computation we never take into account that upon emitting a graviton, a particle would lose part of its energy. A first difference  with respect to our derivation would be a loss of total energy in outgoing states. That energy should be transferred to and revealable in the field $h^{++}$. A computation of that energy could be compared with the theory of gravitational radiation emission. A comparison with gravitational waves properties could yield interesting result even at the level of our computation, even though it ignores the question of that kind of energy conservation. 
			
	
		Apart from going further in higher--order approximations of the calculation we present, more fundamental studies are necessary. Generally speaking, every comparison of the results yielded by the various approaches can shed some light on the subject. Obviously, systems that are solvable with more than one model would be needed.
		Being GR an established theory, to make comparisons it would reasonable to look for systems that are investigated by semiclassical models and solvable and known in GR. An example is the axisymmetric scattering of a particle and a ring we introduce in Section \ref{particlevsring}.
Another detail to investigate could be a test particle deflection.
On the quantum side one can take the special case $E_1\gg E_2$ in which one of the ingoing particles has a test particle behavior. ACV also supply a method to estimate the semiclassical deflection angle \cite{acv}. In the analogous RG context then it should be determined the deflection to be compared, due to an energy distribution.

\pagebreak

However, also distinct semiclassical theories could be compared. For example, a comparison between ACV and Lipatov metrics, which were originally derived independently, would lead to uncover the possible deeper connections between the two models. Every step in that direction would yield  insight into the whole subject of gravitation.






\begin{appendix}
\chapter{Steepest descent method calculations}\label{steepest}\label{chaptersteepest}


We perform a more detailed approximate calculation of
\[
I_T \equiv \int \ddiff{2}{\transverse{P}}~\ddiff{2}{\transverse{p}}~
\ddiff{2}{\transverse{Q}''}~\ddiff{2}{\transverse{Q}'}~
\ddiff{2}{\transverse{b}_1}~\ddiff{2}{\transverse{b}_2}~
  \exp\{\Phi\}
\]
with
\begin{eqnarray*}
\Phi &\equiv& -\frac{1}{\sigma^2}\left[ \left( \transverse{P}^2+\transverse{p}^2\right) + 
\left( \transverse{Q}''-\transverse{Q}'\right)^2\right]+\\
&&-i\bigg\{\left( \transverse{Q}''-\transverse{Q}'\right)\cdot\left( \transverse{b}_{01}-\transverse{b}_{02}\right)-\left( \transverse{Q}''\cdot\transverse{b}_{1}-\transverse{Q}'\cdot\transverse{b}_{2}\right)+\\
&&\phantom{-i\bigg\{}-Gp^-\left[ \left( P^+ +\frac{q^+}{2}\right) \ln \frac{L^2}{\transverse{b}_1^2} -\left( P^+ -\frac{q^+}{2}\right) \ln \frac{L^2}{\transverse{b}_2^2}\right] \bigg\} 
\end{eqnarray*}
using steepest descent method.

That is an approximation method based on the fact that an exponential integrand in first approximation contributes significantly only in points where the phase is stationary. 
Strictly speaking there are a few different versions of that method, on the reality or complexity of the quantities of interest. In our treatment we use it finding a stationary point for the $\Phi$ complex quantity w.r.t.\ $\transverse{Q}$ and $\transverse{b}$, treated as complex quantities themselves.

The vanishing of first derivatives
\[
\left\lbrace
\begin{array}{lcccl}
0&=&\vec \nabla_{\transverse{P}} \Phi &=& -\frac{2}{\sigma^2}\transverse{P}\\
\\
0&=&\vec \nabla_{\transverse{p}} \Phi &=& -\frac{2}{\sigma^2}\transverse{p}\\
\\
0&=&\vec \nabla_{\transverse{Q}''} \Phi &=& -\frac{\transverse{2}}{\sigma^2}\left( \transverse{Q}''-\transverse{Q}'\right) +i \left( \transverse{b}_{01}-\transverse{b}_{02}+\transverse{b}_{1} \right)\\
\\
0&=&\vec \nabla_{\transverse{Q}'} \Phi &=& +\frac{\transverse{2}}{\sigma^2}\left( \transverse{Q}''-\transverse{Q}'\right) -i \left( \transverse{b}_{01}-\transverse{b}_{02}+\transverse{b}_{2} \right)\\
\\
0&=&\vec \nabla_{\transverse{b}_1} \Phi &=& -i \left[ \transverse{Q}''-2Gp^-\left( P^+ +\frac{q^+}{2}\right) \frac{\transverse{b}_1}{\transverse{b}_1^2}\right] \\
\\
0&=&\vec \nabla_{\transverse{b}_2} \Phi &=& +i \left[ \transverse{Q}'-2Gp^-\left( P^+ -\frac{q^+}{2}\right) \frac{\transverse{b}_2}{\transverse{b}_2^2}\right] \\
\end{array}
\right.
\]
is solved by (solution values denoted by a $c$ subscript, for ``center'')
\begin{eqnarray*}
\transverse{P}_c = \transverse{P}_c &=& 0\\
 \transverse{b}_{1c} = \transverse{b}_{2c} &=&+ \frac{\transverse{b}_{02}-\transverse{b}_{01}}{2} \left( 1+\sqrt{1+i~16~\Xi} \right)\\
\transverse{Q}'_c &=& - \frac{\transverse{b}_{02}-\transverse{b}_{01}}{|\transverse{b}_{02}-\transverse{b}_{01}|^2} \left( 1-\sqrt{1+i~16~\Xi} \right) \frac{i}{8}  \frac{2P^+-q^+}{q^+}\\
\transverse{Q}''_c &=& - \frac{\transverse{b}_{02}-\transverse{b}_{01}}{|\transverse{b}_{02}-\transverse{b}_{01}|^2} \left( 1-\sqrt{1+i~16~\Xi} \right) \frac{i}{8}  \frac{2P^++q^+}{q^+}\\
\end{eqnarray*}
with
\[
\Xi \equiv \frac{Gp^-q^+}{\sigma^2 |\transverse{b}_{02}-\transverse{b}_{01}|^2}\qquad.
\]
In our treatment, a work hypothesis is $\Xi \ll 1$ (see also Section \ref{conditions}), thus we can write
\[
\sqrt{1+i~16~\Xi} \simeq i~8~\frac{Gp^-q^+}{\sigma^2 |\transverse{b}_{02}-\transverse{b}_{01}|^2}
\]
and
\begin{eqnarray*}
 \transverse{b}_{1c} = \transverse{b}_{2c} &\simeq& \left(\transverse{b}_{02}-\transverse{b}_{01}\right) \left( 1 +  i~4~\frac{Gp^-q^+}{\sigma^2 |\transverse{b}_{02}-\transverse{b}_{01}|^2} \right)\\
\transverse{Q}'_c &\simeq&  \frac{\transverse{b}_{02}-\transverse{b}_{01}}{|\transverse{b}_{02}-\transverse{b}_{01}|^2} Gp^- \left( 2P^+-q^+\right)\\
\transverse{Q}''_c &\simeq&  \frac{\transverse{b}_{02}-\transverse{b}_{01}}{|\transverse{b}_{02}-\transverse{b}_{01}|^2} Gp^- \left( 2P^++q^+\right)\qquad.
\end{eqnarray*}
Other quantities of interest for the estimate of $\Phi$ are
\begin{eqnarray*}
 \transverse{b}_{1c} - \transverse{b}_{2c} &=& 0\\
\transverse{Q}''_c - \transverse{Q}'_c &\simeq&  \frac{\transverse{b}_{02}-\transverse{b}_{01}}{|\transverse{b}_{02}-\transverse{b}_{01}|^2} 2Gp^- q^+\\
\end{eqnarray*}
and we also note that
\[
\transverse{Q}''_c + \transverse{Q}'_c \simeq  \frac{\transverse{b}_{02}-\transverse{b}_{01}}{|\transverse{b}_{02}-\transverse{b}_{01}|^2} 4Gp^- P^+\\
\]
which yields a magnitude order for exchanged transverse momenta
\[
|\transverse{Q}| \sim Gs~|\transverse{b}_{02}-\transverse{b}_{01}|^{-1}\qquad.
\]
Non-vanishing second derivatives are ($n=1,2$, $j=x,y$)
\begin{eqnarray*}
\pardev{{}^2}{P_j{}^2}\Phi = \pardev{{}^2}{p_j{}^2}\Phi &=& -\frac{2}{\sigma^2}\\
\pardev{{}^2}{Q_j''{}^2}\Phi = -\pardev{{}^2}{Q_j'{}^2}\Phi &=& -\frac{2}{\sigma^2}\\
\pardev{{}^2}{b_{nj}{}^2}\Phi &=& i2Gp^- \left( 2P^+ +q^+ \right) \frac{1}{\transverse{b}_{n}^2} \left( \frac{b_{nj}^2}{\transverse{b}^2_{n}} - \frac{1}{2}\right)\\
\pardev{{}^2}{b_{n,x}\partial b_{n,y}}\Phi = \pardev{{}^2}{b_{n,y}\partial b_{n,x}}\Phi &=& i2Gp^- \left( 2P^+ +q^+\right) \frac{b_{n,x}b_{n,y}}{\transverse{b}_{n}^4}\\
\pardev{}{b_{1,j}}\pardev{}{Q_{j}''}\Phi = -\pardev{}{b_{2,j}}\pardev{}{Q_{j}'}\Phi &=& i\qquad.
\end{eqnarray*}
Then the Hessian determinant $H$ of the system is
\[
H=\det\left(
\begin{array}{c|c}
\Lambda & \mathcal{I}\\
\hline
\mathcal{I} & \mathcal{G}\\
\end{array}
\right)
\]
with
\begin{eqnarray*}
\Lambda &\equiv& \left(\begin{array}{cccc}
		-1&0&1&0\\
		0&-1&0&1\\
		1&0&-1&0\\
		0&1&0&-1\\
		\end{array}\right) \frac{2}{\sigma^2}, \qquad
		\mathcal{I} \equiv \left(\begin{array}{cccc}
		i&0&0&0\\
		0&i&0&0\\
		0&0&-i&0\\
		0&0&0&-i\\
		\end{array}\right)\\
		\\
		\\
				\mathcal{G} &\equiv& \left(\begin{array}{cccc}
		\frac{1}{\transverse{b_{1}}^2} \left( \frac{b_{1x}^2}{\transverse{b}_{1}^2} - \frac{1}{2}\right)& \frac{b_{1x}b_{1y}}{\transverse{b}_{1}^4}&0&0\\
		\frac{b_{1x}b_{1y}}{\transverse{b}_{1}^4}&\frac{1}{\transverse{b_{1}}^2} \left( \frac{b_{1y}^2}{\transverse{b}_{1}^2} - \frac{1}{2}\right)&0&0\\
		0&0&\frac{1}{\transverse{b_{2}}^2} \left( \frac{b_{2x}^2}{\transverse{b}_{2}^2} - \frac{1}{2}\right)& \frac{b_{2x}b_{2y}}{\transverse{b}_{2}^4}\\
		0&0&\frac{b_{2x}b_{2y}}{\transverse{b}_{2}^4}&\frac{1}{\transverse{b_{2}}^2} \left( \frac{b_{2y}^2}{\transverse{b}_{2}^2} - \frac{1}{2}\right)\\
		\end{array}\right) \\
		\\
		&&\times i2Gp^- \left( 2P^+ +q^+\right)\qquad.
\end{eqnarray*}
Using standard algebra it is straightforward to expand $H$ in terms of $\Xi$ as
\[
H = \frac{2^4}{\sigma^8}\left( 1\cdot \Xi^0 + 0 \cdot \Xi^1 + 8 \cdot \Xi^2 + \ldots \right)
\]
that is,
\[
\frac{H}{(2\sigma^{-2})^4} = 1 + 8 \left( \frac{Gp^-q^+}{\sigma^2 |\transverse{b}_{02}-\transverse{b}_{01}|^2} \right)^2 + \mathcal{O} \left( \frac{Gp^-q^+}{\sigma^2 |\transverse{b}_{02}-\transverse{b}_{01}|^2} \right)^3\qquad.
\]

In all of our computations, we consider $\Xi\ll 1$
. We find the Hessian matrix 
is positive--definite and it grants the existence of a stationary point for the complex exponent $\Phi$: 
\[
\Phi_c=iGp^-q^+\ln\frac{L^2}{\left(\transverse{b}_{01}-\transverse{b}_{02}\right)^2}+o \left(\Xi\right)\qquad.
\]

\chapter*{Notation and conventions}


\begin{itemize}
\item $\hbar=c=G=1$ except where otherwise stated.
\item $m_P \equiv \sqrt{\hbar c/G}$ is the Planck mass.
\item $\lambda_P \equiv  \hbar/(m_P c)
$ is the Planck length.
\item $\kappa^2 = 8\pi G$ is the Einstein's constant.
\item $(-)$, $(+)$ sub-- or superscripts denote quantities respectively in regions before and after an AS shock wave.
\item $x^+=t+z
$ and $x^-=t-z
$ are light--cone coordinates. Except where otherwise stated, we boost reference systems along positive $z$ direction.
\item $\diff{x^+}\diff{x^-} = (\diff{x^0})^2 - (\diff{x^3})^2$ when standing for vectorial differentials $\diff{x^+}\otimes\diff{x^-}$ appearing in metric forms.
\item $\diff{x^+}\diff{x^-} = 2~\diff{x^0}~ \diff{x^3}$ when standing for measure differentials appearing in integrals as \mbox{$|\diff{x^+}\wedge\diff{x^-}|$}.
\item $\partial \equiv \partial_{x^3} = \frac{1}{2}\left(\partial_{x^1}-i\partial_{x^2}\right)$.
\item $\bar{\partial} = \partial^*= \frac{1}{2}\left(\partial_{x^1}+i\partial_{x^2}\right)$.
\item $\partial^+ \equiv \partial_{x^+} = \frac{1}{2}\left(\partial_{x^0}+\partial_{x^3}\right)$.
\item $\arcsinh(x)$ expresses the hyperbolic arcsine of $x$.
\item $\delta_+ (x^2) \equiv \theta(x_0)\delta(x^2)$.
\item $ R \equiv 
2GE\equiv2GE_{tot}(CM)=2G\sqrt{s}$ for an elastic scattering.
\item $\tau \equiv r^2 = x^2$ in axisymmetric scattering.
\item $\lambda$ and $\sigma$ represent gaussian widths of wave packets respectively in coordinate and momentum space.
\item Bold symbols like $\transverse{p}$, $\transverse{x}$, etc.\  usually represent transverse components of 4--vectors.
\end{itemize}



\chapter*{Acknowledgements}
I would like to thank my supervisor Dott.~Dimitri Colferai for his support given with great generosity, dedication and patience. Thanks also to Dott.~Francesco Coradeschi for enlightening discussions on both conceptual and technical aspects of 
the subject, and also for additional proofreading. Much kudos to both of them for their highly scientific attitude and the kind disposition they always showed in listening to my questions or discussion proposals.
I owe to fellow graduating students Dott.~Riccardo Iraso and Dott.~Rodolfo Panerai for valuable aid in proofreading and typesetting part of this work, always given with passionate and disinterested readiness.

\end{appendix}

\end{document}